\documentclass[format=acmsmall,review=false,screen=true]{acmart}

\usepackage[ruled,norelsize]{algorithm2e}
\usepackage{amsmath}
\usepackage{booktabs}
\usepackage{multirow}
\usepackage{listings}
\usepackage{array,longtable}
\usepackage{adjustbox}
\renewcommand{\algorithmcfname}{ALGORITHM}

\ifdefined\textln\relax\else\newcommand{\textln}[1]{#1}\fi

\newsavebox\IBoxA \newsavebox\IBoxB \newlength\IHeight
\newcommand\TwoFig[6]{
  \sbox\IBoxA{\includegraphics[width=0.5\textwidth]{#1}}
  \sbox\IBoxB{\includegraphics[width=0.5\textwidth]{#4}}%
  \ifdim\ht\IBoxA>\ht\IBoxB
    \setlength\IHeight{\ht\IBoxB}\else\setlength\IHeight{\ht\IBoxA}\fi%
  \begin{figure}[!htb]
  \minipage[t]{0.49\textwidth}\centering
  \includegraphics[height=\IHeight]{#1}
  \caption{#2}\label{#3}
  \endminipage\hfill
  \minipage[t]{0.49\textwidth}\centering
  \includegraphics[height=\IHeight]{#4}
  \caption{#5}\label{#6}
  \endminipage
  \end{figure}%
}

\newcommand\TwoFigOneCaption[4]{
  \sbox\IBoxA{\includegraphics[width=0.5\textwidth]{#1}}
  \sbox\IBoxB{\includegraphics[width=0.5\textwidth]{#2}}%
  \ifdim\ht\IBoxA>\ht\IBoxB
    \setlength\IHeight{\ht\IBoxB}\else\setlength\IHeight{\ht\IBoxA}\fi%
  \begin{figure}[!htb]
  \minipage[t]{0.49\textwidth}\centering
  \includegraphics[height=\IHeight]{#1}
  \endminipage\hfill
  \minipage[t]{0.49\textwidth}\centering
  \includegraphics[height=\IHeight]{#2}
  \endminipage
  \caption{#3}\label{#4}
  \end{figure}%
}

\acmJournal{TOPC}
\acmVolume{9}
\acmNumber{4}
\acmArticle{39}
\acmYear{2017}
\acmMonth{9}
\copyrightyear{2016}

\acmDOI{0000001.0000001}

\received{January 2017}
\received[revised]{March 2017}
\received[accepted]{April 2017}

\begin{document}

\title{Gunrock: GPU Graph Analytics}
\author{Yangzihao Wang}
\affiliation{
\institution{University of California, Davis}
\department{Computer Science}
\streetaddress{}
\city{}
\state{}
\postcode{}
\country{}
}
\author{Yuechao Pan}
\affiliation{
\institution{University of California, Davis}
\department{Electrical and Computer Engineering}
}
\author{Andrew Davidson}
\affiliation{
\institution{University of California, Davis}
\department{Electrical and Computer Engineering}
}
\author{Yuduo Wu}
\affiliation{
\institution{University of California, Davis}
\department{Electrical and Computer Engineering}
}
\author{Carl Yang}
\affiliation{
\institution{University of California, Davis}
\department{Electrical and Computer Engineering}
}
\author{Leyuan Wang}
\affiliation{
\institution{University of California, Davis}
\department{Computer Science}
}
\author{Muhammad Osama}
\affiliation{
\institution{University of California, Davis}
\department{Electrical and Computer Engineering}
}
\author{Chenshan Yuan}
\affiliation{
\institution{University of California, Davis}
\department{Electrical and Computer Engineering}
}
\author{Weitang Liu}
\affiliation{
\institution{University of California, Davis}
\department{Electrical and Computer Engineering}
}
\author{Andy T. Riffel}
\affiliation{
\institution{University of California, Davis}
\department{Electrical and Computer Engineering}
}
\author{John D. Owens}
\affiliation{
\institution{University of California, Davis}
\department{Electrical and Computer Engineering}
}

\begin{abstract}
For large-scale graph analytics on the GPU, the irregularity of data
access and control flow, and the complexity of programming GPUs, have
presented two significant challenges to developing a programmable
high-performance graph library. ``Gunrock'', our graph-processing
system designed specifically for the GPU, uses a high-level,
bulk-synchronous, data-centric abstraction focused on operations on a
vertex or edge frontier. Gunrock achieves a balance between
performance and expressiveness by coupling high performance GPU
computing primitives and optimization strategies with a high-level
programming model that allows programmers to quickly develop new graph
primitives with small code size and minimal GPU programming knowledge.
We characterize the performance of various optimization strategies and
evaluate Gunrock's overall performance on different GPU architectures
on a wide range of graph primitives that span from traversal-based
algorithms and ranking algorithms, to triangle counting and
bipartite-graph-based algorithms. The results show that on a single
GPU, Gunrock has on average at least an order of magnitude speedup
over Boost and PowerGraph, comparable performance to the fastest GPU
hardwired primitives and CPU shared-memory graph libraries such as
Ligra and Galois, and better performance than any other GPU high-level
graph library.
\end{abstract}

%
%

%
%

\maketitle

\section{Introduction}
\label{sec:intro}
Graphs are ubiquitous data structures that can represent relationships
between people (social networks), computers (the Internet), biological
and genetic interactions, and elements in unstructured meshes. Many
practical problems in social networks, physical simulations,
bioinformatics, and other applications can be modeled in their
essential form by graphs and solved with appropriate graph primitives.
Various types of such graph primitives that compute and exploit
properties of particular graphs are collectively known as graph
analytics. In the past decade, as graph problems have grown larger in
scale and become more computationally complex, the research of
parallel graph analytics has raised great interest to overcome the
computational resource and memory bandwidth limitations of single
processors. In this paper, we describe ``Gunrock,'' our graphics
processor (GPU)-based system for graph processing that delivers high
performance in computing graph analytics with its high-level,
data-centric parallel programming model. Unlike previous GPU graph
programming models that focus on sequencing computation steps, our
data-centric model's key abstraction is the \emph{frontier}, a subset
of the edges or vertices within the graph that is currently of
interest. All Gunrock operations are bulk-synchronous and manipulate
this frontier, either by computing on values within it or by computing
a new frontier from it.

At a high level, Gunrock targets graph primitives that are iterative,
convergent processes. Among the graph primitives we have implemented
and evaluated in Gunrock, we focus in this paper on breadth-first
search (BFS), single-source shortest path (SSSP), betweenness
centrality (BC), PageRank, connected components (CC), and triangle
counting (TC). Though the GPU's excellent peak throughput and energy
efficiency~\cite{Keckler:2011:GAT} have been demonstrated across many
application domains, these applications often exploit regular,
structured parallelism. The inherent irregularity of graph data
structures leads to irregularity in data access and control flow,
making an efficient implementation on GPUs a significant challenge.

Our goal with Gunrock is to deliver both performance and
programmability. Gunrock's performance is similar to customized,
complex GPU hardwired graph primitives, and its high-level programming
model allows programmers to quickly develop new graph primitives. To
do so, we must address the key challenge in a highly parallel graph
processing system: managing irregularity in work distribution. Gunrock
integrates sophisticated load-balancing and work-efficiency strategies
into its core. These strategies are hidden from the programmer; the
programmer instead expresses \emph{what} operations should be
performed on the frontier rather than \emph{how} those operations
should be performed. Programmers can assemble complex and
high-performance graph primitives from operations that manipulate the
frontier (the ``what'') without knowing the internals of the
operations (the ``how'').

Our contributions, extending those from our previous
work~\cite{Wang:2016:GAH}, are as follows:
\begin{enumerate}
\item We present a novel data-centric abstraction for graph operations
  that allows programmers to develop graph primitives at a high level
  of abstraction while delivering high performance. This abstraction,
  unlike the abstractions of previous GPU programmable frameworks, is
  able to elegantly incorporate profitable optimizations---kernel
  fusion, push-pull traversal, idempotent traversal, and priority
  queues---into the core of its implementation.
\item We design and implement a set of simple and flexible APIs that
  can express a wide range of graph processing primitives at a high
  level of abstraction (at least as simple, if not more so, than other
  programmable GPU frameworks).
\item We describe several GPU-specific optimization strategies for
  memory efficiency, load balancing, and workload management that
  together achieve high performance. All of our graph primitives
  achieve comparable performance to their hardwired counterparts and
  significantly outperform previous programmable GPU abstractions.
\item We provide a detailed experimental evaluation of our graph
  primitives with performance comparisons to several CPU and GPU
  implementations.
\end{enumerate}
Gunrock is currently available to external developers in an
open-source repository at \url{http://gunrock.github.io/}, under an
Apache 2.0 license.

\section{Related Work}
\label{sec:related}
This section discusses the research landscape of large-scale graph
analytics frameworks in four fields:
\begin{enumerate}
\item Single-node CPU-based systems, which are in common use for graph
  analytics today, but whose serial or coarse-grained-parallel
  programming models are poorly suited for a massively parallel
  processor like the GPU\@;
\item Distributed CPU-based systems, which offer scalability
  advantages over single-node systems but incur substantial
  communication cost, and whose programming models are also poorly
  suited to GPUs;
\item GPU ``hardwired,'' low-level implementations of specific graph
  primitives, which provide a proof of concept that GPU-based graph
  analytics can deliver best-in-class performance. However,
  best-of-class hardwired primitives are challenging to even the most
  skilled programmers, and their implementations do not generalize
  well to a variety of graph primitives; and
\item High-level GPU programming models for graph analytics, which
  often recapitulate CPU programming models. The best of these systems
  incorporate generalized load-balance strategies and optimized GPU
  primitives, but they generally do not compare favorably in
  performance with hardwired primitives due to the overheads inherent
  in a high-level framework and the lack of primitive-specific
  optimizations.
\end{enumerate}
\subsection{Single-node and Distributed CPU-based Systems}
\label{subsec:cpu}
Parallel graph analytics frameworks provide high-level, programmable,
high-performance abstractions. The Boost Graph Library (BGL) is among
the first efforts towards this goal, though its serial formulation and
C++ focus together make it poorly suited for a massively parallel
architecture like a GPU\@. Designed using the generic programming
paradigm, the parallel BGL~\cite{Gregor:2005:PBG} separates the
implementation of parallel algorithms from the underlying data
structures and communication mechanisms. While many BGL
implementations are specialized per algorithm, its
breadth\_first\_visit pattern (for instance) allows sharing common
operators between different graph algorithms.

There are two major framework families for CPU-based large-scale graph
processing system: Pregel and GraphLab.

Pregel~\cite{Malewicz:2010:PSL} is a Google-initiated programming
model and implementation for large-scale graph computing that follows
the BSP model. A typical application in Pregel is an iterative
convergent process consisting of global synchronization barriers
called super-steps. The computation in Pregel is vertex-centric and
based on message passing. Its programming model is good for
scalability and fault tolerance. However, standard graph algorithms in
most Pregel-like graph processing systems suffer slow convergence on
large-diameter graphs and load imbalance on scale-free graphs. Apache
Giraph is an open source implementation of Google's Pregel. It is a
popular graph computation engine in the Hadoop ecosystem initially
open-sourced by Yahoo!\@.
GraphLab~\cite{Low:2010:GAN} is an open-source large scale graph
processing library that relies on the shared memory abstraction and
the gather-apply-scatter (GAS) programming model. It allows
asynchronous computation and dynamic asynchronous scheduling. By
eliminating message-passing, its programming model isolates the
user-defined algorithm from the movement of data, and therefore is
more consistently expressive. PowerGraph~\cite{Gonzalez:2012:PDG} is
an improved version of GraphLab for power-law graphs. It supports both
BSP and asynchronous execution. For the load imbalance problem, it
uses vertex-cut to split high-degree vertices into equal degree-sized
redundant vertices. This exposes greater parallelism in natural
graphs. GraphChi~\cite{Kyrola:2012:GLG} is a centralized system that
can process massive graphs from secondary storage in a single machine.
It introduces a novel graph partitioning method called Parallel
Sliding Windows (PSW), which sorts the edges by their source node IDs
to provide load balancing.

Beyond these two families, several other graph libraries have
influenced the direction of graph analytics research, including
shared-memory-based systems and domain-specific languages.
Ligra~\cite{Shun:2013:LAL} is a CPU-based graph processing framework
for shared memory. It uses a similar operator
abstraction to Gunrock for doing graph traversal. Its lightweight
implementation is targeted at shared memory architectures and uses
CilkPlus for its multi-threading implementation.
Galois~\cite{Pingali:2011:TTO,Nguyen:2013:ALI} is a graph system for
shared memory based on a different operator abstraction that supports
priority scheduling and dynamic graphs and processes on subsets of
vertices called active elements. However, their model does not
abstract the internal details of the loop from the user. Users have to
generate the active elements set directly for different graph
algorithms.
Green-Marl~\cite{Hong:2012:GDE} is a domain-specific language for
writing graph analysis algorithms on shared memory with built-in
breadth-first search (BFS) and depth-first search (DFS) primitives in
its compiler. Its language approach provides graph-specific
optimizations and hides complexity. However, the language does not
support operations on arbitrary sets of vertices for each iteration,
which makes it difficult to use for traversal algorithms that cannot
be expressed using a BFS or DFS\@.
GraphX~\cite{Gonzalez:2014:GGP} is a distributed graph computation
framework that unifies graph-parallel and data-parallel computation.
It provides a small, core set of graph-parallel operators expressive
enough to implement the Pregel and PowerGraph abstractions, yet is
simple enough to be cast in relational algebra.
Help is a library that provides high-level primitives for large-scale
graph processing~\cite{Salihoglu:2014:HHP}. Using the primitives in
Help is more intuitive and faster than using the APIs of existing
distributed systems.

\subsection{Specialized Parallel Graph Algorithms}
\label{subsec:gpu1}
Recent work has developed numerous best-of-breed, hardwired
implementations of many graph primitives.

\paragraph{BFS} Breadth-first search is among the first few graph
primitives researchers developed on the GPU due to its representative
workload pattern and its fundamental role as the building block
primitive to several other traversal-based graph primitives. Harish
and Narayanan~\shortcite{Harish:2007:ALG} first proposed a quadratic
GPU BFS implementation that maps each vertex's neighbor list to one
thread. Hong et al.~\shortcite{Hong:2011:ACG} improved on this
algorithm by mapping workloads to a series of virtual warps and
letting an entire warp cooperatively strip-mine the corresponding
neighbor list. Merrill et al.'s linear parallelization of the BFS
algorithm on the GPU~\cite{Merrill:2012:SGG} had significant influence
in the field. They proposed an adaptive strategy for load-balancing
parallel work by expanding one node's neighbor list to one thread, one
warp, or a whole block of threads. With this strategy and a
memory-access-efficient data representation, their implementation
achieves high throughput on large scale-free graphs. Beamer et al.'s
recent work on a very fast BFS for shared memory
machines~\cite{Beamer:2012:DBS} uses a hybrid BFS that switches
between top-down and bottom-up neighbor-list-visiting algorithms
according to the size of the frontier to save redundant edge visits.
Enterprise~\cite{Liu:2015:EBG}, a GPU-based BFS system, introduces a
very efficient implementation that combines the benefits of direction
optimization, Merrill et al.'s adaptive load-balancing workload
mapping strategy, and a status-check array.
BFS-4K~\cite{Busato:2015:BAE} is a GPU BFS system that improves the
virtual warp method to a per-iteration dynamic one and uses dynamic
parallelism for better load balancing.

\paragraph{CC} Connected components can be implemented as a BFS-based
primitive. The current fastest connected-component algorithm on the
GPU is Soman et al.'s work~\shortcite{Soman:2010:AFG} based on a PRAM
connected-component algorithm~\cite{Greiner:1994:CPA} that initializes
each vertex in its own component, and merges component IDs by
traversing in the graph until no component ID changes for any vertex.

\paragraph{BC} Several parallel betweenness centrality implementations
on the GPU are based on the work from
Brandes~\shortcite{Brandes:2001:AFA}. Pande and
Bader~\shortcite{Pande:2011:CBC} proposed the first BC implementation
on the GPU using a quadratic BFS\@. Sariy\"{u}ce et
al.~\shortcite{Sariyuce:2013:BCO} adopted PowerGraph's vertex-cut
method (which they termed vertex virtualization) to improve load
balancing and proposed a stride-CSR representation to reorganize
adjacency lists for better memory coalescing. McLaughlin and
Bader~\shortcite{McLaughlin:2014:SAH} developed a work-efficient
betweenness centrality algorithm on the GPU that combines a
queue-based linear multi-source BFS-based work-efficient BC
implementation with an edge-parallel BC implementation. It dynamically
chooses between the two according to frontier size and scales linearly
up to 192 GPUs.

\paragraph{SSSP} Davidson et al.~\shortcite{Davidson:2014:WPG}
proposed a delta-stepping-based~\cite{Meyer:2003:DAP} work-efficient
single-source shortest path algorithm on the GPU that explores a
variety of parallel load-balanced graph traversal and work
organization strategies to outperform previous parallel methods that
are either based on quadratic BFS~\cite{Harish:2007:ALG} or the
Bellman-Ford algorithm (LonestarGPU 2.0~\cite{Burtscher:2012:QSI}).

\paragraph{TC} Green et al.'s GPU
triangle counting algorithm~\cite{Green:2014:FTC} computes set
intersections for each edge in the undirected graph, and uses a
modified intersection path method based on merge
path~\cite{Green:2012:GMP}, which is considered to have the highest
performance for large-sorted-array set intersection on the GPU\@.

After we discuss the Gunrock abstraction in
Section~\ref{sec:abstraction}, we will discuss how to map these
specialized graph algorithms to Gunrock and the differences in
Gunrock's implementations.

\subsection{High-level GPU Programming Models}
\label{subsec:gpu2}
Several works target the construction of a high-level GPU graph
processing library that delivers both good performance and good
programmability. We categorize these into two groups by programming
model: (1) BSP/Pregel's message-passing framework and (2) the GAS
model.

In Medusa~\cite{Zhong:2014:MSG}, Zhong and He presented their
pioneering work on parallel graph processing using a message-passing
model called Edge-Message-Vertex (EMV). It is the first high-level GPU
graph analytics system that can automatically execute different
user-defined APIs for different graph primitives, which increases the
programmability to some extent. However, its five types of
user-defined API---namely, \emph{ELIST, EDGE, MLIST, MESSAGE,
  VERTEX}---are still vertex-centric and need to be split into several
source files. Moreover, Medusa does not have a fine-grained load
balancing strategy for unevenly distributed neighbor lists during
graph traversal, which makes its performance on scale-free graphs
uncompetitive compared to specialized graph primitives.

Two more recent works that follow the message-passing approach both
improve the execution model. Totem~\cite{Gharaibeh:2014:ELS} is a
graph processing engine for GPU-CPU hybrid systems. It either
processes the workload on the CPU or transmits it to the GPU according
to a performance estimation model. Its execution model can potentially
solve the long-tail problem (where the graph has a large diameter with
a very small amount of neighbors to visit per iteration) on GPUs, and
overcome GPU memory size limitations. Totem's programming model allows
users to define two functions---\textit{algo\_compute\_func} and
\textit{msg\_combine\_func}---to apply to various graph primitives.
However, because its API only allows direct neighbor access, it has
limitations in algorithm generality. Frog~\cite{Shi:2015:OAG} is a
lock-free semi-asynchronous parallel graph processing framework with a
graph coloring model. It has a preprocessing step to color the graph
into sets of conflict-free vertices. Although vertices with the same
color can be updated lock-free in a parallel stage called color-step,
on a higher level the program still needs to process color-steps in a
BSP style. Within each color-step, their streaming execution engine is
message-passing-based with a similar API to Medusa. The preprocessing
step of the color model does accelerate the iterative convergence and
enables Frog to process graphs that do not fit in a single GPU's
memory. However, the CPU-based coloring algorithm is inefficient.
Because of the limitation of its programming model, which requires
that it visit all edges in each single iteration, its performance is
restricted.

The GAS abstraction was first applied on distributed
systems~\cite{Gonzalez:2012:PDG}. PowerGraph's vertex-cut splits large
neighbor lists, duplicates node information, and deploys each partial
neighbor list to different machines. Working as a load balancing
strategy, it replaces the large synchronization cost in edge-cut into
a single-node synchronization cost. This is a productive strategy for
multi-node implementations. GAS offers the twin benefits of simplicity
and familiarity, given its popularity in the CPU world.
VertexAPI2~\cite{Elsen:2013:AVC} is the first GPU high-level graph
analytics system that strictly follows the GAS model. It defines two
graph operators: \emph{gatherApply}, the combination of GAS's gather
step and apply step, and \emph{scatterActivate}, GAS's scatter step.
Users build different graph primitives by defining different functors
for these operators. VertexAPI2 has four types of functors:
\emph{gatherReduce}, \emph{gatherMap}, \emph{scatter}, and
\emph{apply}. Underneath their abstraction, VertexAPI2 uses
state-of-the-art GPU data primitives based on two-phase
decomposition~\cite{MGPU:2016}. It shows both better performance and
better programmability compared to message-passing-based GPU
libraries. MapGraph~\cite{Fu:2014:MAH} echoes VertexAPI2's framework
and integrates both moderngpu's load-balanced search~\cite{MGPU:2016}
and Merrill et al.'s dynamic grouping workload mapping
strategy~\cite{Merrill:2012:SGG} to increase its performance.
CuSha~\cite{Khorasani:2014:CVG} is also a GAS model-based GPU graph
analytics system. It solves the load imbalance and GPU
underutilization problem with a GPU adoption of GraphChi's PSW\@. They
call this preprocessing step ``G-Shard'' and combine it with a
concatenated window method to group edges from the same source IDs.

nvGRAPH\footnote{nvGRAPH is available at
  \url{https://developer.nvidia.com/nvgraph}.} is a high-performance
GPU graph analytics library developed by NVIDIA\@. It views graph
analytics problems from the perspective of linear algebra and matrix
computations~\cite{Kepner:2016:MFO}, and uses semi-ring SpMV
operations to express graph computation. It currently supports three
algorithms: PageRank, SSSP, and Single Source Widest Path.

Table~\ref{tab:gpulib} compares Gunrock with the above high-level
GPU-based graph analytics systems on various metrics. Our novel
data-centric programming model and its efficient implementation makes
Gunrock the only high-level GPU-based graph analytics system with
support for both vertex-centric and edge-centric operations, as well
as runtime fine-grained load balancing strategies, without requiring
any preprocessing of input datasets.

\begin{table}
  \resizebox{\columnwidth}{!}{
  \small
  \centering
  \begin{tabular}{*{8}{c}} \toprule Metrics &Medusa&Totem&Frog& VertexAPI2& MapGraph&CuSha&Gunrock \\
    \midrule
    programming model & m-p & m-p & m-p & GAS & GAS & GAS & data-centric
    \\ operator flexibility & v-c, e-c & v-c & v-c & v-c & v-c & v-c & v-c, e-c
    \\ load balancing & no & limited & limited & fine-grained & fine-grained & fine-grained & fine-grained
    \\ preprocessing & no & no & coloring & no & no & G-shard & no
    \\ execution model & BSP & BSP,hybrid & semi-async & BSP & BSP & BSP & BSP
    \\ \bottomrule
  \end{tabular}
  }
  \caption[GPU graph analytics system comparison table.]{Detailed
    comparison of different high-level GPU graph analytics systems.
    m-p means message-passing based model, v-c and e-c mean
    vertex-centric and edge-centric respectively. Note that part of
    load balancing work in Frog and CuSha are done offline during the
    coloring model and G-shard generation process respectively. This
    table focuses on graph-centric abstractions, so we leave out
    libraries that follow a linear algebra abstraction such as
    nvGRAPH\@.\label{tab:gpulib}}
\end{table}

From a programmability perspective, Gunrock aims to 1)~define a
programming model that can abstract most common operations in graph
analytics at a high level; 2)~be flexible enough to express various
types of graph analytics tasks; and 3)~match the GPU's
high-throughput, data-centric, massively parallel execution model,
balancing on the generality of the model with its performance.

From a performance perspective, Gunrock attempts to 1)~build its
low-level optimizations on top of the state-of-the-art basic data
parallel primitives on the GPU and 2)~design optimizations to allow
the usage of different combinations and parameter-tuning methods in
our graph operations. Together these goals enable performance
that is comparable to specialized GPU implementations.

\section{Data-Centric Abstraction}
\label{sec:abstraction}
A common finding from most CPU and GPU graph analytics systems is that
most graph analytics tasks can be expressed as iterative convergent
processes. By ``iterative,'' we mean operations that may require
running a series of steps repeatedly; by ``convergent,'' we mean that
these iterations allow us to approach the correct answer and terminate
when that answer is reached. Mathematically, an iterative convergent
process is a series of operations $f_0, f_1, ..., f_{n-1}$ that
operate on graph data $G$, where $G_i=f(G_{i-1})$, until at iteration
$i$, the stop condition function $C(G_i, i)$ returns true.

\begin{figure}
  \centering
  \includegraphics[width=0.9\textwidth]{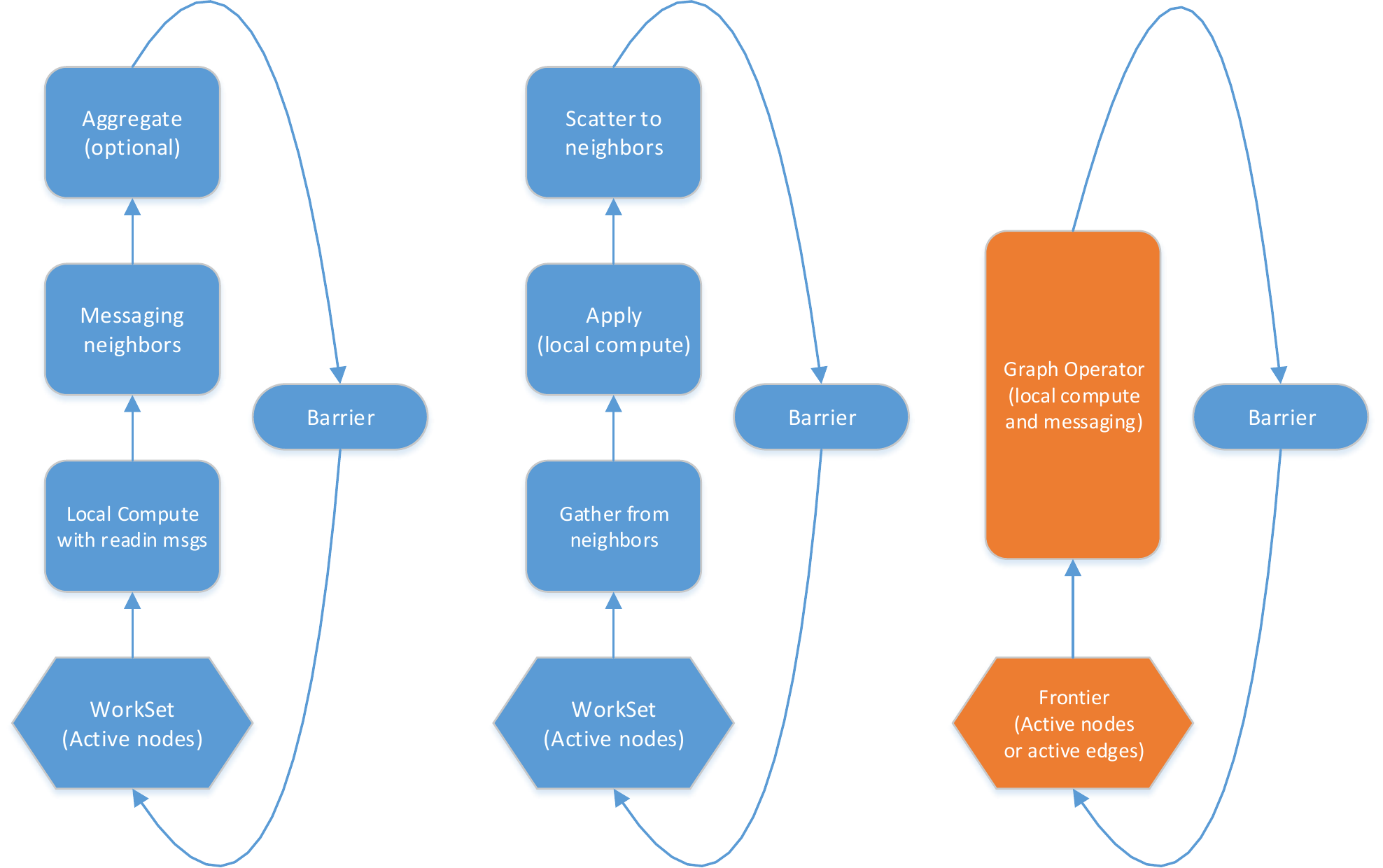}
  \centering
  \caption[Iterative convergence process illustration.]{Iterative
    convergence process presented using Pregel's message passing
    model, PowerGraph's GAS model, and Gunrock's data-centric model.}
  \label{fig:icp}
\end{figure}

Both Pregel and PowerGraph focus on sequencing steps of
\emph{computation}. Where Gunrock differs from them and their variants
is our abstraction. Rather than focusing on sequencing steps of
\emph{computation}, we instead focus on manipulating a data structure,
the \emph{frontier} of vertices or edges. The frontier represents the
subset of vertices or edges that is actively participating in the
computation. Gunrock's data-centric framework design not only provides
the features of other frameworks but also provides high performance.
It is flexible enough to be expanded by new graph operators, as long
as they operate on one or more input frontiers with graph data and
generate one or more output frontiers. Because of this design, we
claim that thinking about graph processing in terms of manipulations
of frontiers is the right abstraction for the GPU\@. We support this
statement qualitatively in this section and quantitatively in
section~\ref{sec:perf}.

One important consequence of designing our abstraction with a
data-centered focus is that Gunrock, from its very beginning, has
supported both node and edge frontiers, and can easily switch between
them within the same graph primitive. We can, for instance, generate a
new frontier of neighboring edges from an existing frontier of
vertices. In contrast, gather-apply-scatter (PowerGraph) and
message-passing (Pregel) abstractions are focused on operations on
vertices and either cannot support edge-centric operations or could
only support them with heavy redundancy within their abstractions.

In our abstraction, we expose bulk-synchronous ``steps''
that manipulate the frontier, and programmers build graph primitives
from a sequence of steps. Different steps may have dependencies
between them, but individual operations within a step can be processed
in parallel. For instance, a computation on each vertex within the
frontier can be parallelized across vertices, and updating the
frontier by identifying all the vertices neighboring the current
frontier can also be parallelized across vertices.

\begin{figure}
  \centering
  \includegraphics[width=0.9\textwidth]{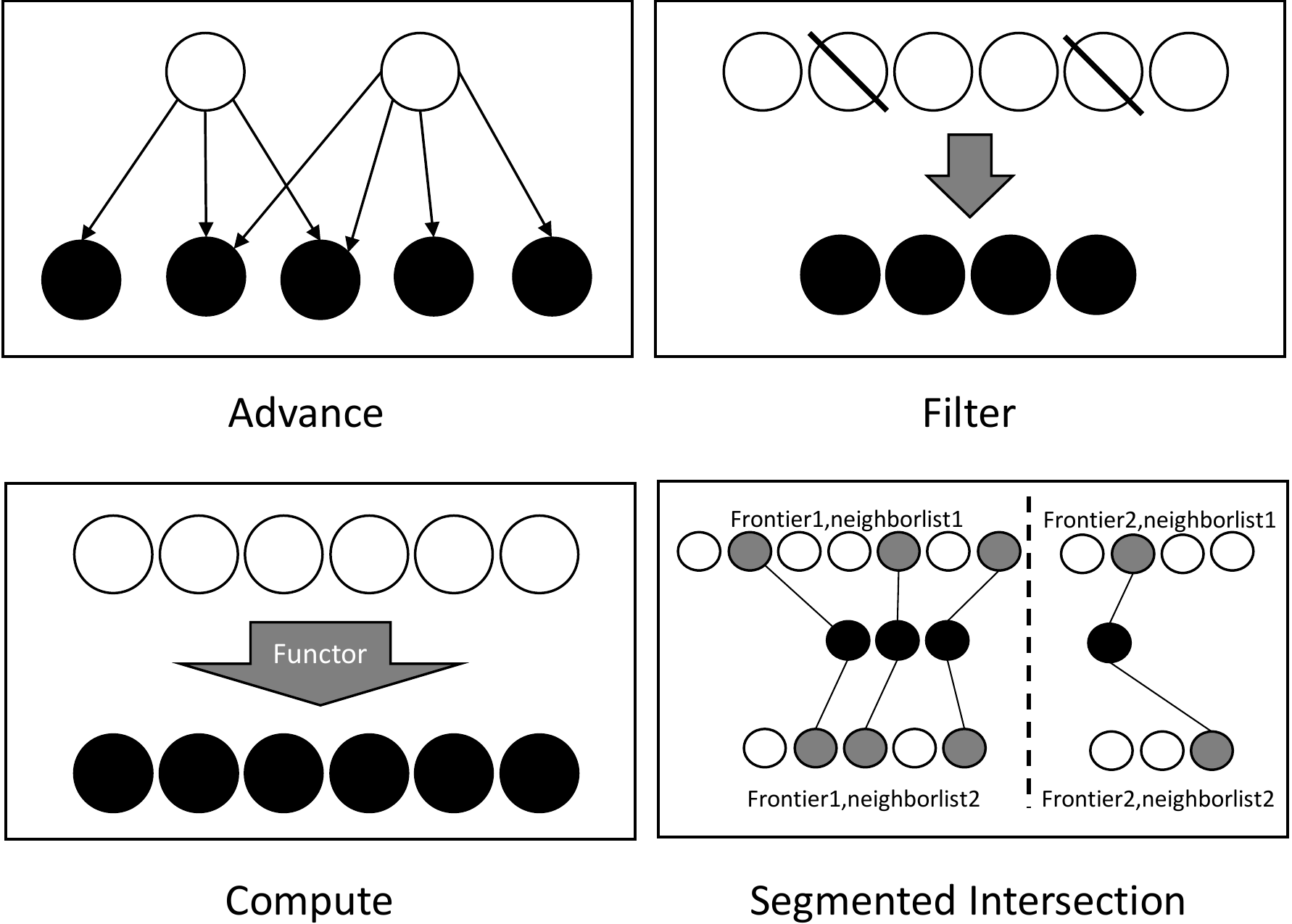}
  \centering
  \caption[Four operators in Gunrock's data-centric abstraction.]{Four
    operators in Gunrock's data-centric abstraction convert a current
    frontier (in white) into a new frontier (in black).}
  \label{fig:data-centric}
\end{figure}

The graph primitives we describe in this paper use three traversal
operators: advance, filter, and segmented intersection. They may also
use one compute operator, which is often fused with one of the
traversal operators (Figure~\ref{fig:data-centric}). Each graph
operator manipulates the frontier in a different way. The input
frontier of each operator contains either node IDs or edge IDs that
specify on which part of the graph we are going to perform our
computations. The traversal operators traverse the graph and generate
an output frontier. Within a traversal operator, each input item can
potentially map to zero, one, or more output items; efficiently
handling this irregularity is the principal challenge in our
implementation. In this section, we discuss the functionality of the
operators, then discuss how we implement them in
Section~\ref{sec:imp_opt}.

\begin{description}
\item[Advance] An \emph{advance} operator generates a new frontier
  from the current frontier by visiting the neighbors of the current
  frontier. Each input item maps to multiple output items from the
  input item's neighbor list. A frontier can consist of either
  vertices or edges, and an advance step can input and output either
  kind of frontier. Advance is an irregularly-parallel operation for
  two reasons: 1)~different vertices in a graph have different numbers
  of neighbors and 2)~vertices share neighbors. An efficient advance
  is the most significant challenge of a GPU implementation.

  The generality of Gunrock's advance allows us to use the same
  advance implementation across a wide variety of interesting graph
  operations. According to the type of input frontier and output
  frontier, Gunrock supports 4 kinds of advance: V-to-V, V-to-E,
  E-to-V, and E-to-E, where E represents an edge frontier and V
  represents a vertex frontier. For instance, we can utilize Gunrock
  advance operators to 1)~visit each element in the current frontier
  while updating local values and/or accumulating global values (e.g.,
  BFS distance updates); 2)~visit the node or edge neighbors of all
  the elements in the current frontier while updating source vertex,
  destination vertex, and/or edge values (e.g., distance updates in
  SSSP); 3)~generate edge frontiers from vertex frontiers or vice
  versa (e.g., BFS, SSSP, SALSA, etc.); or 4)~pull values from all
  vertices 2 hops away by starting from an edge frontier, visiting all
  the neighbor edges, and returning the far-end vertices of these
  neighbor edges.

\item[Filter] A \emph{filter} operator generates a new frontier from
  the current frontier by choosing a subset of the current frontier
  based on programmer-specified criteria. Each input item maps to zero
  or one output item. Though filtering is an irregular operation,
  using parallel scan for efficient filtering is well-understood on
  GPUs. Gunrock's filters can either 1)~split vertices or edges based
  on a filter (e.g., SSSP's 2-bucket delta-stepping), or 2)~compact
  out filtered items to throw them away (e.g., duplicated vertices in
  BFS or edges where both end nodes belong to the same component in
  CC).

\item[Segmented Intersection] A
  \emph{segmented intersection} operator takes two input frontiers
  with the same length, or an edge frontier, and generates both the
  number of total intersections and the intersected node IDs as the
  new frontier. We call input items with the same index in two input
  frontiers a pair. If the input is an edge frontier, we treat each
  edge's two nodes as an input item pair. One input item pair maps to
  multiple output items, which are the intersection of the neighbor
  lists of two input items. Segmented intersection is the key operator
  in TC\@. It is useful for both global triangle counting and
  computing the clustering coefficient. The output frontier could be
  combined with the two input frontiers to enumerate all the triangles
  in the graph.

\item[Compute] A \emph{compute} operator defines an operation on all
  elements (vertices or edges) in its input frontier. A
  programmer-specified compute operator can be used together with all
  three traversal operators. Gunrock performs that operation in
  parallel across all elements without regard to order. The
  computation can access and modify global memory through a device
  pointer of a structure of array that may store per-node and/or
  per-edge data. Because this parallelism is regular, computation is
  straightforward to parallelize in a GPU implementation. Many simple
  graph primitives (e.g., computing the degree distribution of a
  graph) can be expressed with a single Gunrock computation operator.
\end{description}

\noindent
Gunrock primitives are assembled from a sequence of these four
operators, which are executed sequentially: one step completes all of
its operations before the next step begins. Typically, Gunrock graph
primitives run to convergence, which on Gunrock usually equates to an
empty frontier; as individual elements in the current frontier reach
convergence, they can be filtered out of the frontier. Programmers can
also use other convergence criteria such as a maximum number of
iterations or volatile flag values that can be set in a computation
step.

\paragraph{Example: Expressing SSSP in programmable GPU frameworks}
Now we use an example to show how different programmable CPU/GPU
frameworks express a graph primitive to further study the key
difference between Gunrock's data-centric abstraction and other
frameworks. We choose SSSP because it is a reasonably complex graph
primitive that computes the shortest path from a single node in a
graph to every other node in the graph. We assume weights between
nodes are all non-negative, which permits the use of Dijkstra's
algorithm and its parallel variants. Efficiently implementing SSSP
continues to be an interesting problem in the GPU world
~\cite{Burtscher:2012:AQS,Davidson:2014:WPG,Delling:2010:PHS}.

The iteration starts with an input frontier of active vertices (or a
single vertex) initialized to a distance of zero. First, SSSP
enumerates the sizes of the frontier's neighbor list of edges and
computes the length of the output frontier. Because the neighbor edges
are unequally distributed among the frontier's vertices, SSSP next
redistributes the workload across parallel threads. This can be
expressed within an advance operator. In the final step of the advance
operator, each edge adds its weight to the distance value at its
source value and, if appropriate, updates the distance value of its
destination vertex. Finally, SSSP removes redundant vertex IDs
(specific filter), decides which updated vertices are valid in the new
frontier, and computes the new frontier for the next iteration.

  \begin{algorithm}
    \SetKwProg{Fn}{Function}{}{}
    \Fn{Set\_Problem\_Data\,$(G,P,root)$}{
    Set $P.labels$ to $\infty$\;
    Set $P.preds$ to -1\;
    $P.labels$[$root$] $\leftarrow$ 0\;
    $P.preds$[$root$] $\leftarrow$ -1\;
    Insert root to $P.frontier$\;
}
\Fn{Update\_Label\,$(s\_id,d\_id,e\_id,P)$}{
    $new\_label$ $\leftarrow$ $P.labels$[$s\_id$]+$P.weights$[$e\_id$]\;
    \KwRet{$new\_label$ $<$ atomicMin($P.labels$[$d\_id$],$new\_label$)}\;
}
\Fn{Set\_Pred\,$(s\_id,d\_id,P)$}{
    $P.preds$[$d\_id$] $\leftarrow$ $s\_id$\;
    $P.output\_queue\_ids$[$d\_id$] $\leftarrow$ $output\_queue\_id$\;
}
\Fn{Remove\_Redundant\,$(node\_id,P)$}{
    \KwRet{$P.output\_queue\_id$[$node\_id$] == $output\_queue\_id$}\;
}
\Fn{SSSP\_Enactor\,$(G,P,root)$}{
  $\textit{Set\_Problem\_Data}\,(G,P,\textit{root})$\;
  \While{$\textit{P.frontier.Size}\,() > 0$}{
    $\textit{Advance}\,(G,P,\textit{Update\_Label},\textit{Set\_Pred})$\;
    $\textit{Filter}\,(G,P,\textit{Remove\_Redundant})$\;
    $\textit{PriorityQueue}\,(G,P)$\;
  }
}
\caption{Single-Source Shortest Path, expressed in Gunrock's
  abstraction.\label{alg:ssspcode}}
\end{algorithm}

Gunrock maps one SSSP iteration onto three Gunrock operators: (1)
\emph{advance}, which computes the list of edges connected to the
current vertex frontier and (transparently) load-balances their
execution; (2) \emph{compute}, to update neighboring vertices with new
distances; and (3) \emph{filter}, to generate the final output
frontier by removing redundant nodes, optionally using a 2-level
priority queue, whose use enables delta-stepping (a binning strategy
to reduce overall workload~\cite{Davidson:2014:WPG,Meyer:2003:DAP}).
With this mapping in place, the traversal and computation of path
distances is simple and intuitively described, and Gunrock is able to
create an efficient implementation that fully utilizes the GPU's
computing resources in a load-balanced way.

\begin{figure*}
    \centering
    \includegraphics[width=0.9\textwidth]{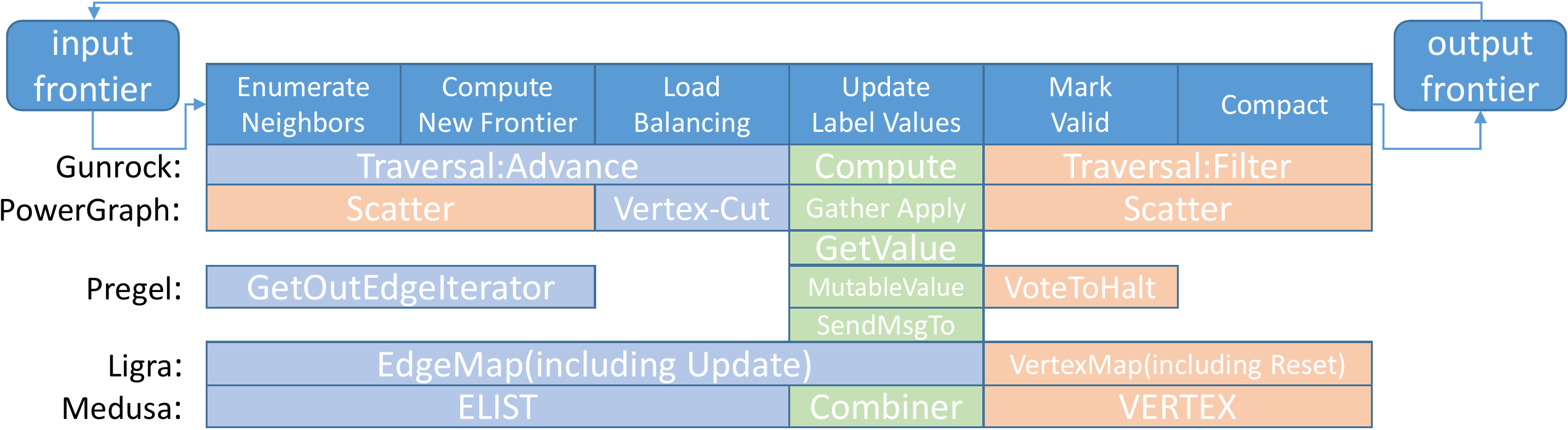}
    \centering
    \caption{Operations that make up one iteration of SSSP and their
      mapping to Gunrock, PowerGraph
      (GAS)~\protect\cite{Gonzalez:2012:PDG},
      Pregel~\protect\cite{Malewicz:2010:PSL},
      Ligra~\protect\cite{Shun:2013:LAL}, and
      Medusa~\protect\cite{Zhong:2014:MSG} abstractions.}
    \label{fig:abstraction}
\end{figure*}

\subsection{Alternative Abstractions}
In this section we discuss Gunrock's design choices compared to
several alternative abstractions designed for graph processing on
various architectures.

\begin{description}
\item[Gather-apply-scatter (GAS) abstraction] Recently, Wu et
  al.~\cite{Wu:2015:PCF} compared Gunrock vs.\ two GPU GAS frameworks,
  VertexAPI2 and MapGraph, demonstrating that Gunrock had appreciable
  performance advantages over the other two frameworks. One of the
  principal performance differences they identified comes from the
  significant fragmentation of GAS programs across many kernels that
  we discuss in more detail in section~\ref{sec:imp_opt}. Applying
  automatic kernel fusion~\cite{Pai:2016:ACT} to GAS+GPU
  implementations could potentially help close their performance gap.

  At a more fundamental level, we found that a compute-focused
  programming model like GAS was not flexible enough to manipulate the
  core frontier data structures in a way that enabled powerful
  features and optimizations such as direction-optimizing traversal
  and two-level priority queues; both fit naturally into Gunrock's
  abstraction. We believe bulk-synchronous operations on frontiers are
  a better fit than GAS for forward-looking GPU graph programming
  frameworks.

\item[Message-passing] Pregel~\cite{Malewicz:2010:PSL} is a
  vertex-centric programming model that only provides data parallelism
  on vertices. For graphs with significant variance in vertex degree
  (e.g., power-law graphs), this would cause severe load imbalance on
  GPUs. The traversal operator in Pregel is general enough to apply to
  a wide range of graph primitives, but its vertex-centric design only
  achieves good parallelism when nodes in the graph have small and
  evenly-distributed neighborhoods. For real-world graphs that often
  have an uneven distribution of node degrees, Pregel suffers from
  severe load imbalance. The Medusa authors note the complexity of
  managing the storage and buffering of these messages, and the
  difficulty of load-balancing when using segmented reduction for
  per-edge computation. Though they address both of these challenges
  in their work, the overhead of \emph{any} management of messages is
  a significant contributor to runtime. Gunrock prefers the less
  costly direct communication between primitives and supports both
  push-based (scatter) communication and pull-based (gather)
  communication during traversal steps.

\item[CPU strategies] Ligra's powerful load-balancing strategy is
  based on CilkPlus, a fine-grained task-parallel library for CPUs.
  Despite promising GPU research efforts on task
  parallelism~\cite{Cederman:2008:ODL,Tzeng:2012:AGT}, no such
  equivalent is available on GPUs, thus we implement our own
  load-balancing strategies within Gunrock. Galois, like Gunrock,
  cleanly separates data structures from computation; their key
  abstractions are ordered and unordered set iterators that can add
  elements to sets during execution (such a dynamic data structure is
  a significant research challenge on GPUs). Galois also benefits from
  speculative parallel execution whose GPU implementation would also
  present a significant challenge. Both Ligra and Galois scale well
  within a node through inter-CPU shared memory; inter-GPU
  scalability, both due to higher latency and a lack of hardware
  support, is a much more manual, complex process.

\item[Help's Primitives] Help~\cite{Salihoglu:2014:HHP} characterizes
  graph primitives as a set of functions that enable special
  optimizations for different primitives at the cost of losing
  generality. Its Filter, Local Update of Vertices (LUV), Update
  Vertices Using One Other Vertex (UVUOV), and Aggregate Global Value
  (AGV) are all Gunrock filter operations with different computations.
  Aggregating Neighbor Values (ANV) maps to the advance operator in
  Gunrock. We also successfully implemented Form Supervertices (FS) in
  Gunrock using two filter passes, one advance pass, and several other
  GPU computing primitives (sort, reduce, and scan).

\item[Asynchronous execution] Many CPU and GPU frameworks (e.g.,
  Galois, GraphLab, and Frog) efficiently incorporate asynchronous
  execution, but the GPU's expensive synchronization or locking
  operations would make this a poor choice for Gunrock. We do recover
  some of the benefits of prioritizing execution through our two-level
  priority queue.

\end{description}

\TwoFigOneCaption{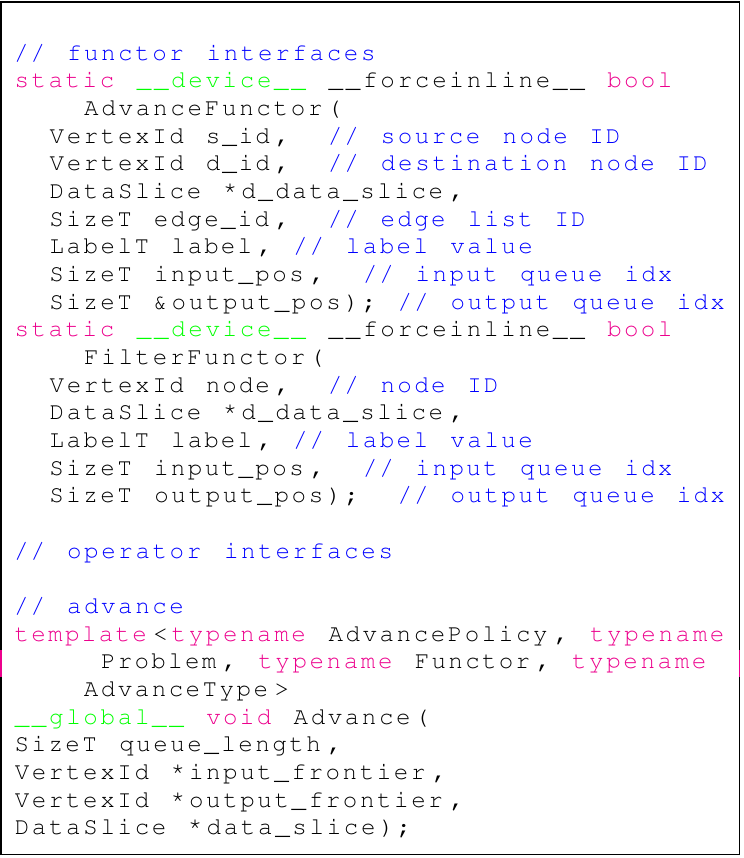}
{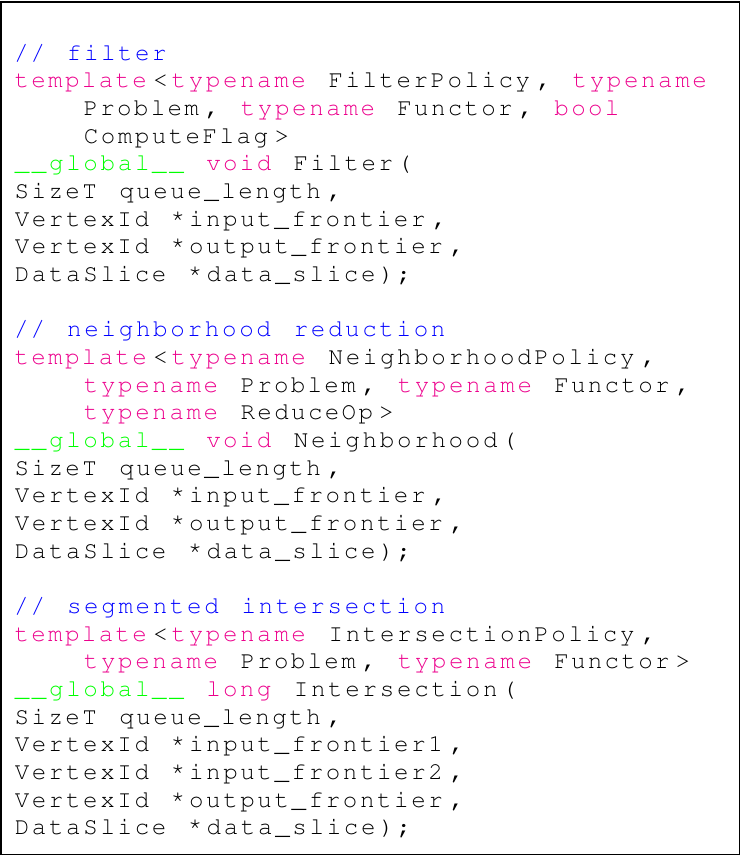}
{Gunrock's Graph Operator and Functor APIs.}
{fig:api}

Gunrock's software architecture is divided into two parts. Above the
traversal-compute abstraction is the application module. This is where
users define different graph primitives using the high-level APIs
provided by Gunrock. Under the abstraction are the utility functions,
the implementation of operators used in traversal, and various
optimization strategies.

Gunrock programs specify three components: the \emph{problem}, which
provides graph topology data and an algorithm-specific data management
interface; the \emph{functors}, which contain user-defined computation
code and expose kernel fusion opportunities that we discuss below; and
an \emph{enactor}, which serves as the entry point of the graph
algorithm and specifies the computation as a series of graph operator
kernel calls with user-defined kernel launching settings.

Given Gunrock's abstraction, the most natural way to specify Gunrock
programs would be as a sequence of bulk-synchronous steps, specified
within the enactor and implemented as kernels, that operate on
frontiers. Such an enactor is in fact the core of a Gunrock program,
but an enactor-only program would sacrifice a significant performance
opportunity. We analyzed the techniques that hardwired
(primitive-specific) GPU graph primitives used to achieve high
performance. One of their principal advantages is leveraging
producer-consumer locality between operations by integrating multiple
operations into single GPU kernels. Because adjacent kernels in CUDA
or OpenCL share no state, combining multiple logical operations into a
single kernel saves significant memory bandwidth that would otherwise
be required to write and then read intermediate values to and from
memory. In the CUDA C++ programming environment, we have no ability to
automatically fuse neighboring kernels together to achieve this
efficiency.

We summarize the interfaces for these operations in
figure~\ref{fig:api}. Our focus on kernel fusion enabled by our API
design is absent from other programmable GPU graph libraries, but it
is crucial for performance.

To conclude this section, we list the benefits of Gunrock's
data-centric programming model for graph processing on the GPU\@:

\begin{itemize}
\item The data-centric programming model allows more flexibility on
  the operations, since frontier is a higher level abstraction of
  streaming data, which could represent nodes, edges, or even
  arbitrary sub-graph structures. In Gunrock, every operator can be
  vertex-centric or edge-centric for any iteration. As a result, we
  can concentrate our effort on solving one problem---implementing
  efficient operators---and see that effort reflected in better
  performance on various graph primitives.

\item The data-centric programming model decouples a compute operator
  from traversal operators. For different graph primitives, compute
  operators can be very different in terms of complexity and the way
  they interact with graph operators. Decoupling Gunrock's compute
  operator from its traversal operators gives Gunrock more flexibility
  in implementing a general set of graph primitives.

\item The data-centric programming model allows an implementation that
  both leverages state-of-the-art data-parallel primitives and enables
  various types of optimizations. The result is an implementation with
  better performance than any other GPU graph-processing system,
  achieving comparable performance to specialized GPU graph
  algorithms.

\item The data-centric programming model uses high level graph
  operator interfaces to encapsulate complicated implementations. This
  makes Gunrock's application code smaller in size and clearer in
  logic compared to other GPU graph libraries. Gunrock's Problem class
  and kernel enactor are both template-based C++ code; Gunrock's
  functor code that specifies per-node or per-edge computation is
  C-like device code without any CUDA-specific keywords.

\item The data-centric programming model eases the job of extending
  Gunrock' single-GPU execution model to multiple GPUs. It fits with
  other execution models such as a semi-asynchronous execution model
  and a multi-GPU single-node/multi-node execution model. Our
  multi-GPU graph processing framework~\cite{Pan:2016:MGA} is built on
  top of our data-centric programming model with an unchanged core
  single-GPU implementation coupled with advanced communication and
  partition modules designed specifically for multi-GPU execution.
\end{itemize}

\section{Efficient Graph Operator Design}
\label{sec:graph_operators}
In this section, we analyze the implementation of each graph operator,
and briefly discuss how these implementations enable optimizations
(with more details in section~\ref{sec:imp_opt}).
Figure~\ref{fig:toygraph} shows the sample graph we use to illustrate
our graph operator design, and Figure~\ref{fig:toygraph_data} shows
the corresponding arrays for graph storage in compressed sparse row
(CSR) format (discussed in section~\ref{subsec:mem_optimization}) in
Gunrock.

\TwoFig{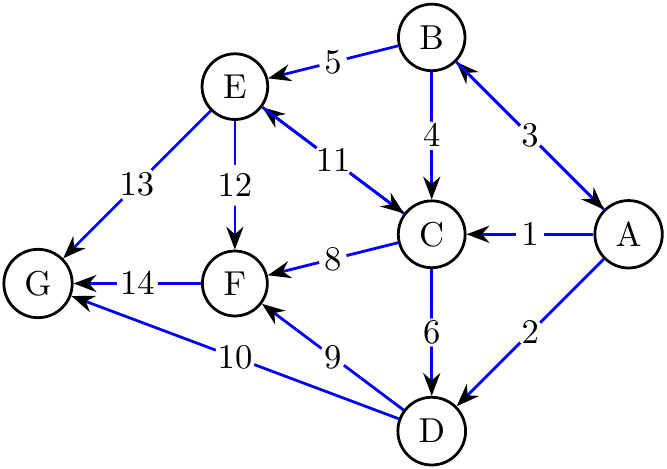}
{A sample directed graph with 7 nodes and 15 edges.}
{fig:toygraph}
{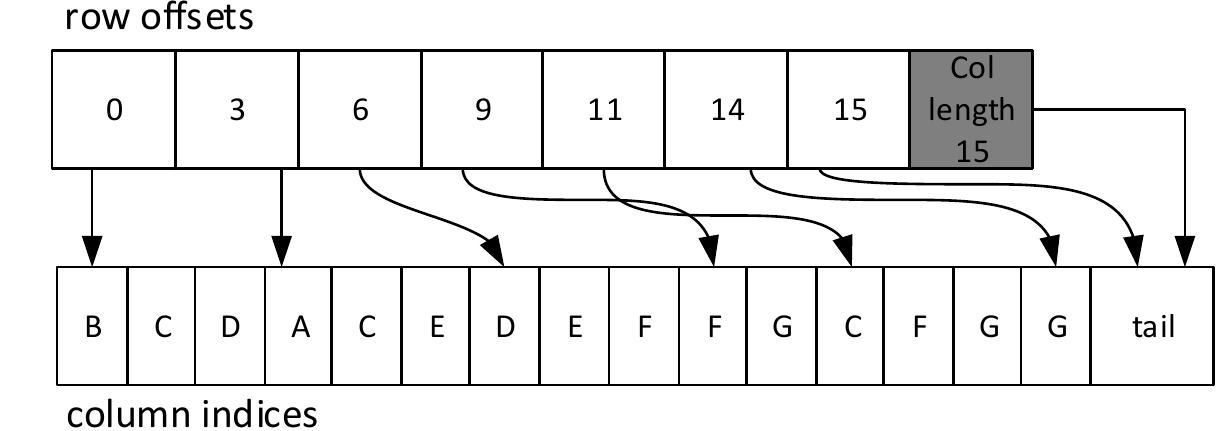}
{CSR format of sample graph in Gunrock.}
{fig:toygraph_data}

\subsection{Advance}
The advance operator takes the input frontier, visits the neighbor
list of each item in the input frontier, then writes those neighbor
lists to the output frontier. The size of these neighbor lists may
differ significantly between different input items. Thus an efficient
implementation of advance needs to reduce the task granularity to a
homogeneous size and then evenly distribute these smaller tasks among
threads~\cite{Merrill:2012:SGG} so that we can efficiently parallelize
this operator.

At a high level, advance can be seen as a vectorized device memory
assignment and copy, where parallel threads
place dynamic data (neighbor lists with various lengths) within shared
data structures (output frontier). The efficient parallelization of
this process requires two stages: 1)~for the allocation part, given a
list of allocation requirements for each input item (neighbor list
size array computed from row offsets), we need the scatter offsets to
write the output frontier; 2)~for the copy part, we need to
load-balance parallel scatter writes with various lengths over a
single launch.

The first part is typically implemented with prefix-sum. For the
second part, there are several implementation choices, which differ in
two primary ways:
\begin{description}
\item[load balancing] A coarse-grained load-balancing strategy divides
  the neighbor lists by size into multiple groups, then processes each
  group independently with a strategy that is optimized for the sizes
  in that group. A fine-grained load-balancing strategy instead
  rebalances work so that the same number of input items or the same
  number of output items are assigned to a thread or group of threads.
\item[traversal direction] A push-based advance expands the neighbor
  lists of the current input frontier; a pull-based advance instead
  intersects the neighbor lists of the unvisited node frontier with
  the current frontier.
\end{description}
We provide more details in section~\ref{sec:graph-traversal}.

\begin{figure}
  \centering
  \includegraphics[width=0.9\textwidth]{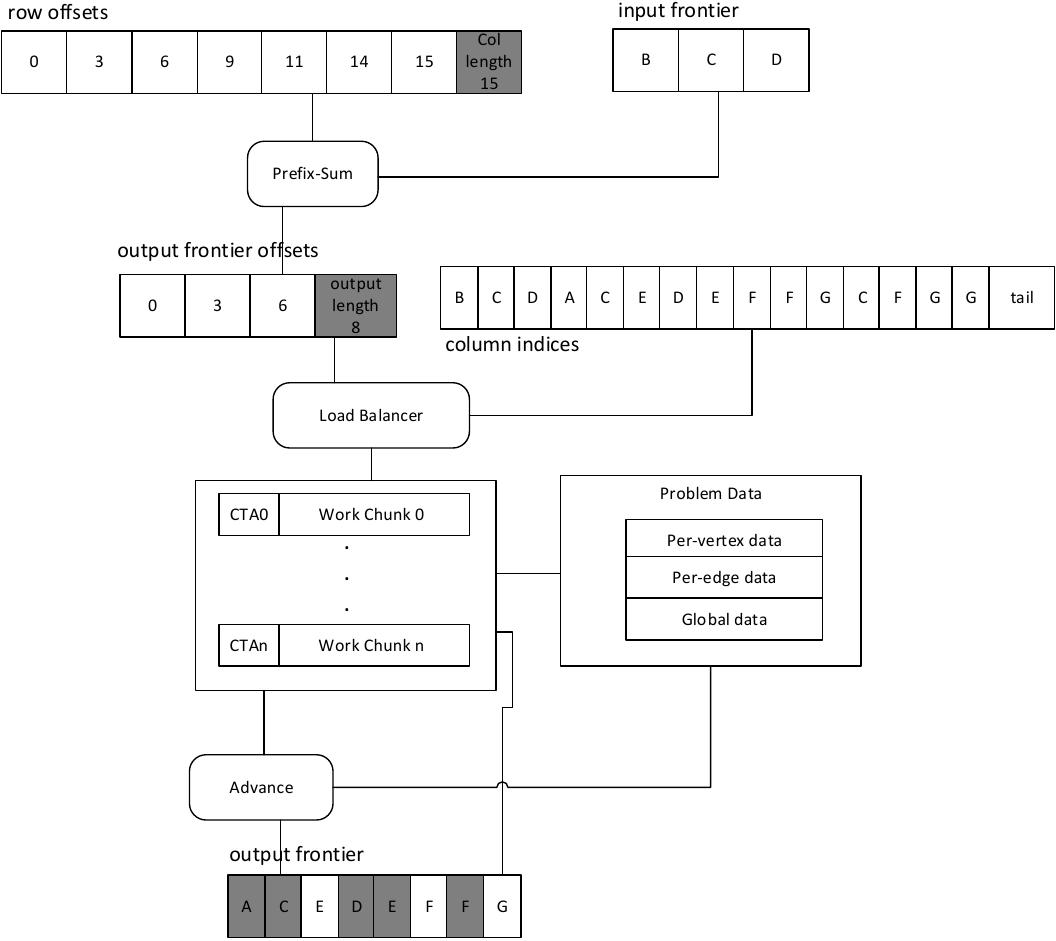}
  \centering
  \caption[Workflow of advance operator.]{Advance has a prefix-sum
    part and a load balancer part.}
  \label{fig:advance}
\end{figure}

\subsection{Filter}
Gunrock's filter operator is in essence a stream compaction operator
that transforms a sparse representation of an array (input frontier)
to a compact one (output frontier), where the sparsity comes from the
different returned values for each input item in the validity test
function. (In graph traversal, multiple redundant nodes will fail the
validity test.) Efficient stream compaction implementation also
typically use on prefix-sum; this is a well-studied
problem~\cite{Billeter:2009:ESC,Harris:2016:CUDPP}. In Gunrock, we
adopt Merrill et al.'s filtering
implementation~\cite{Merrill:2012:SGG}, which is based on local
prefix-sums with various culling heuristics. The byproduct of this
implementation is a uniquification feature that does not strictly
require the complete removal of undesired items in the input frontier.
We provide more detail in Section~\ref{sec:sync}.

\begin{figure}
  \centering
  \includegraphics[width=0.7\textwidth]{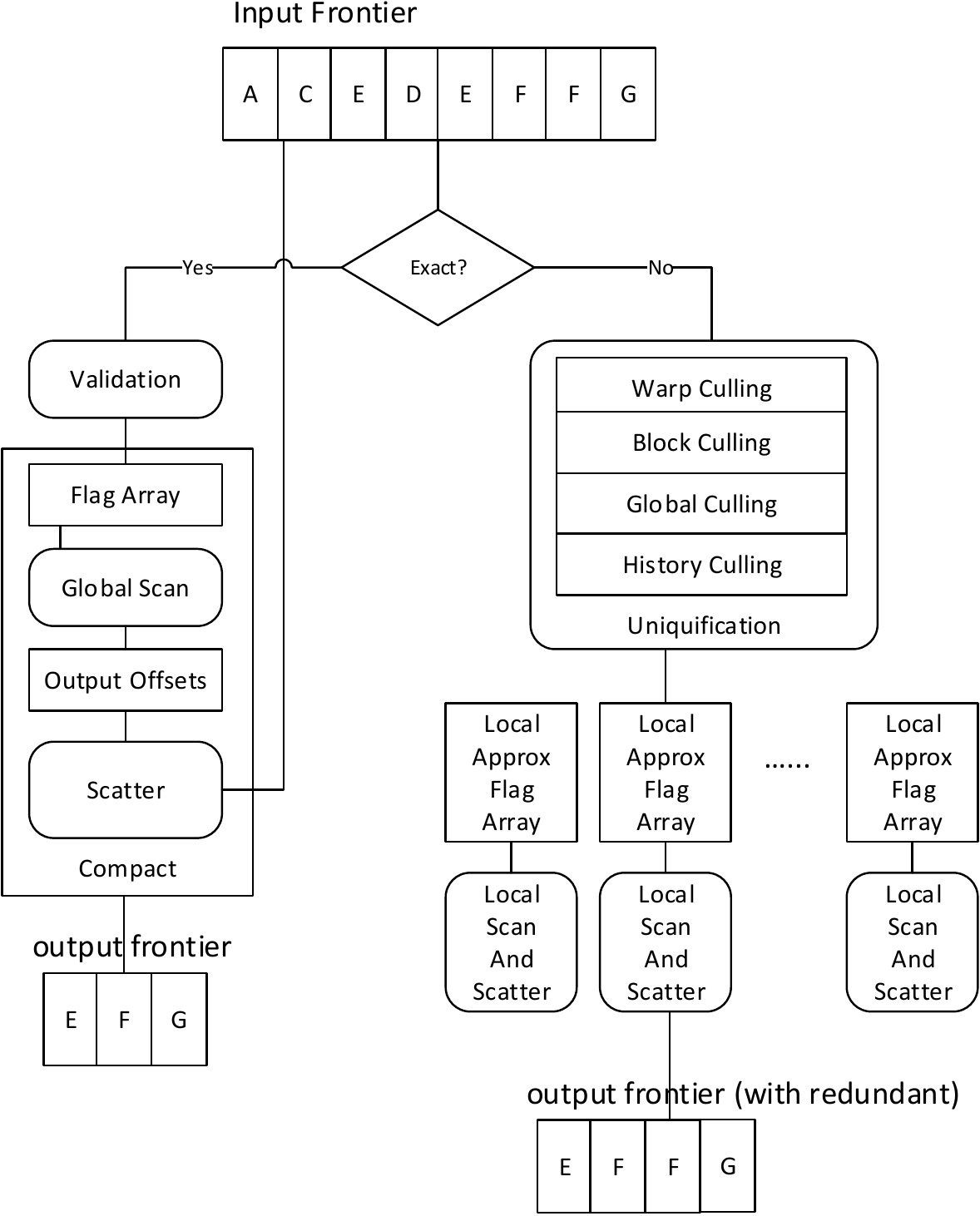}
  \centering
  \caption[Workflow of filter operator.]{Filter is based on compact,
    which uses either a global scan and scatter (for exact filtering)
    or a local scan and scatter after heuristics (for inexact
    filtering).}
  \label{fig:filter}
\end{figure}

\subsection{Segmented Intersection}
Gunrock's segmented intersection operator takes two input frontiers.
For each pair of input items, it computes the intersection of two
neighbor lists, then outputs the segmented items. It is known that for
intersection computation on two large frontiers, a modified merge-path
algorithm would achieve high performance because of its load
balance~\cite{Baxter:2013:MGM}. However, for segmented intersection,
the workload per input item pair depends on the size of each item's
neighbor list. For this reason, we still use prefix-sum for
pre-allocation, then perform a series of load-balanced intersections
according to a heuristic based on the sizes of the neighbor list
pairs. Finally, we use a stream compaction to generate the output
frontier, and a segmented reduction as well as a global reduction to
compute segmented intersection counts and the global intersection
count.

\begin{figure}
  \centering
  \includegraphics[width=0.9\textwidth]{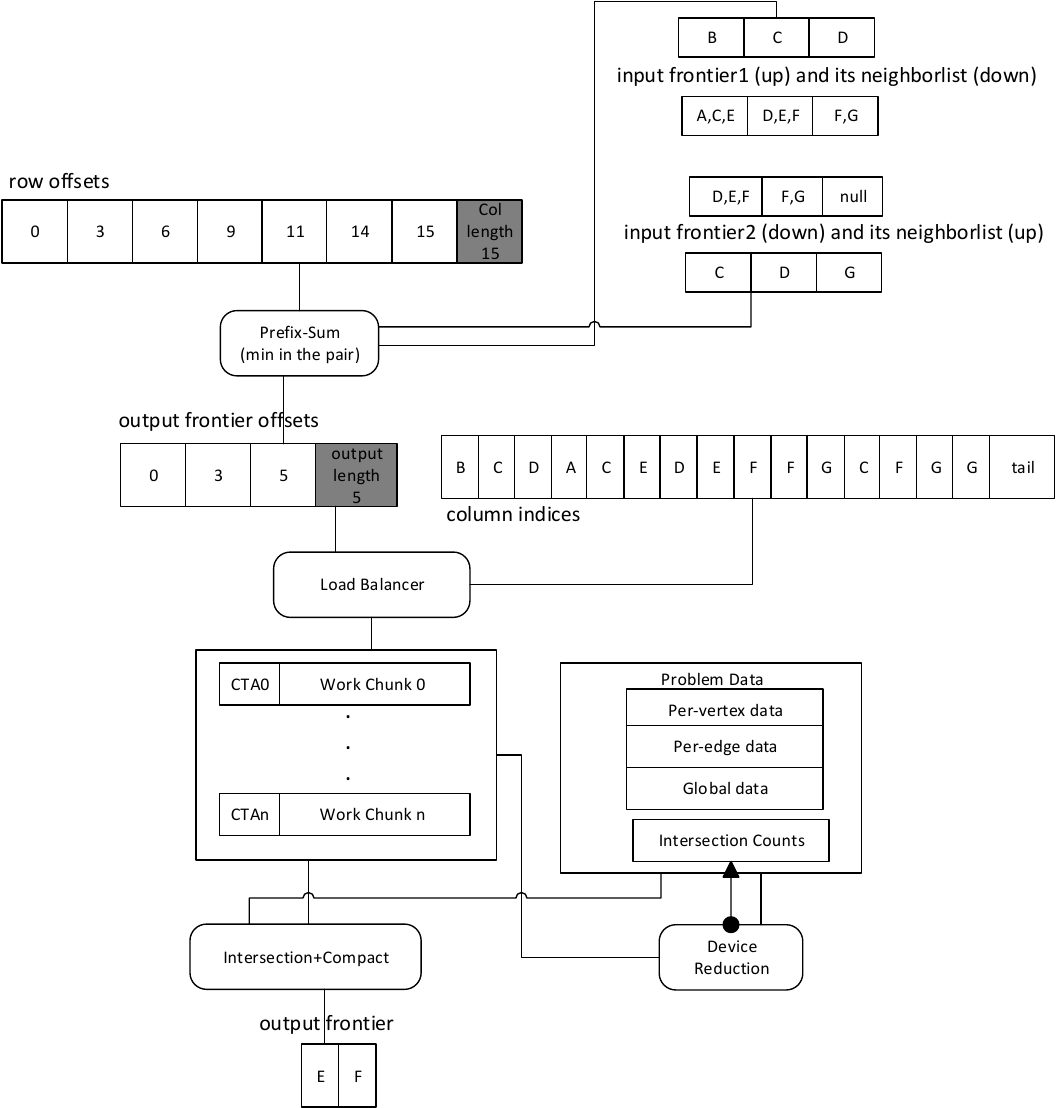}
  \centering
  \caption[Workflow of segmented intersection operator.]{Segmented
    intersection implementation that uses prefix-sum, compact,
    merge-based intersection, and reduction.}
  \label{fig:intersection}
\end{figure}

High-performance segmented intersection requires a similar focus to
high-performance graph traversal: effective load-balancing and GPU
utilization. In our implementation, we use the same dynamic grouping
strategy proposed in Merrill's BFS work~\cite{Merrill:2012:SGG}. We
divide the edge lists into two groups: 1)~small neighbor lists and
2)~one small and one large neighbor list. We implement two kernels
(Two-Small and Small-Large) that cooperatively compute intersections.
Our TwoSmall kernel uses one thread to compute the intersection of a
node pair. Our SmallLarge kernel starts a binary search for each node
in the small neighbor list on the large neighbor list. By using this
2-kernel strategy and carefully choosing a threshold value to divide
the edge list into two groups, we can process intersections with the
same level of workload together to gain load balancing and higher GPU
resource utilization. Currently, if two neighbor lists of a pair are
both large ones, we use the SmallLarge kernel. Ideally a third kernel
that utilizes all threads within a block to work on two large neighbor
lists would yield better performance.

\section{System Implementation and Optimizations}
\label{sec:imp_opt}
Choosing the right abstraction is one key component in achieving high
performance within a graph framework. The second component is
optimized implementations of the primitives within the framework. One
of the main goals in designing the Gunrock abstraction was to easily
allow integrating existing and new alternatives and optimizations into
our primitives to give more options to programmers. In general, we
have found that our data-centric abstraction, and our focus on
manipulating the frontier, have been an excellent fit for these
alternatives and optimizations, compared to a more difficult
implementation path for other GPU computation-focused abstractions. In
this section, we offer examples by discussing optimizations that help
increase the performance in four different categories:
\begin{itemize}
\item Graph traversal throughput 
\item Synchronization throughput 
\item Kernel launch throughput 
\item Memory access throughput 
\end{itemize}

\subsection{Graph Traversal Throughput Optimizations}
\label{sec:graph-traversal}

One of Gunrock's major contributions is generalizing different types
of workload-distribution and load-balance strategies. These strategies
previously only appeared in specialized GPU graph primitives. We
implement them in Gunrock's general-purpose advance operators. As a
result, any existing or new graph primitive that uses an advance
operator benefits from these strategies.

In this section, we define the workload for graph primitives as
per-edge and/or per-node computation that happens during graph
traversal. Gunrock's advance step generates an irregular workload.
Consider an advance that generates a new vertex frontier from the
neighbors of all vertices in the current frontier. If we parallelize
over input vertices, graphs with a variation in vertex degree (with
different-sized neighbor lists) will generate a corresponding
imbalance in per-vertex work. Thus, mapping the workload of each
vertex onto the GPU so that all vertex work can be processed in a
load-balanced way is essential for efficiency. Table~\ref{tab:advance}
shows our traversal throughput optimization strategies and their
corresponding module name in Gunrock implementation. Later in this
section, we summarize the pros and cons, with guidelines for usage, of
these optimizations in Table~\ref{tab:opt_pros_cons}.

\begin{table}
  \resizebox{\columnwidth}{!}{
  \small
  \centering
  \begin{tabular}{m{0.6\columnwidth} >{\arraybackslash}m{0.4\columnwidth}} \toprule Optimization Strategy & Module Name in Gunrock \\
    \midrule
    Static Workload Mapping & ThreadExpand
    \\ Dynamic Grouping Workload Mapping & TWC\_FORWARD
    \\ Merge-based Load-Balanced Partitioning Workload Mapping & LB, LB\_LIGHT, and LB\_CULL
    \\ Pull Traversal & Inverse\_Expand
    \\ \bottomrule
  \end{tabular}
  }
  \caption[Traversal throughput optimization strategies in Gunrock.]{Four graph
  traversal throughput optimization strategies and their corresponding module
  names in Gunrock implementation, where LB\_LIGHT processes load balance over
  input frontier, LB processes load balance over output frontier, and LB\_CULL
  combines LB and LB\_LIGHT with a follow-up filter into a fused
  kernel.
    \label{tab:advance}}
\end{table}

The most significant previous work in this area balances load by
cooperating between threads. Targeting BFS, Hong et
al.~\shortcite{Hong:2011:ACG} map the workload of a single vertex to a
series of virtual warps. Merrill et al.~\shortcite{Merrill:2012:SGG}
use a more flexible strategy that  maps the workload of a single
vertex to a thread, a warp, or a block, according to the size of its
neighbor list. Targeting SSSP, Davidson et
al.~\shortcite{Davidson:2014:WPG} use two load-balanced workload
mapping strategies, one that groups input work and the other that
groups output work. The first load-balances over the input
frontier, the second load-balances over the output frontier. Every
strategy has a tradeoff between computation overhead and load-balance
performance. We show in Gunrock how we integrate and redesign these
two strategies, use heuristics to choose from them to achieve the best
performance for different datasets, and generalize this optimization
to various graph operators.

\subsubsection{Static Workload Mapping Strategy}
One straightforward approach to map the workload is to map one
frontier vertex's neighbor list to one thread. Each thread loads the
neighbor list offset for its assigned node, then serially processes
edges in its neighbor list. Though this simple solution will cause
severe load imbalance for scale-free graphs with unevenly distributed
degrees and extremely small diameter, it has the significant advantage
of negligible load balancing overhead and works well for
large-diameter graphs with a relatively even degree distribution. Thus
in Gunrock, we keep this static strategy but provide several
improvements in the following aspects:
\begin{description}
\item[cooperative process] We load all the neighbor list offsets into
  shared memory, then use a block of threads to cooperatively process
  per-edge operations on the neighbor list.
\item[loop strip mining] We split the neighbor list of a node so that
  multiple threads within the same SIMD lane can achieve better
  utilization.
\end{description}
This ``ThreadExpand'' method performs better when used for
large-diameter graphs with a relatively even degree distribution since
it balances thread work within a block, but not across blocks. For
graphs with a more uneven degree distribution (e.g., scale-free social
graphs), we turn to a second strategy.

\subsubsection{Dynamic Grouping Workload Mapping Strategy}
\label{sec:opt:twc}
Gunrock uses a load balancing strategy called TWC (Thread/Warp/CTA
Expansion) based on Merrill et al's
BFS implementation~\cite{Merrill:2012:SGG} but with more flexible launch
settings and user-specific computation functor support. As a graph
operator building block, it now can be generalized to support the
traversal steps in other traversal-based graph primitives. It is
implemented to solve the performance bottleneck when ThreadExpand is
applied to frontiers with significant differences in neighbor list
sizes. Like Merrill et al., we directly address the variation in size
by grouping neighbor lists into three categories based on their size,
then individually processing each category with a strategy targeted
directly at that size. Our three sizes are 1)~lists larger than a
block, 2)~lists larger than a warp (32 threads) but smaller than a
block, and 3)~lists smaller than a warp. We begin by assigning a
subset of the frontier to a block. Within that block, each thread owns
one node. The threads that own nodes with large lists arbitrate for
control of the entire block. All the threads in the block then
cooperatively process the neighbor list of the winner's node. This
procedure continues until all nodes with large lists have been
processed. Next, all threads in each warp begin a similar procedure to
process all the nodes with medium-sized lists. Finally, the remaining
nodes are processed using our ThreadExpand method. As Merrill et al.\
noted in their paper, this strategy can guarantee a high utilization
of resource and limit various types of load imbalance such as SIMD
lane underutilization (by using per-thread mapping), intra-thread
imbalance (by using warp-wise mapping), and intra-warp imbalance (by
using block-wise mapping) (Figure~\ref{fig:workload1}).

\TwoFig{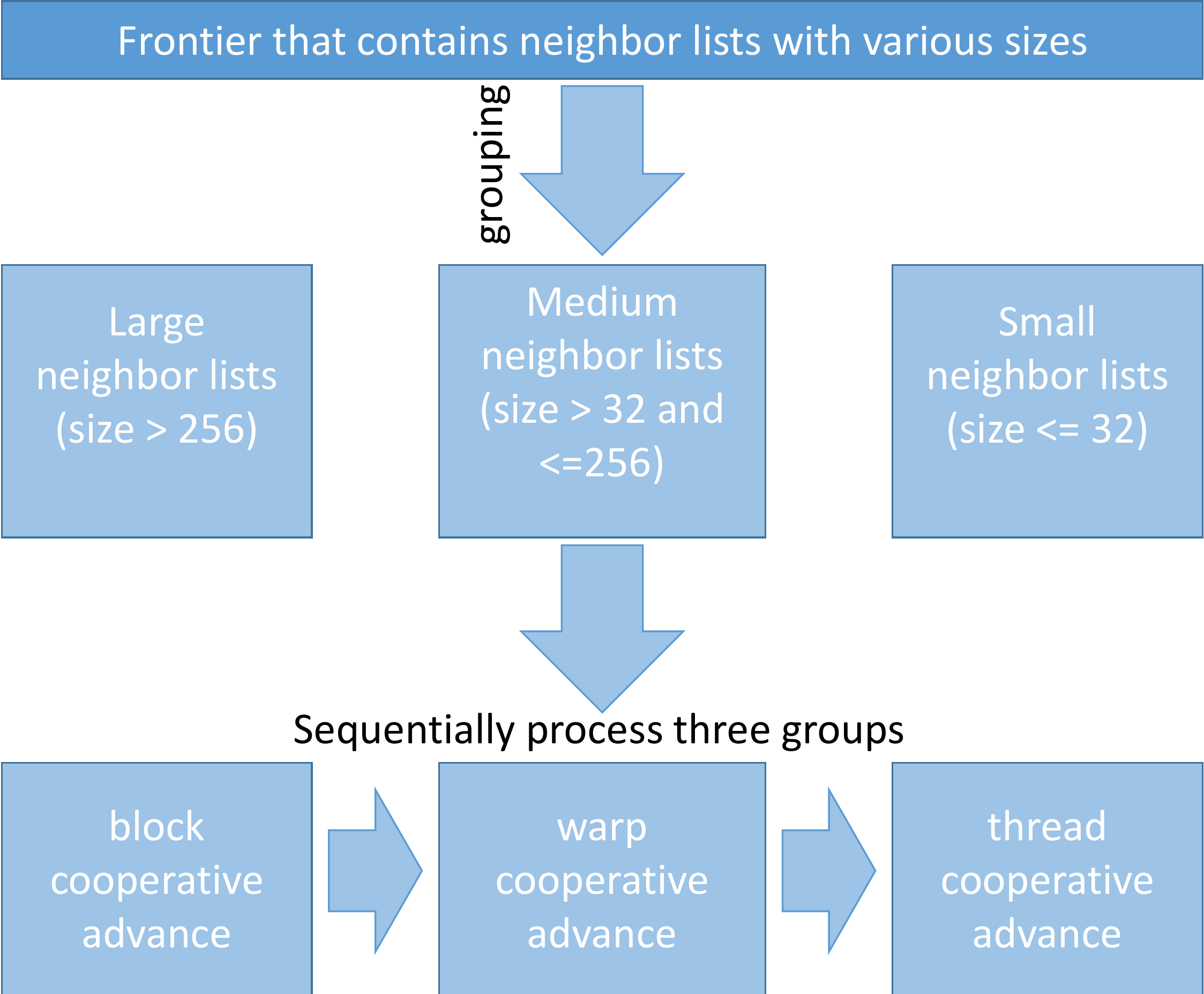}
{Dynamic grouping workload mapping strategy~\cite{Merrill:2012:SGG}.}
{fig:workload1}
{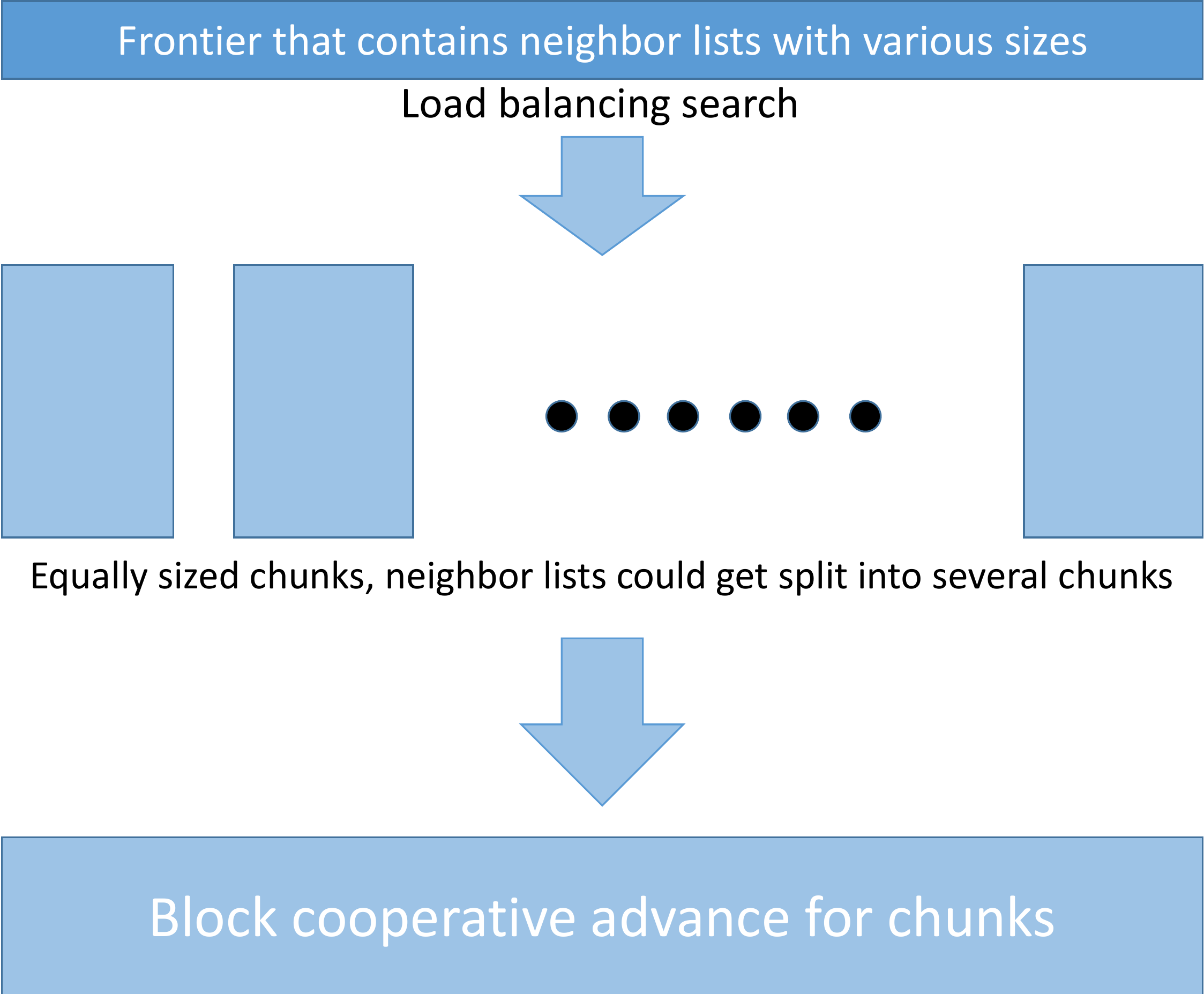}
{Merge-based load-balanced partitioning workload mapping strategy~\cite{Davidson:2014:WPG}.}
{fig:workload2}

\subsubsection{Merge-based Load-Balanced Partitioning workload mapping strategy}
\label{subsubsec:lb}
The specialization of dynamic grouping workload mapping strategy
allows higher throughput on frontiers with a high variance in degree
distribution, but at the cost of higher overhead due to the sequential
processing of the three different sizes. Also, to deal with
intra-block load imbalance, additional scheduling and work stealing
method must be applied, which further adds to the load-balancing
overhead. A global one-pass load balancing strategy would potentially
solve the intra-block load imbalance problem and thus bring better
performance. Davidson et al.\ and Gunrock improve on the dynamic
grouping workload mapping strategy by globally load-balancing over
either the input frontier or the output frontier. This introduces
load-balancing overhead, but significantly increases the traversal
throughput for scale-free graphs.
\begin{description}
\item[Input frontier load balance] maps the same number of input items
  to a block, then puts the output offset for each input item computed
  by prefix-sum into shared memory. Just as in ThreadExpand, Gunrock
  uses cooperative processing and loop strip mining here as well. All
  threads within a single block will cooperatively visit all the
  neighbor lists of the input items that belong to this block. When a
  thread starts to process a new neighbor list, it requires a binary
  search to find the corresponding source node ID\@.
\item[Output frontier load balance] first uses a global prefix-sum
  to compute all output offsets, then forms an arithmetic
  progression of $0, N, 2N, \ldots, |\textit{Input}|$ where $N$ is the
  number of edges each block processes. A global sorted
  search~\cite{MGPU:2016} of this arithmetic progression in the output
  offset array will find the starting indices for all the blocks
  within the frontier. After organizing groups of edges into
  equal-length chunks, all threads within one block cooperatively
  process edges. When we start to process a neighbor list of a new
  node, we use binary search to find the node ID for the edges that
  are going to be processed. Using this method, we ensure both
  inter-and-intra-block load-balance (Figure~\ref{fig:workload2}).
\end{description}

At a high level, Gunrock makes its load-balancing strategy decisions
depending on graph topology. We note that our load-balancing workload
mapping method performs better on social graphs with irregularly
distributed degrees, while the dynamic grouping method is superior for
graphs where most nodes have small degrees. For this reason, in
Gunrock we implement a hybrid of both methods on both vertex and edge
frontiers, using the dynamic grouping strategy for nodes with
relatively smaller neighbor lists and the load-balancing strategy for
nodes with relatively larger neighbor lists. We pick average degree as
the metric for choosing between these two strategies. When the graph
has an average degree of 5 or larger, we use the load-balancing
strategy, otherwise we use dynamic grouping strategy. Within the
load-balancing strategy, we set a static threshold. When the frontier
size is smaller than the threshold, we use coarse-grained load balance
over nodes (load balance on the input frontier), otherwise
coarse-grained load balance over edges (load balance on the output
frontier). We have found that setting this threshold to 4096 yields
consistent high performance for tests across all Gunrock-provided
graph primitives. Users can also change this value easily in the
Enactor module for their own datasets or graph primitives. Superior
load balancing is one of the most significant reasons why Gunrock
outperforms other GPU frameworks~\cite{Wu:2015:PCF}.

\subsubsection{Push vs.\ Pull Traversal}
\label{subsec:pull}
Certainly other GPU programmable graph frameworks also support an
advance step. However, because they are centered on vertex operations
on an implicit frontier, they generally only support ``push''-style
advance: the current frontier of active vertices ``pushes'' active
status to its neighbors, which creates a new frontier of newly active
vertices. Beamer et al.~\shortcite{Beamer:2012:DBS} described a
``pull''-style advance on CPUs: instead of starting with a frontier of
active vertices, pull starts with a frontier of \emph{unvisited}
vertices, filters it for vertices with neighbors in the current
frontier, and generates a new frontier with the output of the filter.

\TwoFig{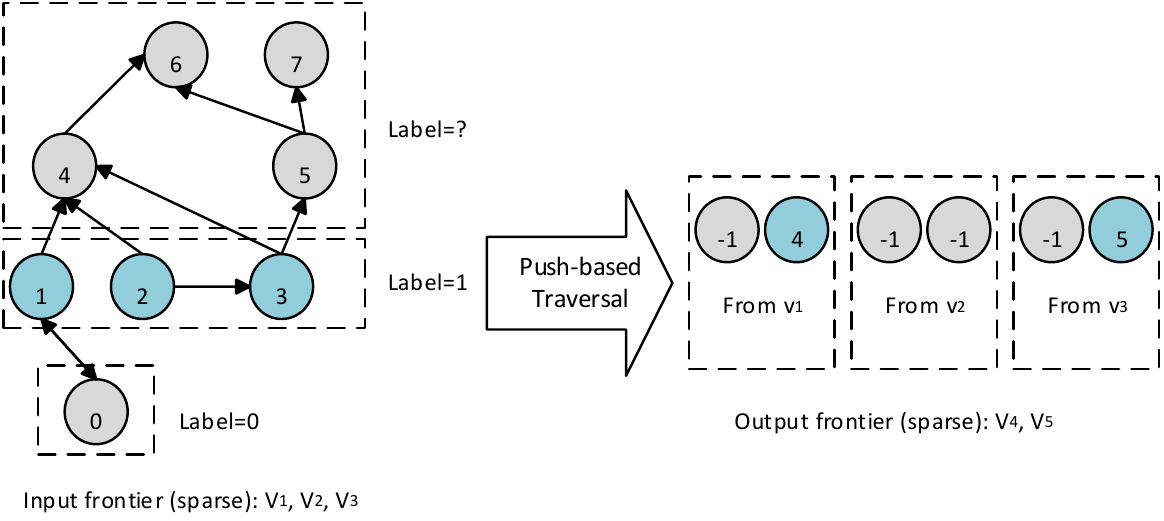}
{Push-based graph traversal.}
{fig:push}
{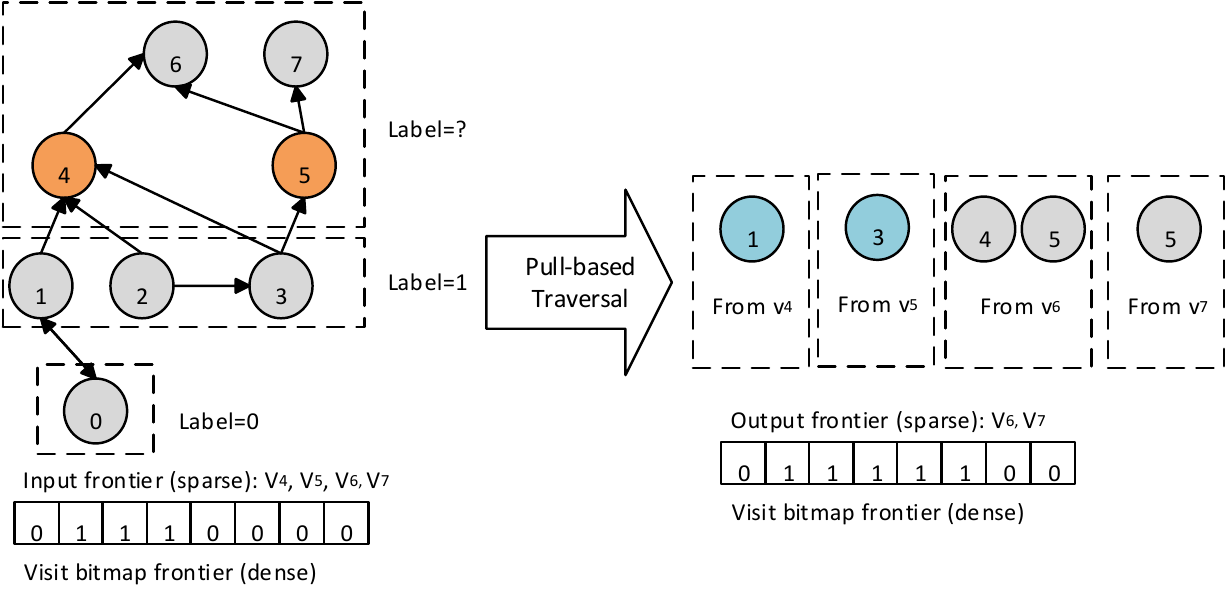}
{Pull-based graph traversal.}
{fig:pull}

Beamer et al.\ showed this approach is beneficial when the number of
unvisited vertices drops below the size of the current frontier.
Vertex-centered GPU frameworks have found it challenging to integrate
this optimization into their abstraction. Compared to them, our
data-centric abstraction is a much more natural fit, because we can
easily perform more flexible operations on frontiers. Gunrock achieves
this with two frontiers. During the ``push'' phase, it uses the active
frontier as usual in the advance step. When switching to the ``pull''
phase, it first generates a unvisited frontier with all the unvisited
nodes, Then it uses the unvisited frontier in the advance step,
visiting all unvisited nodes that have visited predecessors, and
generates both a new active frontier and a new unvisited frontier. The
capability of keeping two active frontiers differentiates Gunrock from
other GPU graph processing programming models. For better performance,
Gunrock may use per-node bitmaps to indicate whether a node has been
visited.

\begin{algorithm}
    \caption{Direction optimizing BFS}\label{alg:push_pull}
    \SetKwComment{Comment}{$\triangleright$\ }{}
  \KwIn{$G$\Comment*[r]{Csr format of graph storage}}
  \KwIn{$I\_active$\Comment*[r]{The input active frontier}}
  \KwIn{$I\_unvisited$\Comment*[r]{The input unvisited frontier}}
  \KwOut{$O\_active$\Comment*[r]{The output active frontier}}
  \KwOut{$O\_unvisited$\Comment*[r]{The output unvisited frontier}}
  \If{DirectionDecision() == push}{
    Advance(G,I\_active,O\_active)\;
    advance\_mode $\leftarrow$ push;
  }
  \Else{
    \If{advance\_mode == push}{
      I\_unvisited $\leftarrow$ GenerateUnvisitedFrontier(G.labels,V);
    }
    Reverse\_Advance(G,I\_unvisited,O\_active,O\_unvisited)\;
    advance\_mode $\leftarrow$ pull;
  }
\end{algorithm}

For pull-based traversal on the GPU, we modified Beamer et al.'s
shared-memory-CPU-based heuristics to switch between push and
pull-based traversal. Given the number of edges to check from the
frontier ($m_f$), the number of vertices in the frontier ($n_f$), the
edges to check from unexplored vertices ($m_u$), and two tuning
parameters $\alpha$ and $\beta$, Beamer et al.\ define two equations:
\begin{align}
    & m_f > \frac{m_u}{\alpha} = C_{TB} \\
    & n_f < \frac{n}{\beta} = C_{BT}
\end{align}
where $n$ is number of nodes in the graph and $C_{TB}$ and $C_{BT}$
are thresholds for the push-to-pull-based traversal switch and the
pull-to-push-based traversal switch separately. However, on the GPU,
because computing $m_f$ and $m_u$ requires two additional passes of
prefix-sum, we estimate them as follows:
\begin{align}
    & m_f = \frac{n_f \times m}{n} \\
    & m_u = \frac{n_u \times n}{n - n_u}
\end{align}
where $m$ is the number of edges in the graph, $n_f$ is the current
frontier length, and $n_u$ is the number of unvisited nodes. Instead
of directly using the size of $I\_unvisited$, we keep subtracting the
size of $I\_active$ from $n$ for each iteration to calculate $n_u$,
because $I\_unvisited$ is not available during the ``push'' phase. We
also modified the switching points:
\begin{align}
    & m_f > m_u \times do\_a = C_{TB} \\
    & m_f < m_u \times do\_b = C_{BT}
\end{align}
With proper selection of $do\_a$ and $do\_b$, our new heuristics find
the optimal iteration to switch.

With this optimization, we see a large speedup on BFS for scale-free
graphs. In an abstraction like Medusa, with its fixed method
(segmented reduction) to construct frontiers, it would be a
significant challenge to integrate a pull-based advance.

\begin{table}
  \resizebox{\columnwidth}{!}{
  \small
  \centering
  \begin{tabular}{p{0.3\columnwidth} >{\arraybackslash}m{0.7\columnwidth}} \toprule Optimization Strategy & Summary \\
    \midrule
    Static Workload Mapping & Low cost for load balancing; bad for varying
    degree distributions. Should use to traverse graphs with a
                              relatively uniform
                              edge distribution per vertex.\\
    \midrule
    Dynamic Grouping Workload Mapping & Moderate cost for load balancing; bad
    for scale-free graphs. Should use to traverse mesh-like graphs with large
    diameters when frontier size gets larger than 1 million.\\
    \midrule
    Merge-based Load-Balanced Partitioning Workload Mapping & High cost for
    load balancing on power-law graphs, but low cost for regular graphs; shows
    consistently better performance on most graphs. Should use as a default traversal strategy choice.\\
    \midrule
    Pull Traversal & Has one-time frontier conversion/preparation cost; good
    for scale-free graphs with a large amount of common neighbors. Should not use
    on regular graphs and when the number of unvisited vertices is either too
    large or too small (for more details please refer to section~\ref{perf:opt}.)
    \\ \bottomrule
  \end{tabular}}
  \caption[Pros and cons of throughput optimizations in Gunrock.]{Four graph
  traversal throughput optimization strategies, with their pros, cons,
  and guidelines for usage in Gunrock.
  \label{tab:opt_pros_cons}}
\end{table}

\subsubsection{Priority Queue}
A straightforward BSP implementation of an operation on a frontier
treats each element in the frontier equally, i.e., with the same
priority. Many graph primitives benefit from prioritizing certain
elements for computation with the expectation that computing those
elements first will save work overall (e.g., delta-stepping for
SSSP~\cite{Meyer:2003:DAP}). Gunrock generalizes the approach of
Davidson et al.~\shortcite{Davidson:2014:WPG} by allowing user-defined
priority functions to organize an output frontier into ``near'' and
``far'' slices. This allows the GPU to use a simple and
high-performance split operation to create and maintain the two
slices. Gunrock then considers only the near slice in the next
processing steps, adding any new elements that do not pass the near
criterion into the far slice, until the near slice is exhausted. We
then update the priority function and operate on the far slice.

Like other graph operators in Gunrock, constructing a priority queue
directly manipulates the frontier data structure. It is difficult to
implement such an operation in a GAS-based programming model since
that programming model has no explicit way to reorganize a frontier.

Gunrock's priority queue implementation is a modified filter operator,
which uses two stream compactions to not only form the output frontier
of input items with true flag values, but also to form a ``far'' pile
of input items with false flag values.

Currently Gunrock uses this specific optimization only in SSSP\@.
However, we believe a workload reorganization strategy based on a more
general multisplit operator~\cite{Ashkiani:2016:GM}, which maps one
input frontier to multiple output frontiers according to an arbitrary
number of priority levels, would fit nicely into Gunrock's
many-frontiers-in, many-frontiers-out data-centric programming model.
By dividing a frontier into multiple subfrontiers and making the computation
of each subfrontier not conflict with the computation of others, we can run
the computation of each subfrontier asynchronously.
This, in turn, offers more opportunity to exploit parallelism between
subfrontiers and will potentially increase the performance of
various types of algorithms, such as node ranking, community
detection, and label propagation, as well as those on graphs with
small ``long tail'' frontiers.

\subsection{Synchronization Throughput Optimization}
\label{sec:sync}
For graph processing on the GPU, the bottlenecks of synchronization
throughput come from two places:
\begin{description}
\item[concurrent discovery] In a tree traversal, there can be only one
  path from any node to any other node. However in graph traversal,
  starting from a source node, there could be several redundant visits
  if no pruning is conducted. Beamer et
  al.~\shortcite{Beamer:2012:DBS} categorize these redundant visits
  into three types: 1)~visited parents; 2)~being-visited peers; and
  3)~concurrently discovered children. Concurrent discovery of child
  nodes contributes to most synchronization overhead when there is
  per-node computation.
\item[dependencies in parallel data-primitives] Sometimes the
  computation during traversal has a reduction (Pagerank, BC) and/or
  intersection (TC) step on each neighbor list.
\end{description}
For synchronization overhead brought by concurrent discovery, the
generalized pull-and-push based traversal discussed in
section~\ref{subsec:pull} will reduce it. This section we present
another optimization that solves this problem from the perspective of
the idempotence of a graph operation.

\subsubsection{Idempotent vs.\ non-idempotent operations}
\label{subsec:idem}
Multiple elements in the frontier may share a common neighbor. This
has two consequences: 1)~it will cause an advance step to generate an
output frontier that has duplicated elements; 2)~it will cause any
computation on the common neighbors to run multiple times. The second
consequence causes a potential synchronization problem by producing
race conditions. A general way to avoid race conditions is to
introduce atomic operations. However, for some graph primitives with
``idempotent'' operations (e.g., BFS's visit status check), repeating
a computation causes no harm. In such a case, Gunrock's advance step
will avoid (costly) atomic operations, repeat the computation multiple
times, and output all redundant items to the output frontier.
Gunrock's filter step has incorporated a series of inexpensive
heuristics~\cite{Merrill:2012:SGG} to reduce, but not eliminate,
redundant entries in the output frontier. These heuristics include a
global bitmask, a block level history hashtable, and a warp level
hashtable. The sizes of each hashtable is adjustable to achieve the
optimal tradeoff between performance and redundancy-reduction rate.

\subsubsection{Atomic Avoidance Reduction Operations}
To reduce the synchronization overhead of reduction, we either
1)~reduce the atomic operations by hierarchical reduction and the
efficient use of shared memory on the GPU or 2)~assign several
neighboring edges to one thread in our dynamic grouping strategy so
that partial results within one thread can be accumulated without
atomic operations.

\subsection{Kernel Launch Throughput Optimization}
Gunrock and all other BSP-model-based GPU graph processing library
launch one or several GPU kernels per iteration, and often copy a
condition check byte from device to host after each iteration to
decide whether to terminate the program. Pai and Pingali also noted
this kernel launch overhead and recently proposed several compiler
optimizations to reduce it~\shortcite{Pai:2016:ACT}. In Gunrock, we
implemented two optimizations that target this overhead:
\begin{description}
\item[Fuse computation with graph operator] Specialized GPU
  implementations fuse regular computation steps together with more
  irregular steps like advance and filter by running a computation
  step (with regular parallelism) on the input or output of the
  irregularly-parallel step, all within the same kernel. To enable
  similar behavior in a programmable way, Gunrock exposes its
  computation steps as \emph{functors} that are integrated into all
  its graph operator kernels at compile time to achieve similar
  efficiency. We support functors that apply to \{edges, vertices\}
  and either return a Boolean value (the ``cond'' functor), useful for
  filtering, or perform a computation (the ``apply'' functor). These
  functors will then be integrated into Gunrock's graph operator
  kernel calls, which hide any complexities of how those steps are
  internally implemented.
\item[Fuse filter step with traversal operators] Several
  traversal-based graph primitives have a filter step immediately
  following an advance or neighborhood-reduction step. Gunrock
  implements a fused single-kernel traversal operator that launches
  both advance and filter steps. Such a fused kernel reduces the data
  movement between double-buffered input and output frontiers.
\end{description}

\subsection{Memory Access Throughput Optimization}
\label{subsec:mem_optimization}
For graph problems that require irregular data accesses, in addition
to exposing enough parallelism, a successful GPU implementation
benefits from the following application characteristics: 1)~coalesced
memory access, 2)~effective use of the memory hierarchy, and 3)
reducing scattered reads and writes. Our choice of graph data
structure helps us achieve these goals.

By default, Gunrock uses a compressed sparse row (CSR) sparse matrix
for vertex-centric operations. Gunrock also allows users to choose an
coordinate list (COO) representation for edge-centric operations. CSR
uses a column-indices array, $C$, to store a list of neighbor vertices
and a row-offsets array, $R$, to store the offset of the neighbor list
for each vertex. It provides compact and efficient memory access, and
allows us to use prefix-sum to reorganize sparse and uneven workloads
into dense and uniform ones in all phases of graph
processing~\cite{Merrill:2012:SGG}. In terms of data structures,
Gunrock represents all per-node and per-edge data as
structure-of-array (SOA) data structures that allow coalesced memory
accesses with minimal memory divergence. In terms of graph operators,
Gunrock implements carefully designed for loops in its kernel to
guarantee coalesced memory access. We also efficiently use shared
memory and local memory to increase memory access throughput in the
following ways:
\begin{description}
\item[In dynamic grouping workload mapping] Gunrock moves chunks of
  input frontiers into local memory, and uses warp scan and warp
  streaming compaction.
\item[In load-balanced partition workload mapping] Gunrock uses shared
  memory to store the resulting indices computed by merge-based load
  balanced search.
\item[In filter operator] Gunrock stores two types of hash tables
  (block-wise history hash tables and warp-wise history hash tables)
  in shared memory.
\end{description}

\section{Graph Applications}
\label{sec:app}
One of the principal advantages of Gunrock's abstraction is that our
advance, filter, segmented intersection and compute steps can be composed to
build new graph primitives with minimal extra work. For each primitive
below, we describe the hardwired GPU implementation to which we
compare, followed by how we express this primitive in Gunrock.
Section~\ref{sec:perf} compares the performance between hardwired and
Gunrock implementations.

\begin{figure}[htb]
  \centering
  \includegraphics[width=\columnwidth]{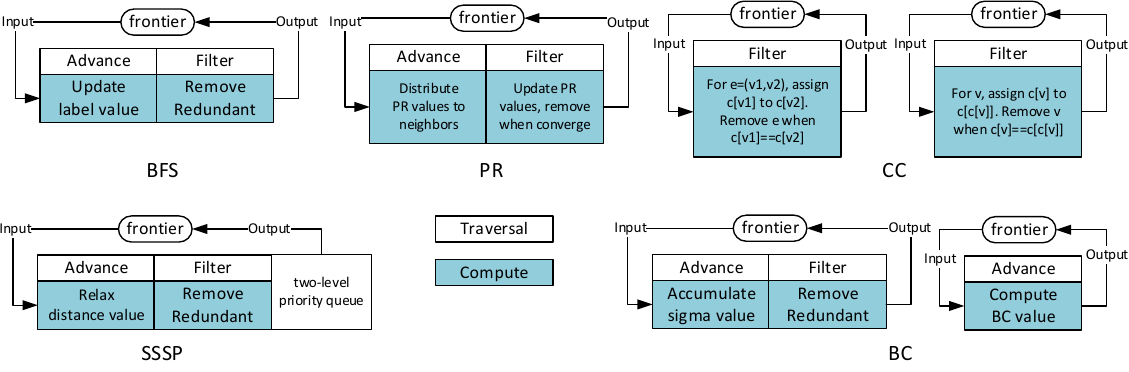}
  \centering
  \caption{Operation flow chart for selected primitives in Gunrock (a
    black line with an arrow at one end indicates a while loop that
    runs until the frontier is empty).\label{fig:flow}}
\end{figure}

\subsection{Breadth-First Search (BFS)}

BFS initializes its vertex frontier with a single source vertex. On
each iteration, it generates a new frontier of vertices with all
unvisited neighbor vertices in the current frontier, setting their
depths and repeating until all vertices have been visited. BFS is one
of the most fundamental graph primitives and serves as the basis of
several other graph primitives.

\begin{description}

\item[Hardwired GPU Implementation] The well-known BFS implementation
  of Merrill et al.~\cite{Merrill:2012:SGG} achieves its high
  performance through careful load-balancing, avoidance of atomics,
  and heuristics for avoiding redundant vertex discovery. Its chief
  operations are expand (to generate a new frontier) and contract (to
  remove redundant vertices) phases.

\item[Gunrock Implementation] Merrill et al.'s expand maps nicely to
  Gunrock's advance operator, and contract to Gunrock's filter
  operator. During advance, we set a label value for each vertex to
  show the distance from the source, and/or set a predecessor value
  for each vertex that shows the predecessor vertex's ID\@. We
  implement efficient load-balancing (section~\ref{sec:opt:twc} and
  section~\ref{subsubsec:lb}) and both push- and pull-based advance
  (section~\ref{subsec:pull}) for more efficient traversal. Our base
  implementation uses atomics during advance to prevent concurrent
  vertex discovery. When a vertex is uniquely discovered, we set its
  label (depth) and/or predecessor ID\@. Gunrock's BFS uses the
  idempotent advance operator to avoid the cost of atomics and uses
  heuristics within its filter that reduce the concurrent discovery of
  child nodes (section~\ref{subsec:idem}).

\end{description}

\subsection{Single-Source Shortest Path}
\label{sec:sssp}
Single-source shortest path finds paths between a given source vertex
and all other vertices in the graph such that the weights on the path
between source and destination vertices are minimized. While the
advance mode of SSSP is identical to BFS, the computation mode
differs.

\begin{description}
\item[Hardwired GPU Implementation] Currently the highest performing
  SSSP algorithm implementation on the GPU is the work from Davidson
  et al.~\shortcite{Davidson:2014:WPG}. They provide two key
  optimizations in their SSSP implementation: 1)~a load-balanced graph
  traversal method and 2)~a priority queue implementation that
  reorganizes the workload. Gunrock generalizes both optimization
  strategies into its implementation, allowing them to apply to other
  graph primitives as well as SSSP\@. We implement Gunrock's priority
  queue as an additional filter pass between two iterations.

\item[Gunrock Implementation] We start from a single source vertex in
  the frontier. To compute a distance value from the source vertex, we
  need one advance and one filter operator. On each iteration, we
  visit all associated edges in parallel for each vertex in the
  frontier and relax the distance's value (if necessary) of the
  vertices attached to those edges. We use an AtomicMin to atomically
  find the minimal distance value we want to keep and a bitmap flag
  array associated with the frontier to remove redundant vertices.
  After each iteration, we use a priority queue to reorganize the
  vertices in the frontier.
\end{description}

\subsection{Betweenness Centrality}
\label{sec:app:bc}

The BC index can be used in social network analysis as an indicator of
the relative importance of vertices in a graph. At a high level, the
BC for a vertex in a graph is the fraction of shortest paths in a
graph that pass through that vertex. Brandes's BC
formulation~\shortcite{Brandes:2001:AFA} is most commonly used for GPU
implementations.

\begin{description}
\item[Hardwired GPU Implementation] Brandes's formulation has two
  passes: a forward BFS pass to accumulate sigma values for each node,
  and a backward BFS pass to compute centrality values. Jia et
  al.~\shortcite{Jia:2011:ENP} and Sariy\"{u}ce et
  al.~\shortcite{Sariyuce:2013:BCO} both use an edge-parallel method
  to implement the above two passes. We achieve this in Gunrock using
  two advance operators on an edge frontier with different
  computations. The recent (hardwired) multi-GPU BC algorithm by
  McLaughlin and Bader~\shortcite{McLaughlin:2014:SAH} uses task
  parallelism, dynamic load balancing, and sampling techniques to
  perform BC computation in parallel from different sources on
  different GPU SMXs.

\item[Gunrock Implementation] Gunrock's implementation also contains
  two phases. The first phase has an advance step identical to the
  original BFS and a computation step that computes the number of
  shortest paths from source to each vertex. The second phase uses an
  advance step to iterate over the BFS frontier backwards with a
  computation step to compute the dependency scores. We achieve
  competitive performance on scale-free graphs with the latest
  hardwired BC algorithm~\cite{McLaughlin:2015:AFE}. Within Gunrock,
  we have not yet considered task parallelism since it appears to be
  limited to BC, but it is an interesting area for future work.
\end{description}

\subsection{Connected Component Labeling}

The connected component primitive labels the vertices in each
connected component in a graph with a unique component ID\@.

\begin{description}
\item[Hardwired GPU Implementation] Soman et
  al.~\shortcite{Soman:2010:AFG} base their implementation on two PRAM
  algorithms: hooking and pointer-jumping. Hooking takes an edge as
  the input and tries to set the component IDs of the two end vertices
  of that edge to the same value. In odd-numbered iterations, the
  lower vertex writes its value to the higher vertex, and vice versa
  in the even numbered iteration. This strategy increases the rate of
  convergence. Pointer-jumping reduces a multi-level tree in the graph
  to a one-level tree (star). By repeating these two operators until
  no component ID changes for any node in the graph, the algorithm
  will compute the number of connected components for the graph and
  the connected component to which each node belongs.

\item[Gunrock Implementation] Gunrock uses a filter operator on an
  edge frontier to implement hooking. The frontier starts with all
  edges and during each iteration, one end vertex of each edge in the
  frontier tries to assign its component ID to the other vertex, and
  the filter step removes the edge whose two end vertices have the
  same component ID\@. We repeat hooking until no vertex's component
  ID changes and then proceed to pointer-jumping, where a filter
  operator on vertices assigns the component ID of each vertex to its
  parent's component ID until it reaches the root. Then a filter step
  removes the node whose component ID equals its own node ID\@. The
  pointer-jumping phase also ends when no vertex's component ID
  changes.
\end{description}

\subsection{PageRank and Other Node Ranking Algorithms}

The PageRank link analysis algorithm assigns a numerical weighting to
each element of a hyperlinked set of documents, such as the World Wide
Web, with the purpose of quantifying its relative importance within
the set. The iterative method of computing PageRank gives each vertex
an initial PageRank value and updates it based on the PageRank of its
neighbors until the PageRank value for each vertex converges.
PageRank is one of the simplest graph algorithms to implement on GPUs
because the frontier always contains all vertices, so its computation
is congruent to sparse matrix-vector multiply; because it is simple,
most GPU frameworks implement it in a similar way and attain similar
performance.

In Gunrock, we begin with a frontier that contains all vertices in the
graph and end when all vertices have converged. Each iteration
contains one advance operator to compute the PageRank value on the
frontier of vertices, and one filter operator to remove the vertices
whose PageRanks have already converged. We accumulate PageRank values
with AtomicAdd operations.

\paragraph{Bipartite graphs} Geil et al.~\shortcite{Geil:2014:WGC}
used Gunrock to implement Twitter's who-to-follow algorithm
(``Money''~\cite{Goel:2015:WST}), which incorporated three
node-ranking algorithms based on bipartite graphs (Personalized
PageRank (PPR), Stochastic Approach for Link-Structure Analysis
(SALSA), and Hyperlink-Induced Topic Search (HITS))\@. Their
implementation, the first to use a programmable framework for
bipartite graphs, demonstrated that Gunrock's advance operator is
flexible enough to encompass all three node-ranking algorithms,
including a 2-hop traversal in a bipartite graph.

\subsection{Triangle Counting}
\begin{description}
\item[Hardwired GPU Implementation] The extensive survey by Schank and
  Wagner~\shortcite{Schank:2005:FCL} shows several sequential
  algorithms for counting and listing triangles in undirected graphs.
  Two of the best performing algorithms, \emph{edge-iterator} and
  \emph{forward}, both use edge-based set intersection primitives. The
  optimal theoretical bound of this operation coupled with its high
  potential for parallel implementation make this method the core idea
  behind several GPU
  implementations~\cite{Green:2014:FTC,Polak:2015:CTL}.
\item[Gunrock Implementation] In Gunrock, we view the TC problem as a
  set intersection problem~\cite{Wang:2016:ACS} by the following
  observation: An edge $e=(u,v)$, where $u,v$ are its two end nodes,
  can form triangles with edges connected to both $u$ and $v$. Let the
  intersections between the neighbor lists of $u$ and $v$ be
  $(w_1, w_2, \ldots, w_N)$, where $N$ is the number of intersections.
  Then the number of triangles formed with $e$ is $N$, where the three
  edges of each triangle are
  $(u, v), (w_i, u), (w_i, v), i \in [1,N]$. In practice, computing
  intersections for every edge in an undirected graph is redundant. We
  visit all the neighbor lists using advance. If two nodes $u$ and $v$
  have two edges $(u,v)$ and $(v,u)$ between them, we only keep one
  edge that points from the node with larger degree to the node with
  smaller degree. This halves the number of the edges that we must
  process. Thus, in general, set intersection-based TC algorithms have
  two stages: (1)~forming edge lists; (2)~computing set intersections
  for two neighbor lists of an edge. Different optimizations can be
  applied to either stage. Our GPU implementation follows the
  \emph{forward} algorithm and uses advance, filter, and
  segmented-intersection operators.
\end{description}

\subsection{Subgraph Matching}
\label{sec:app:sm}
The subgraph matching primitive finds all embeddings of a graph
pattern $q$ in a large data graph $g$. It can also be extended to
handle graph homomorphism problems, which can be used in database
pattern searching.
\begin{description}
\item[Hardwired GPU Implementation] Existing subgraph matching
  algorithms are mostly based on a backtracking strategy. First,
  vertices that cannot contribute to the final solutions are filtered
  out based on their labels. Then these candidates are passed to a
  recursive procedure to be further pruned in different matching
  orders based on the query graph and the sub-region where the
  candidate is located in the data graph. The order can significantly
  affect the performance. The recursive subroutine is hard to map
  efficiently to GPUs for two reasons. First, because of different
  matching orders for candidate vertices, warp divergence is a
  problem. Second, due to irregular graph patters, we see uncoalesced
  memory accesses. Some existing GPU implementations parallelize the
  backtracking method to join candidate edges in parallel to form
  partial solutions and repeat the method until the query graph
  pattern is obtained. However, these implementations generate more
  intermediate results, which makes the problem memory-bounded.

\item[Gunrock Implementation] Our implementation using
  Gunrock~\cite{Wang:2016:ACS} follows a filtering-and-joining
  procedure. In the filtering phase, we use a filter operator on a
  vertex frontier to prune out vertices based on both vertex labels
  and vertex degrees. Vertices with degree less than a certain query
  node's degree or with a different label cannot be that node's
  candidate. After that, we use advance and filter operators to
  collect the candidate edges. Then we do a join using our optimized
  set-intersection operator.
\end{description}

\section{Performance Characterization}
\label{sec:perf}
We first show overall performance analysis of Gunrock on nine datasets
including both real-world and generated graphs; the topology of these
datasets spans from regular to scale-free.

\begin{table}
  \small
  \centering
  \setlength{\tabcolsep}{3pt}
  \begin{tabular}{*{6}{c}} \toprule Dataset &Vertices&Edges&Max Degree& Diameter& Type \\
    \midrule
    soc-orkut & 3M & 212.7M & 27,466 & 9 & rs
    \\ soc-Livejournal1 & 4.8M & 85.7M & 20,333 & 16 & rs
    \\ hollywood-09 & 1.1M & 112.8M & 11,467 & 11 & rs
    \\ indochina-04 & 7.4M & 302M & 256,425 & 26 & rs
    \\ rmat\_s22\_e64 & 4.2M & 483M & 421,607 & 5 & gs
    \\ rmat\_s23\_e32 & 8.4M & 505.6M & 440,396 & 6 & gs
    \\ rmat\_s24\_e16 & 16.8M & 519.7M & 432,152 & 6 & gs
    \\ rgg\_n\_24 & 16.8M & 265.1M & 40 & 2622 & gm
    \\ roadnet\_USA & 23.9M & 577.1M & 9 & 6809 & rm
    \\ \bottomrule
  \end{tabular}
  \caption[Dataset description table.]{Dataset Description Table.
    Graph types are: r: real-world, g: generated, s: scale-free, and
    m: mesh-like. All datasets have been converted to undirected
    graphs. Self-loops and duplicated edges are
    removed.\label{tab:dataset}}
\end{table}

We ran all experiments in this paper on a Linux workstation with
2$\times$Intel Xeon E5-2637 v2 CPUs (3.50~GHz, 4-core, hyperthreaded),
256~GB of main memory, and an NVIDIA K40c GPU with 12~GB on-board
memory. GPU programs were compiled with NVIDIA's nvcc compiler
(version~8.0.44) with the -O3 flag. The BGL and PowerGraph code were
compiled using gcc 4.8.4 with the -O3 flag. Ligra was compiled using
icpc 15.0.1 with CilkPlus. For Ligra, elapsed time for best possible
sub\_algorithm was considered (for example, BFSCC was considered instead
of CC when it was the faster implementation for some particular datasets).
All PageRank implementations were executed with maximum iteration set
to 1. All results ignore transfer time (both disk-to-memory and
CPU-to-GPU)\@. All Gunrock tests were run 10 times with the average
runtime and MTEPS used for results.

\paragraph{Datasets}
We summarize the datasets we use for evaluation in
Table~\ref{tab:dataset}. Soc-orkut (soc-ork), soc-livejournal1
(soc-lj), and hollywood-09 (h09) are three social graphs; indochina-04
(i04) is a crawled hyperlink graph from indochina web domains;
rmat\_s22\_e64 (rmat-22), rmat\_s23\_e32 (rmat-23), and rmat\_s24\_e16
(rmat-24) are three generated R-MAT graphs with similar vertex counts.
All seven datasets are scale-free graphs with diameters of less than
30 and unevenly distributed node degrees (80\% of nodes have degree
less than 64). Both rgg\_n\_24 (rgg) and roadnet\_USA (roadnet)
datasets have large diameters with small and evenly distributed node
degrees (most nodes have degree less than 12). soc-ork is from the
Stanford Network Repository; soc-lj, i04, h09, and roadnet are from
the UF Sparse Matrix Collection; rmat-22, rmat-23, rmat-24, and rgg
are R-MAT and random geometric graphs we generated. For R-MAT, we use
16 as the edge factor, and the initiator parameters for the Kronecker
graph generator are: $a=0.57,b=0.19,c=0.19,d=0.05$. This setting is
the same as in the Graph 500 Benchmark. For random geometric graphs,
we set the threshold parameter to 0.000548. The edge weight values
(used in SSSP) for each dataset are uniform random values between 1
and 64.

\paragraph{Measurement methodology}
We report both runtime and traversed edges per second (TEPS) as our
performance metrics. (In general we report runtimes in milliseconds
and TEPS as millions of traversals per second [MTEPS].) Runtime is
measured by measuring the GPU kernel running time and MTEPS is
measured by recording the number of edges visited during the running
(the sum of neighbor list lengths of all visited vertices) divided by
the runtime. When a library does not report MTEPS, we use the
following equations to compute it for BFS and BC: $\frac{|E|}{t}$
(BFS) and $\frac{2\times|E|}{t}$ (BC), where $E$ is the number of
edges visited and $t$ is runtime. For SSSP, since one edge can be
visited multiple times when relaxing its destination node's distance
value, there is no accurate way to estimate its MTEPS number.

\begin{table}
\small
    \centering
        \begin{tabular}{*{5}{c}} \toprule Algorithm & Galois &BGL&PowerGraph&Medusa \\
        \midrule
        BFS & 8.812 & --- & --- & 22.49
        \\ SSSP & 2.532 & 99.99 & 8.058 & 2.158$^*$
        \\ BC & 1.57 & 32.12 & --- & ---
        \\ PageRank & 2.16 & --- & 17.73 & 2.463$^*$
        \\ CC & 1.745 & 341.1 & 182.7 & ---
        \\ \bottomrule
    \end{tabular}
    \caption[Geomean speedups of Gunrock over frameworks not in
     Table~\ref{tab:exp_largetable}.]{Geometric-mean runtime speedups
      of Gunrock on the datasets from Table~\ref{tab:dataset} over
      frameworks not in Table~\ref{tab:exp_largetable}. $^*$Due to
      Medusa's memory limitations~\cite{Zhong:2014:MSG}, its SSSP and
      PageRank comparisons were measured on four smaller datasets from
      Medusa's original paper.\label{tab:speedup}}
\end{table}

\begin{table}[t]
  \small
  \centering
  \renewcommand{\arraystretch}{0.5} 
  \resizebox{\linewidth}{!}{
    \begin{tabular}{*{13}{c}} \toprule
      && \multicolumn{5}{c}{Runtime (ms) [lower is better]} && \multicolumn{5}{c}{Edge throughput (MTEPS) [higher is better]} \\
      \cmidrule{3-7}\cmidrule{9-13}
           &         &       &          & Hardwired &       &         &&       &          & Hardwired &       &         \\
      Alg. & Dataset & CuSha & MapGraph & GPU       & Ligra & Gunrock && CuSha & MapGraph & GPU       & Ligra & Gunrock \\
      \midrule\parbox[t]{2mm}{\multirow{9}{*}{\rotatebox[origin=c]{90}{BFS}}}
           & soc-ork & 244.9 & OOM      & 25.81     & 26.1  & 5.573   && 868.3 & OOM      & 12360     & 8149  & 38165   \\
           & soc-lj  & 263.6 & 116.5    & 36.29     & 42.4  & 14.05   && 519.5 & 1176     & 5661      & 2021  & 6097    \\
           & h09     & 855.2 & 63.77    & 11.37     & 12.8  & 5.835   && 131.8 & 1766     & 14866     & 8798  & 19299   \\
           & i04     & 17609 & OOM      & 67.7      & 157   & 77.21   && 22.45 & OOM      & 8491      & 1899  & 3861    \\
           & rmat-22 & 1354  & OOM      & 41.81     & 22.6  & 3.943   && 369.1 & OOM      & 17930     & 21374 & 122516  \\
           & rmat-23 & 1423  & OOM      & 59.71     & 45.6  & 7.997   && 362.7 & OOM      & 12971     & 11089 & 63227   \\
           & rmat-24 & 1234  & OOM      & 270.6     & 89.6  & 16.74   && 426.4 & OOM      & ---       & 5800  & 31042   \\
           & rgg     & 68202 & OOM      & 138.6     & 918   & 593.9   && 3.887 & OOM      & 2868      & 288.8 & 466.4   \\
           & roadnet & 36194 & 763.4    & 141       & 978   & 676.2   && 3.189 & 151      & 1228      & 59.01 & 85.34   \\
      \midrule\parbox[t]{2mm}{\multirow{9}{*}{\rotatebox[origin=c]{90}{SSSP}}}
           & soc-ork & ---   & OOM      & 807.2     & 595   & 981.6   && ---   & OOM      & 770.6     & ---   & 216.7   \\
           & soc-lj  & ---   & ---      & 369       & 368   & 393.2   && ---   & ---      & 1039      & ---   & 217.9   \\
           & h09     & ---   & 1069     & 143.8     & 164   & 83.2    && ---   & ---      & 1427      & ---   & 1354    \\
           & i04     & ---   & OOM      & ---       & 397   & 371.8   && ---   & OOM      & ---       & ---   & 801.7   \\
           & rmat-22 & ---   & OOM      & ---       & 774   & 583.9   && ---   & OOM      & ---       & ---   & 827.3   \\
           & rmat-23 & ---   & OOM      & ---       & 1110  & 739.1   && ---   & OOM      & ---       & ---   & 684.1   \\
           & rmat-24 & ---   & OOM      & ---       & 1560  & 884.5   && ---   & OOM      & ---       & ---   & 587.5   \\
           & rgg     & ---   & OOM      & ---       & 80800 & 115554  && ---   & OOM      & ---       & ---   & 2.294   \\
           & roadnet & ---   & OOM      & 4860      & 29200 & 11037   && ---   & OOM      & 25.87     & ---   & 5.229   \\
      \midrule\parbox[t]{2mm}{\multirow{9}{*}{\rotatebox[origin=c]{90}{BC}}}
           & soc-ork & ---   & ---      & 1029      & 186   & 397.8   && ---   & ---      & 413.3     & 4574  & 1069    \\
           & soc-lj  & ---   & ---      & 492.8     & 180   & 152.7   && ---   & ---      & 347.7     & 1904  & 1122    \\
           & h09     & ---   & ---      & 441.3     & 59    & 73.36   && ---   & ---      & 510.3     & 7635  & 3070    \\
           & i04     & ---   & ---      & 1270      & 362   & 117     && ---   & ---      & 469       & 3294  & 5096    \\
           & rmat-22 & ---   & ---      & 1867      & 399   & 742.6   && ---   & ---      & 517.5     & 4840  & 1301    \\
           & rmat-23 & ---   & ---      & 2102      & 646   & 964.4   && ---   & ---      & 481.3     & 3130  & 1049    \\
           & rmat-24 & ---   & ---      & 2415      & 978   & 1153    && ---   & ---      & 430.3     & 2124  & 901.2   \\
           & rgg     & ---   & ---      & 26938     & 2510  & 1023    && ---   & ---      & 19.69     & 422.5 & 518.4   \\
           & roadnet & ---   & ---      & 15803     & 2490  & 1204    && ---   & ---      & 7.303     & 92.7  & 95.85   \\
      \midrule\parbox[t]{2mm}{\multirow{9}{*}{\rotatebox[origin=c]{90}{PageRank}}}
           & soc-ork & 52.54 & OOM      & ---       & 476   & 173.1   \\
           & soc-lj  & 33.61 & 250.7    & ---       & 200   & 54.1    \\
           & h09     & 34.71 & 93.48    & ---       & 77.4  & 20.05   \\
           & i04     & 164.6 & OOM      & ---       & 210   & 41.59   \\
           & rmat-22 & 188.5 & OOM      & ---       & 1250  & 304.5   \\
           & rmat-23 & 147   & OOM      & ---       & 1770  & 397.2   \\
           & rmat-24 & 128   & OOM      & ---       & 2180  & 493.2   \\
           & rgg     & 53.93 & OOM      & ---       & 247   & 181.3   \\
           & roadnet & --- & 123.2    & ---       & 209   & 24.11   \\
      \midrule\parbox[t]{2mm}{\multirow{9}{*}{\rotatebox[origin=c]{90}{CC}}}
           & soc-ork & ---   & OOM      & 46.97     & 260   & 211.7   \\
           & soc-lj  & ---   & OOM      & 43.51     & 184   & 93.27   \\
           & h09     & ---   & 547.1    & 24.63     & 90.8  & 96.15   \\
           & i04     & ---   & OOM      & 130.3     & 315   & 773.7   \\
           & rmat-22 & ---   & OOM      & 149.4     & 563   & 429.8   \\
           & rmat-23 & ---   & OOM      & 212       & 1140  & 574.3   \\
           & rmat-24 & ---   & OOM      & 256.7     & 1730  & 664.1   \\
           & rgg     & ---   & OOM      & 103.9     & 6000  & 355.2   \\
           & roadnet & ---   & OOM      & 124.9     & 50500 & 208.9   \\
      \bottomrule
    \end{tabular}}
  \caption[Gunrock's performance comparison with other graph libraries.]
  {Gunrock's performance comparison (runtime and edge throughput) with other
  graph libraries (CuSha, MapGraph, Ligra) and hardwired GPU implementations on
  a Tesla K40c GPU~\@.
  All PageRank times are normalized to one iteration. Hardwired GPU
  implementations for each primitive are Enterprise (BFS)~\protect\cite{Liu:2015:EBG},
  delta-stepping SSSP~\protect\cite{Davidson:2014:WPG},
  gpu\_BC (BC)~\protect\cite{Sariyuce:2013:BCO}, and conn
  (CC)~\protect\cite{Soman:2010:AFG}. OOM means out-of-memory.
  A missing data entry means either there is a runtime error, or the specific
  primitive for that library is not available.
  \label{tab:exp_largetable}}
\end{table}

\begin{figure}
    \centering
    \includegraphics[width=\textwidth]{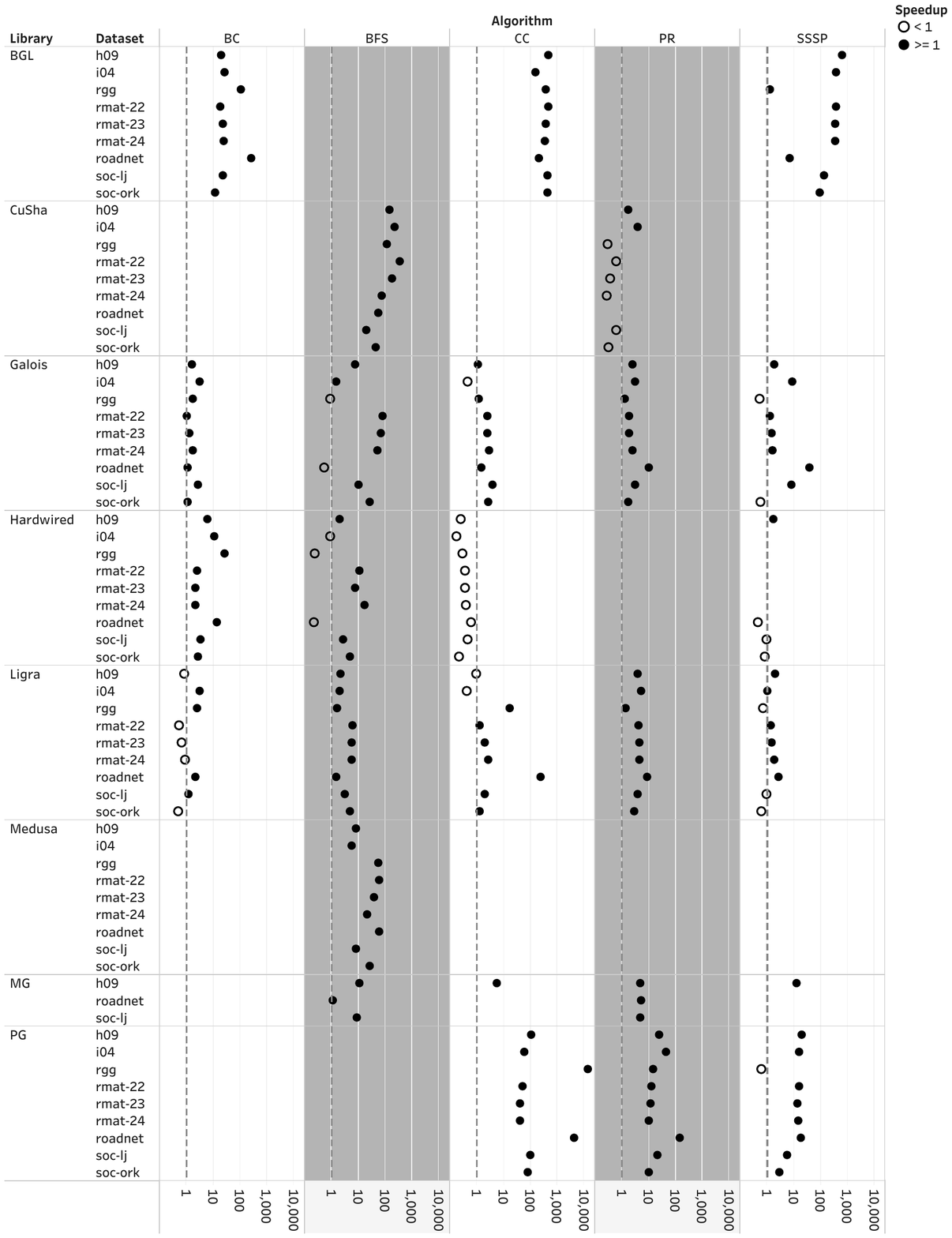}
    \centering
    \caption[Speedup for Gunrock.]{Execution-time speedup for Gunrock
      vs.\ five other graph processing libraries/hardwired algorithms
      on nine different graph inputs. Data is from
      Table~\ref{tab:exp_largetable}. Black dots indicate Gunrock is
      faster, white dots slower.\label{fig:speedup}}
\end{figure}

\begin{figure}
    \centering
    \includegraphics[width=\textwidth]{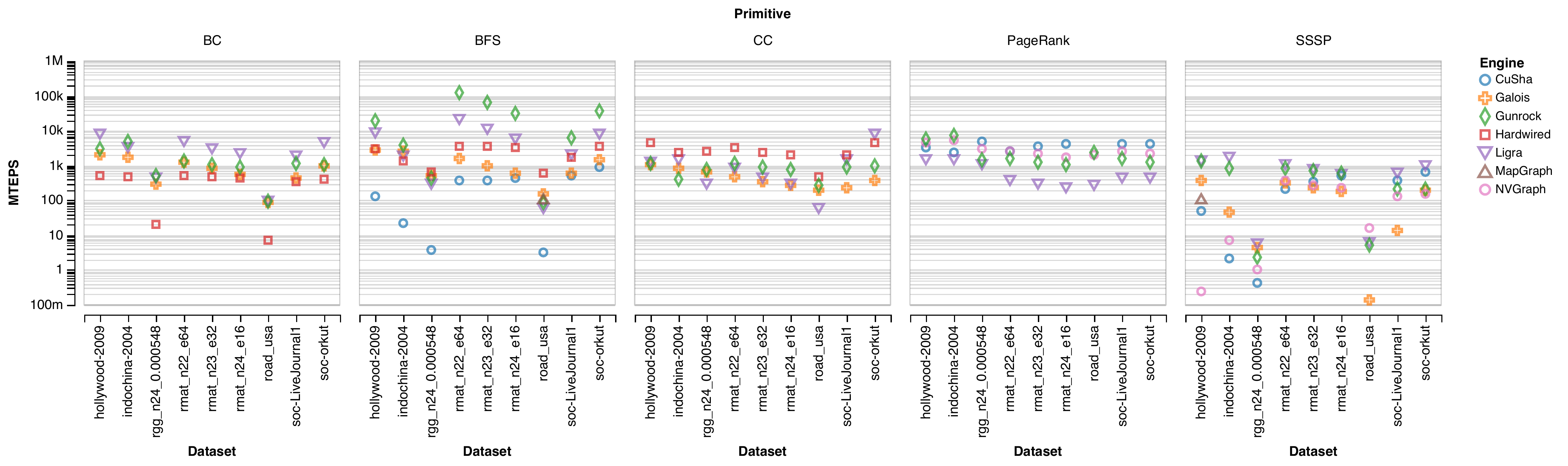}
    \centering
    \caption[MTEPS comparison for Gunrock and others.]{Performance in
      MTEPS for Gunrock vs.\ six other graph processing
      libraries/hardwired algorithms on nine different graph inputs.
      Data is from Table~\ref{tab:exp_largetable}.\label{fig:mteps}}
\end{figure}

\begin{figure}
    \centering
    \includegraphics[width=\textwidth]{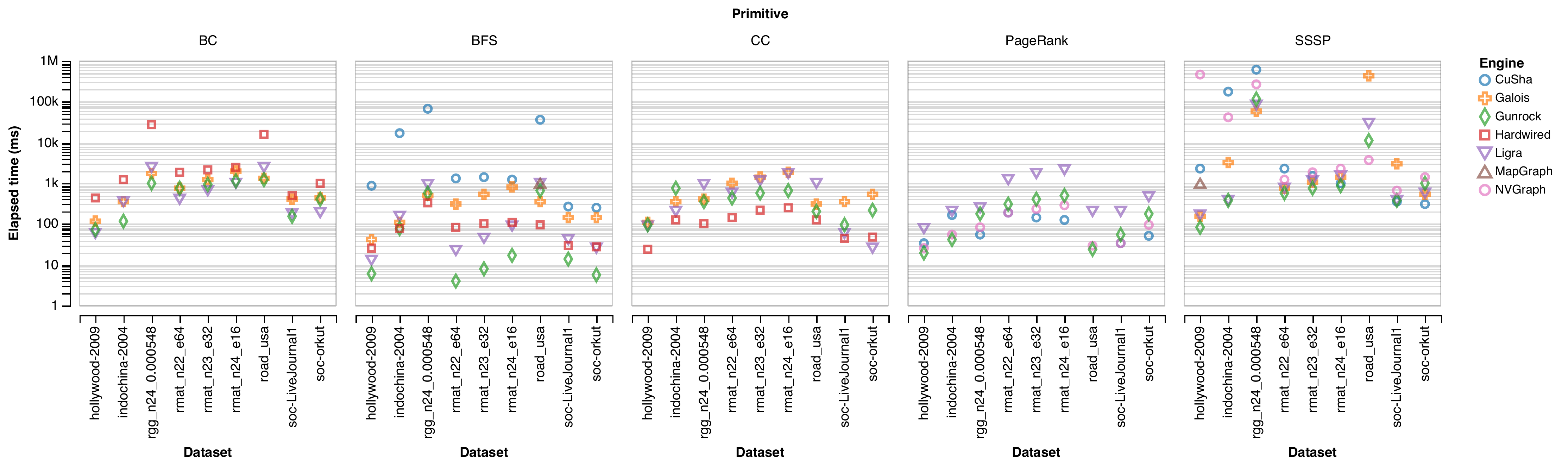}
    \centering
    \caption[Runtime comparison for Gunrock and others.]{Performance
      in runtime for Gunrock vs.\ six other graph processing
      libraries/hardwired algorithms on nine different graph inputs.
      Data is from Table~\ref{tab:exp_largetable}.\label{fig:runtime}}
\end{figure}

\subsection{Performance Summary}
Tables~\ref{tab:speedup} and~\ref{tab:exp_largetable}, and
Figures~\ref{fig:speedup}, ~\ref{fig:mteps}, and~\ref{fig:runtime},
compare Gunrock's performance against several other graph libraries
and hardwired GPU implementations. In general, Gunrock's performance
on BFS-based primitives (BFS, BC, and SSSP) shows comparatively better
results when compared to other graph libraries on seven scale-free
graphs (soc-orkut, soc-lj, h09, i04, and rmats), than on two
small-degree large-diameter graphs (rgg and roadnet). The primary
reason is our load-balancing strategy during traversal
(Table~\ref{tab:wee} shows Gunrock's superior performance on warp
efficiency, a measure of load-balancing quality, across GPU frameworks
and datasets) and particularly our emphasis on good performance for
highly irregular graphs. As well, graphs with uniformly low degree
expose less parallelism and would tend to show smaller gains in
comparison to CPU-based methods. Table~\ref{tab:scalability} shows
Gunrock's scalability. In general, runtimes scale roughly linearly
with graph size for BFS, but primitives with heavy use of atomics on
the frontier (e.g., BC, SSSP, and PR) show increased atomic contention
within the frontier as graph sizes increase and thus do not scale
ideally. For CC, the reason of non-ideal scalability is mainly due to
the increase of race conditions when multiple edges try to hook their
source node to a common destination node, which happens more often
when power-law graphs get bigger and degree numbers become more
unevenly distributed.

\subsection{vs.\ CPU Graph Libraries} We compare Gunrock's performance
with four CPU graph libraries: the Boost Graph Library
(BGL)~\cite{Siek:2001:TBG}, one of the highest-performing CPU
single-threaded graph libraries~\cite{McColl:2014:APE}; PowerGraph, a
popular distributed graph library~\cite{Gonzalez:2012:PDG}; and
Ligra~\cite{Shun:2013:LAL} and
Galois~\cite{Nguyen:2013:ALI,Pingali:2011:TTO}, two of the
highest-performing multi-core shared-memory graph libraries. Against
both BGL and PowerGraph, Gunrock achieves 6x--337x speedup on average
on all primitives. Compared to Ligra, Gunrock's performance is
generally comparable on most tested graph primitives; note Ligra
results are on a 2-CPU machine. The performance inconsistency for SSSP
vs.\ Ligra is due to comparing our delta-stepping-based
method~\cite{Meyer:2003:DAP} with Ligra's Bellman-Ford algorithm. Our
SSSP's edge throughput is smaller than BFS but similar to BC because
of similar computations (atomicMin vs.\ atomicAdd) and a larger number
of iterations for convergence. The performance inconsistency for BC
vs.\ Ligra on four scale-free graphs is because Ligra applies
pull-based traversal on BC while Gunrock has not yet done so. Compared
to Galois, Gunrock shows more speedup on traversal-based graph
primitives (BFS, SSSP, and BC) and less performance advantage on
PageRank and CC due to their dense computation and more regular
frontier structures.

\subsection{vs.\ Hardwired GPU Implementations and GPU Libraries}
Compared to hardwired GPU implementations, depending on the dataset,
Gunrock's performance is comparable or better on BFS, BC, and SSSP\@.
For CC, Gunrock is 5x slower (geometric mean) than the hardwired GPU
implementation due to irregular control flow because each Gunrock
iteration starts with full edge lists in both hooking and
pointer-jumping phases. The alternative is extra steps to perform
additional data reorganization. This tradeoff is not typical of our
other primitives. While still achieving high performance, Gunrock's
application code is smaller in size and clearer in logic compared to
other GPU graph libraries.

Gunrock's Problem class (that defines problem data used for the graph
algorithm) and kernel enactor are both template-based C++ code;
Gunrock's functor code that specifies per-node or per-edge computation
is C-like device code without any CUDA-specific keywords. Writing
Gunrock code may require parallel programming concepts (e.g., atomics)
but neither details of low-level GPU programming nor optimization
knowledge.\footnote{We believe this assertion is true given our
  experience with other GPU libraries when preparing this evaluation
  section, but freely acknowledge this is nearly impossible to
  quantify. We invite readers to peruse our annotated code for BFS and
  SALSA at \url{http://gunrock.github.io/gunrock/doc/annotated_primitives/annotated_primitives.html}.}

Gunrock compares favorably to existing GPU graph libraries.
\begin{description}
\item[vs.\ MapGraph] MapGraph is faster than Medusa on all but one
  test~\cite{Fu:2014:MAH} and Gunrock is faster than MapGraph on all
  tests: the geometric mean of Gunrock's speedups over MapGraph on
  BFS, SSSP, PageRank, and CC are 4.679, 12.85, 3.076, and 5.69,
  respectively.
\item[vs.\ CuSha] Gunrock outperforms CuSha on BFS and SSSP\@. For
  PageRank, Gunrock achieves comparable performance with no
  preprocessing when compared to CuSha's G-Shard data preprocessing,
  which serves as the main load-balancing module in CuSha.
\item[vs.\ Totem] The 1-GPU Gunrock implementation has 1.83x more
  MTEPS (4731 vs.\ 2590) on direction-optimized BFS on the
  soc-LiveJournal dataset (a smaller scale-free graph in their test
  set) than the 2-CPU, 2-GPU configuration of
  Totem~\cite{Sallinen:2015:ADB}.
\item[vs.\ nvGRAPH] For SSSP, nvGRAPH is faster than Gunrock on the
  roadnet dataset, but slower on the other datasets. Gunrock in
  general performs better on scale-free graphs than it does on regular
  graphs. For PageRank, nvGRAPH is
  faster than Gunrock on six datasets and slower on three (h04, i09,
  and roadnet). nvGRAPH is closed-source and thus a detailed
  comparison is infeasible.
\end{description}
All three GPU BFS-based high-level-programming-model efforts (Medusa,
MapGraph, and Gunrock) adopt load-balancing strategies from Merrill et
al.'s BFS~\cite{Merrill:2012:SGG}. While we would thus expect Gunrock
to show similar performance on BFS-based graph primitives as these
other frameworks, we attribute our performance advantage to two
reasons: 1)~our improvements to efficient and load-balanced traversal
that are integrated into the Gunrock core, and 2)~a more powerful,
GPU-specific programming model that allows more efficient high-level
graph implementations. 1)~is also the reason that Gunrock
implementations can compete with hardwired implementations; we believe
Gunrock's load-balancing and work distribution strategies are at least
as good as if not better than the hardwired primitives we compare
against. Gunrock's memory footprint is at the same level as Medusa and
better than MapGraph (note the OOM test cases for MapGraph in
Table~\ref{tab:exp_largetable}). Our data footprint is
$\alpha|E|+\beta|V|$ for current graph primitives, where $|E|$ is the
number of edges, $|V|$ is the number of nodes, and $\alpha$ and
$\beta$ are both integers where $\alpha$ is usually 1 and at most 3
(for BC) and $\beta$ is between 2 to 8.

\begin{figure*}
    \centering
    \includegraphics[width=1.0\textwidth]{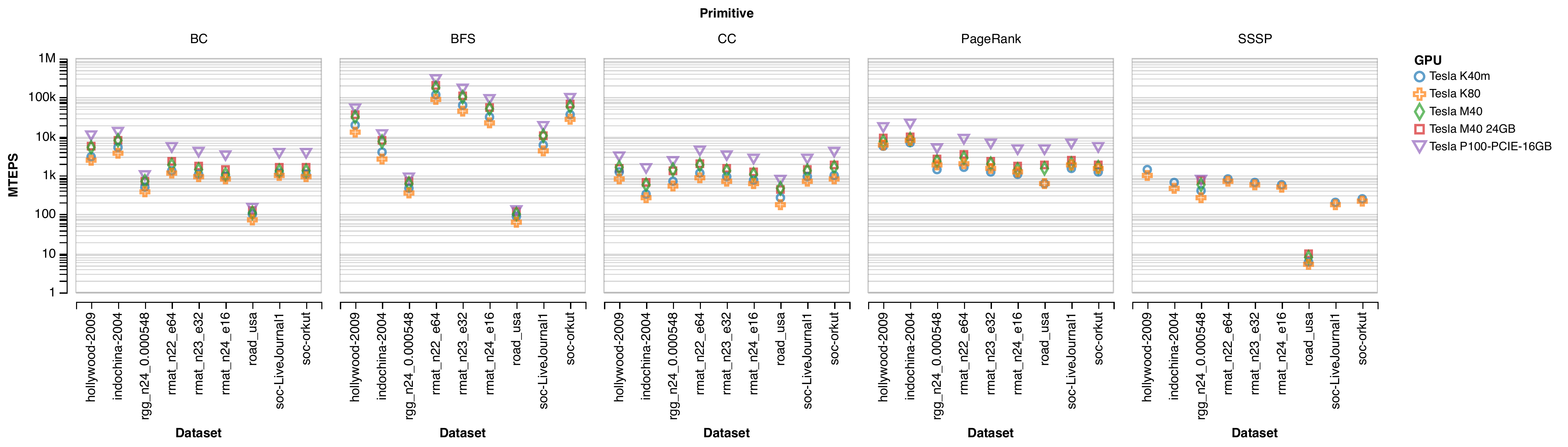}
    \centering
    \caption{Gunrock's performance on different GPU devices. The Tesla
      M40 24GB is attached to a Haswell CPU and has a higher boost
      clock (1328500 Hz). The Tesla M40 is attached to an Ivy Bridge
      CPU with a boost clock of 1112000 Hz.}
    \label{fig:devices}
\end{figure*}

Figure~\ref{fig:devices} shows Gunrock's performance on Gunrock's v0.4
release on four different GPUs: Tesla K40m, Tesla K80, Tesla M40, and
Tesla P100. Programs are compiled by nvcc (8.0.44) with -O3 flag and
GPU SM versions according to the actual hardware.
Across different GPUs, Gunrock's performance generally scales
with memory bandwidth, with the newest Tesla P100 GPU demonstrating
the best performance.

\begin{table}[t]
\centering
 \resizebox{\linewidth}{!}{
 \begin{tabular}{*{10}{c}} \toprule & \multicolumn{5}{c}{Runtime (ms)} && \multicolumn{3}{c}{Edge throughput (MTEPS)} \\
\cmidrule{2-6}\cmidrule{8-10}
Dataset & BFS & BC & SSSP & CC & PageRank && BFS & BC & SSSP \\
  \midrule
  kron\_g500-logn18 ($v = 2^{18}, e = 14.5\text{M}$) & 1.319 & 12.4 & 13.02 & 9.673 & 3.061 && 10993 & 2339 & 1113 \\
  kron\_g500-logn19 ($v = 2^{19}, e = 29.7\text{M}$) & 1.16 & 26.93 & 26.59 & 20.41 & 6.98 && 25595 & 2206 & 1117 \\
  kron\_g500-logn20 ($v = 2^{20}, e = 60.6\text{M}$) & 2.355 & 67.57 & 56.22 & 43.36 & 19.63 && 25723 & 1793 & 1078 \\
  kron\_g500-logn21 ($v = 2^{21}, e = 123.2\text{M}$) & 3.279 & 164.9 & 126.3 & 97.64 & 60.14 && 37582 & 1494 & 975.2 \\
  kron\_g500-logn22 ($v = 2^{22}, e = 249.9\text{M}$) & 5.577 & 400.3 & 305.8 & 234 & 163.9 && 44808 & 1248 & 817.1 \\
  kron\_g500-logn23 ($v = 2^{23}, e = 505.6\text{M}$) & 10.74 & 900 & 703.3 & 539.5 & 397.7 && 47077 & 1124 & 719 \\
  \bottomrule
\end{tabular}}
\centering
\caption[Scalability of five Gunrock primitives on a Tesla K40c
GPU\@.]{Scalability of 5 Gunrock primitives (runtime and edges
  traversed per second) on a single GPU on five differently-sized
  synthetically-generated Kronecker graphs with similar scale-free
  structure.\label{tab:scalability}}
\end{table}

\label{sec:warp-execution-efficiency}
\begin{table}
  \centering
  \resizebox{\linewidth}{!}{
  \begin{tabular}{@{}llccccccccc@{}}
    \toprule
    Alg. & Framework & soc-ork & soc-lj & h09  & i04  & rmat-22 & rmat-23 & rmat-24 & rgg & roadnet \\
    \midrule
    \multirow{3}{*}{BFS}
    & Gunrock & 92.01\% & 87.89\% & 86.52\% & 94.3\% & 90.11\% & 91.21\% & 92.85\% & 79.23\% & 80.65\%\\
    & MapGraph & --- & 95.83\% & 95.99\% & --- & --- & --- & --- & --- & 86.59\% \\
    & CuSha & 69.42\% & 58.85\% & 57.2\% & 52.64\% & 49.8\% & 43.65\% & 42.17\% & 64.77\% & --- \\
    \midrule
    \multirow{3}{*}{SSSP}
    & Gunrock & 96.54\% & 95.96\% & 94.78\% & 96.34\% & 96.79\% & 96.03\% & 96.85\% & 77.55\% & 80.11\%\\
    & MapGraph & --- & --- & --- & --- & --- & --- & --- & --- & 99.82\% \\
    & CuSha & --- & --- & --- & --- & --- & --- & --- & --- & --- \\
    \midrule
    \multirow{3}{*}{PR}
    & Gunrock & 99.54\% & 99.45\% & 99.37\% & 99.54\% & 98.74\% & 98.68\% & 98.29\% & 99.56\% & 96.21\%\\
    & MapGraph & --- & 98.8\% & 98.72\% & --- & --- & --- & --- & --- & 99.82\% \\
    & CuSha & 80.13\% & 64.23\% & 93.23\% & 55.62\% & 79.7\% & 74.87\% & 70.29\% & 75.11\% & --- \\
    \bottomrule
  \end{tabular}}
\caption[Average warp execution efficiency table.]{Average warp
  execution efficiency (fraction of threads active during computation)
  of Gunrock running on a Tesla K40c GPU\@. This figure is a good
  metric for the quality of a framework's load-balancing capability.
  (--- indicates the graph framework failed to run on that dataset.)
  \label{tab:wee}}
\end{table}

\subsection{Optimization Strategies Performance Analysis}
\label{perf:opt}
\begin{figure}
    \centering
    \includegraphics[width=1.0\textwidth]{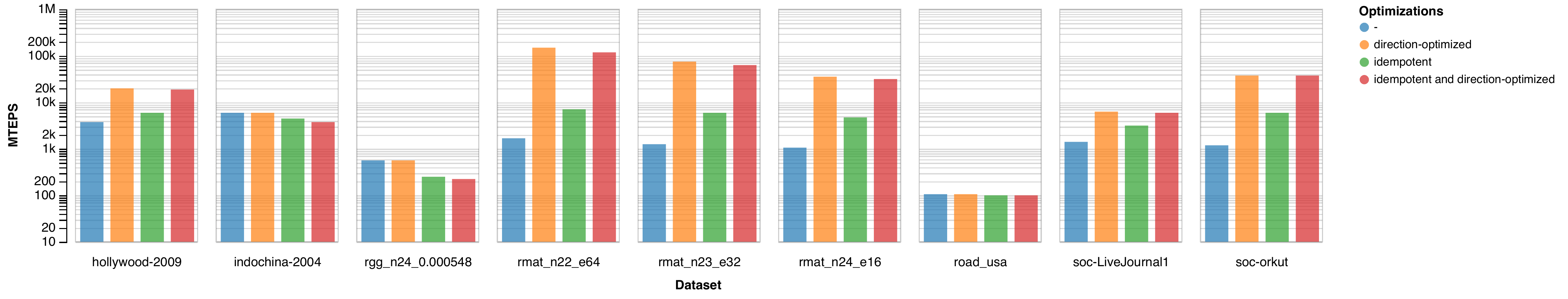}
    \centering \caption[Impact of different optimization strategies on
    Gunrock's performance.]{Gunrock's performance with different
      combinations of idempotence and direction-optimized
      traversal.\label{fig:perf_optimization}}
\end{figure}

\begin{figure}
    \centering
    \includegraphics[width=1.0\textwidth]{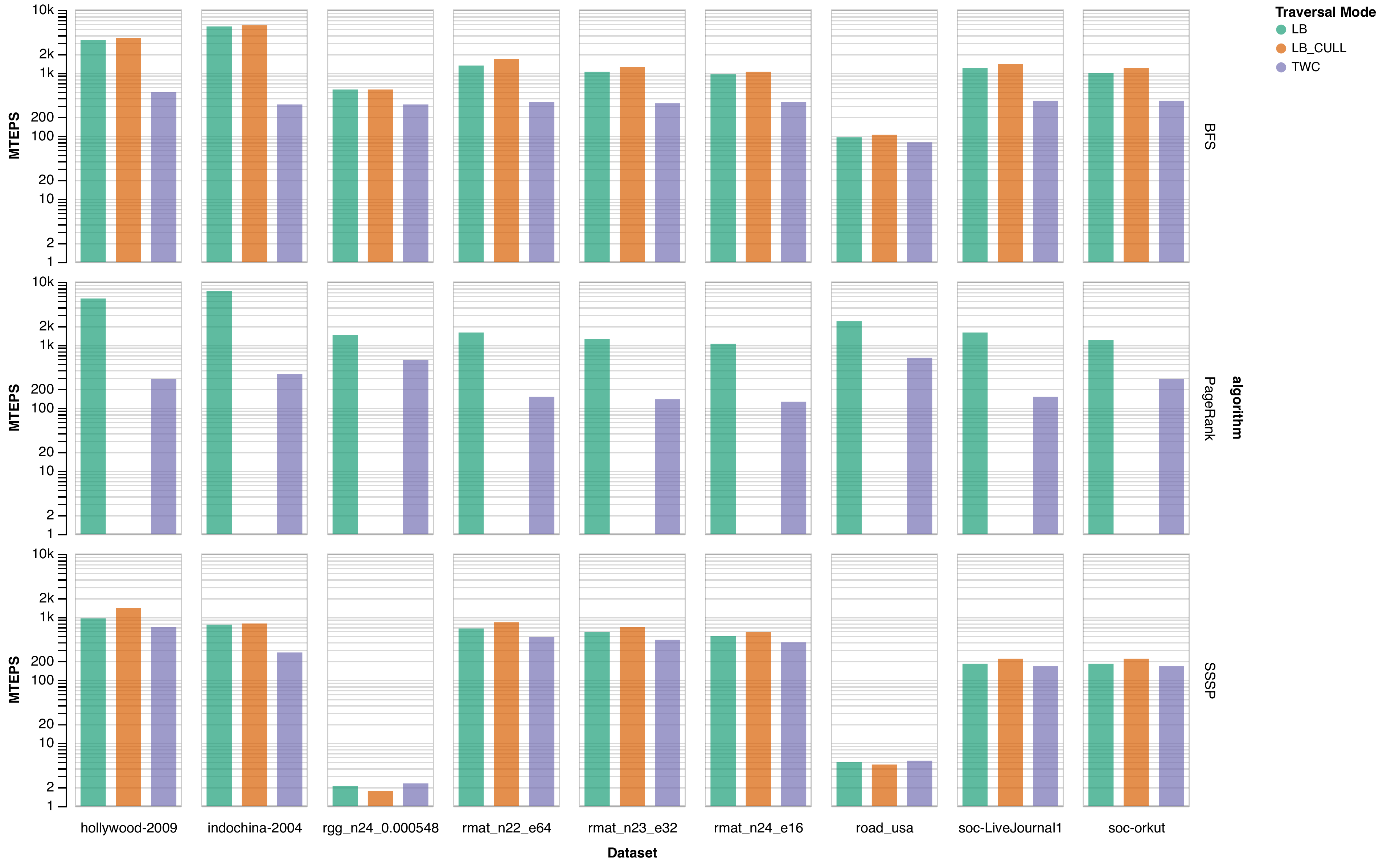}
    \centering \caption[Impact of different traversal modes on
    Gunrock's performance.]{BFS, SSSP, and PR's performance in Gunrock
      using three different workload mapping strategies: LB, LB\_CULL,
      and TWC\@.\label{fig:perf_traversal}}
\end{figure}

Figures~\ref{fig:perf_optimization} and~\ref{fig:perf_traversal} show
how different optimization strategy combinations and different
workload mapping strategies affect the performance of graph traversal.

Without losing generality, for our tests on different optimization
strategies, we use BFS and fix the workload mapping strategy to
LB\_CULL so that we can focus on the impact of different optimization
strategies. Our two key optimizations---idempotence and
direction-optimized traversal---both show performance increases
compared to the baseline LB\_CULL traversal, except for rgg and
roadnet, for which idempotence does not increase performance because
the inflated frontiers cancel out the performance increase for
avoiding atomics. Also, our experiment shows that when using LB\_CULL
for advance, enabling both direction-optimized and idempotence always
yields worse performance than with only direction-optimized enabled.
The reason behind this is that enabling idempotence operation in
direction-optimized traversal iteration causes additional global data
accesses to a visited-status bitmask array. We also note that for
graphs with very small degree standard deviation ($< 5$), using
idempotence will actually hurt the performance. The reason is for
these graphs, the number of common neighbors is so small that avoiding
atomic operations will not provide much performance gain, but using
idempotence introduces an additional pass of filter heuristics, and
the cost of that pass outweights the performance gain from avoiding
atomic operations.

Our tests on different workload mapping strategies (traversal mode)
shows that LB\_CULL constantly outperforms other two strategies on
these 9 datasets. Its better performance compared to TWC is due to its
switching between load-balance over the input frontier and load-
balancing over the output frontier according to the input frontier
size (Section~\ref{subsubsec:lb}). Its better performance compared to
LB is due to its kernel fusion implementation, which reduces the
kernel launch overhead and also some additional data movement between
operators. However, without a more thorough performance
characterization, we cannot make the conclusion that LB\_CULL will
always have better performance. For instance, in our SSSP tests on two
mesh-like graphs with large diameters and small average degrees, TWC
shows better performance. In general, we currently predict which
strategies will be most beneficial based only on the degree
distribution; many application scenarios may allow pre-computation of
this distribution and thus we can choose the optimal strategies before
we begin computation.

\begin{figure}
\centering
\includegraphics[width=.3\textwidth]{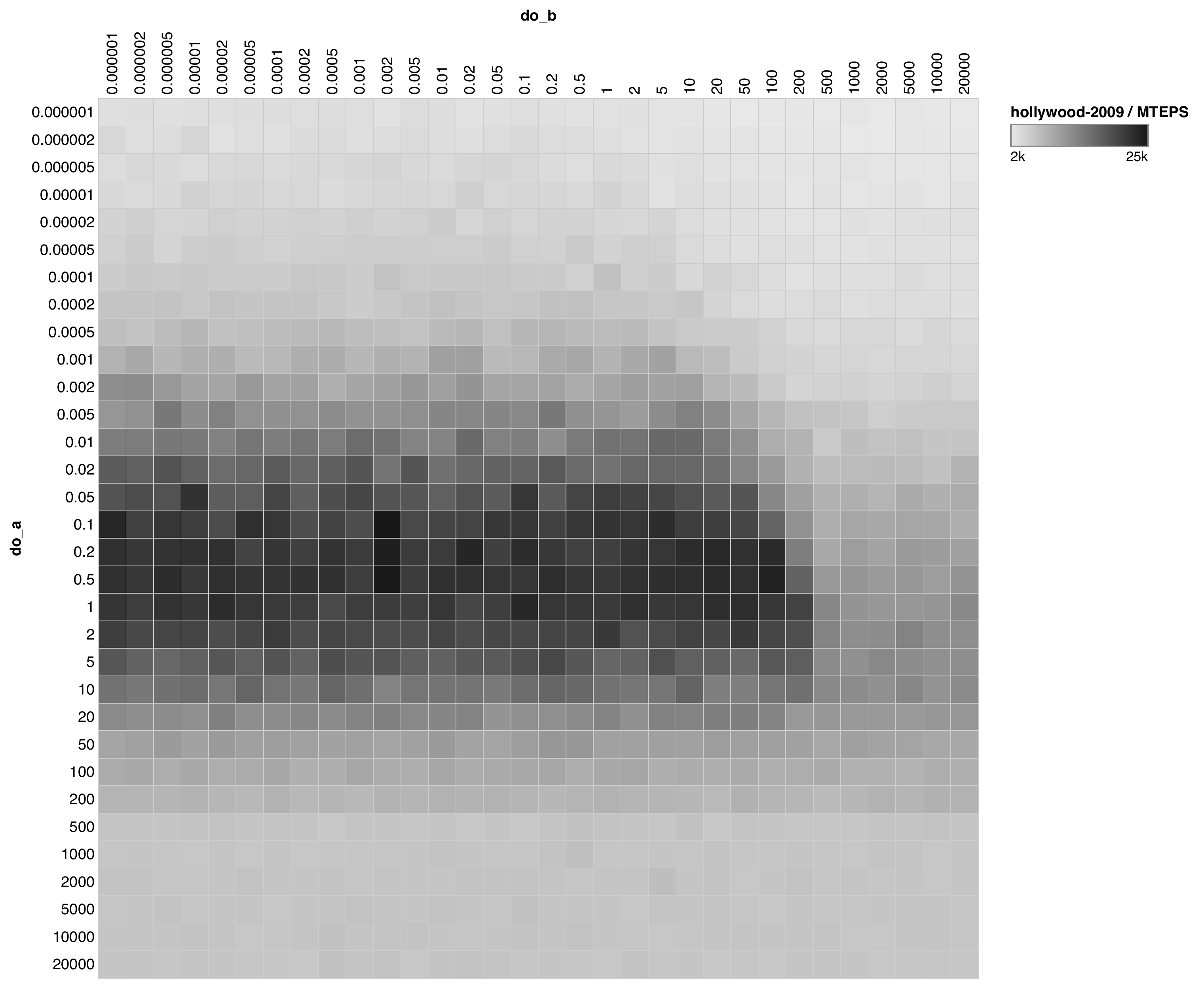}\quad
\includegraphics[width=.3\textwidth]{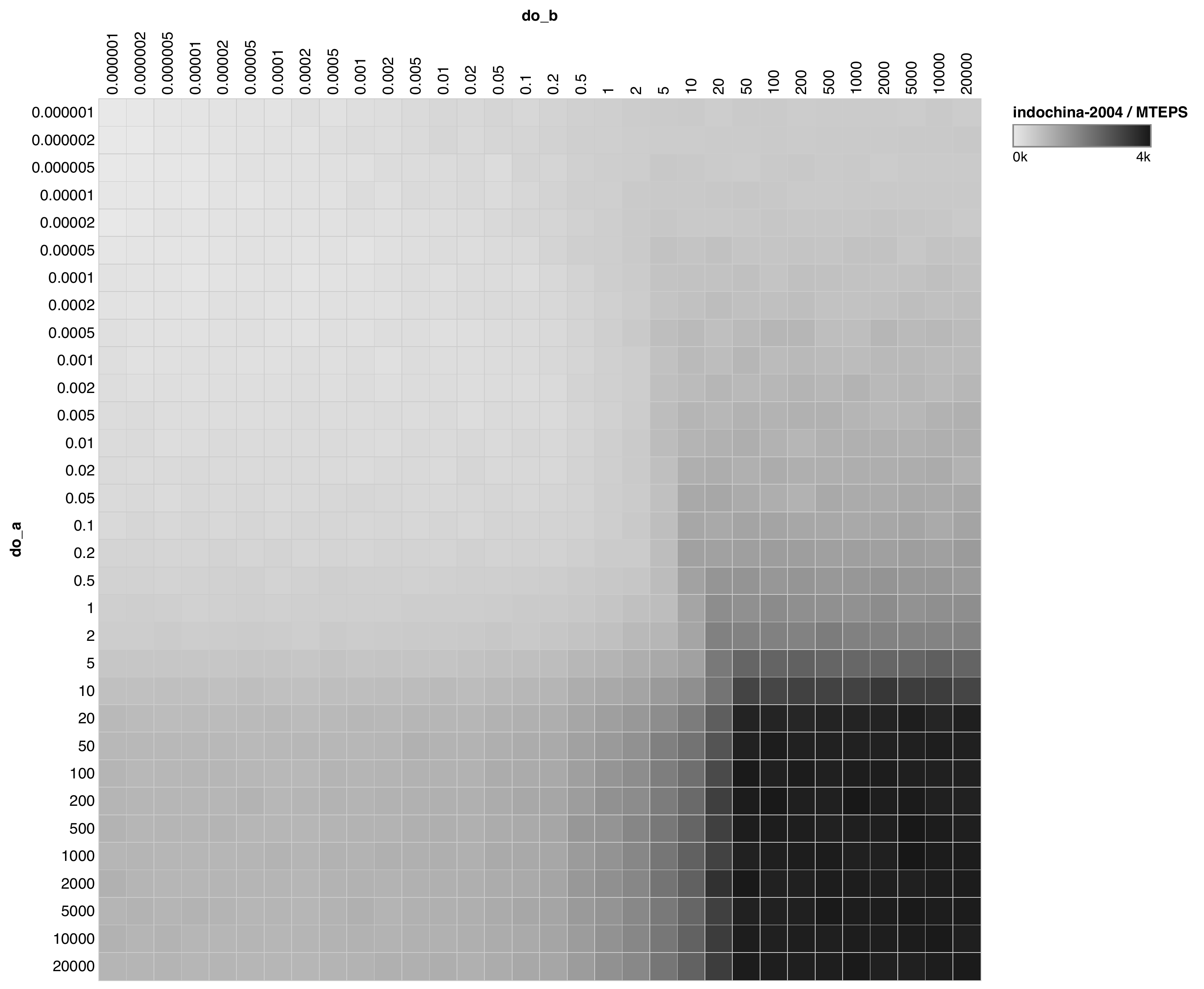}\quad
\includegraphics[width=.3\textwidth]{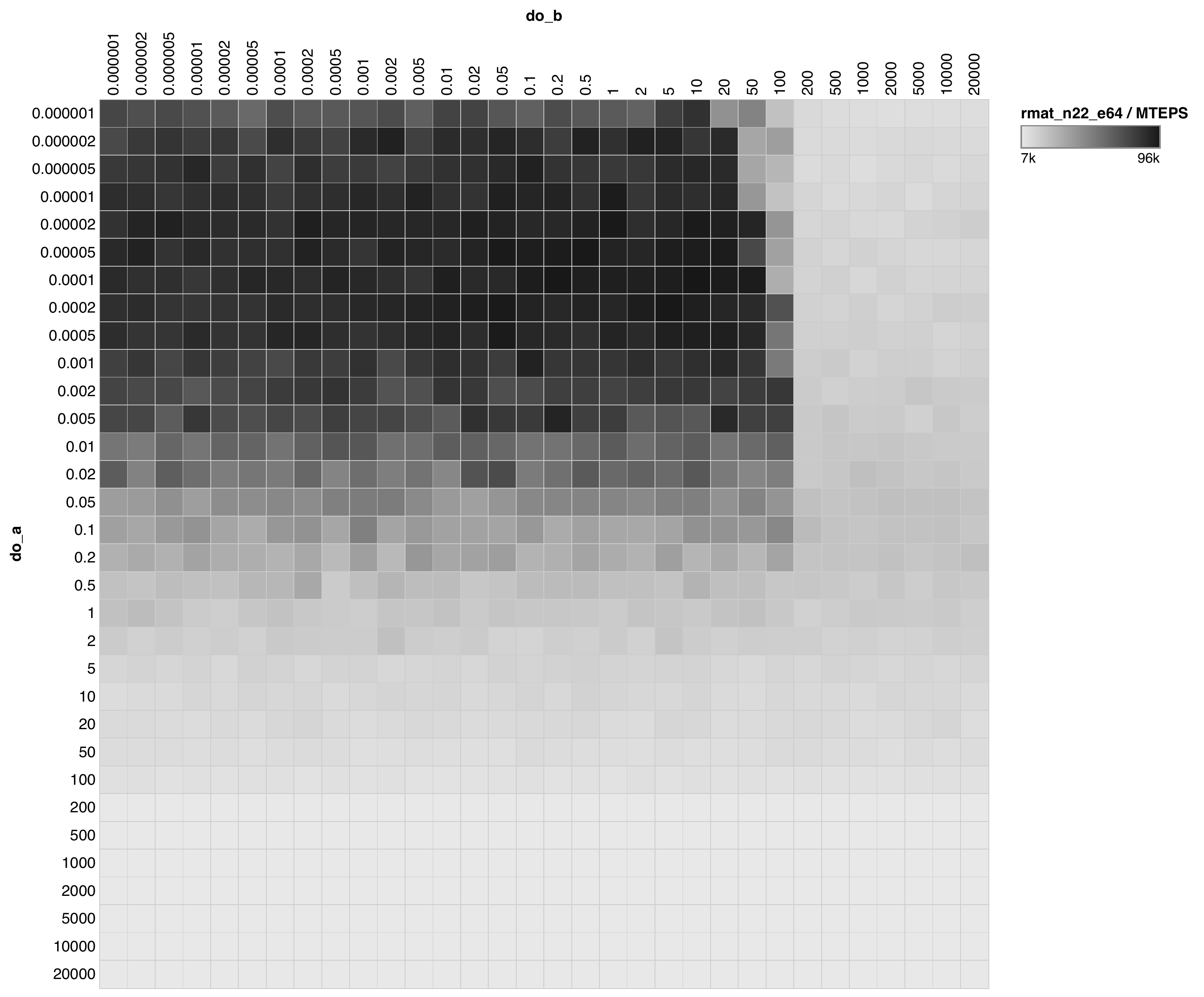}

\medskip

\includegraphics[width=.3\textwidth]{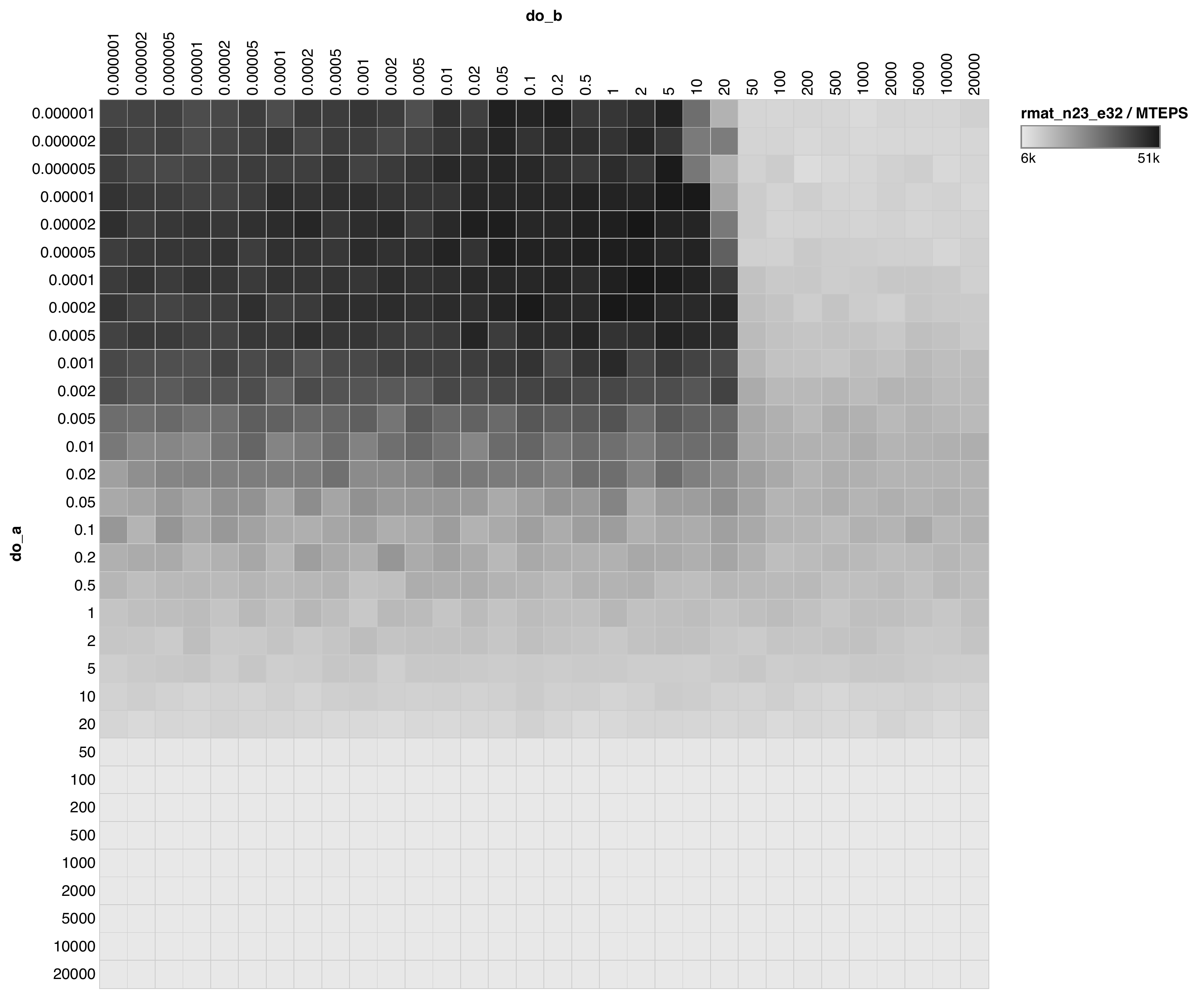}\quad
\includegraphics[width=.3\textwidth]{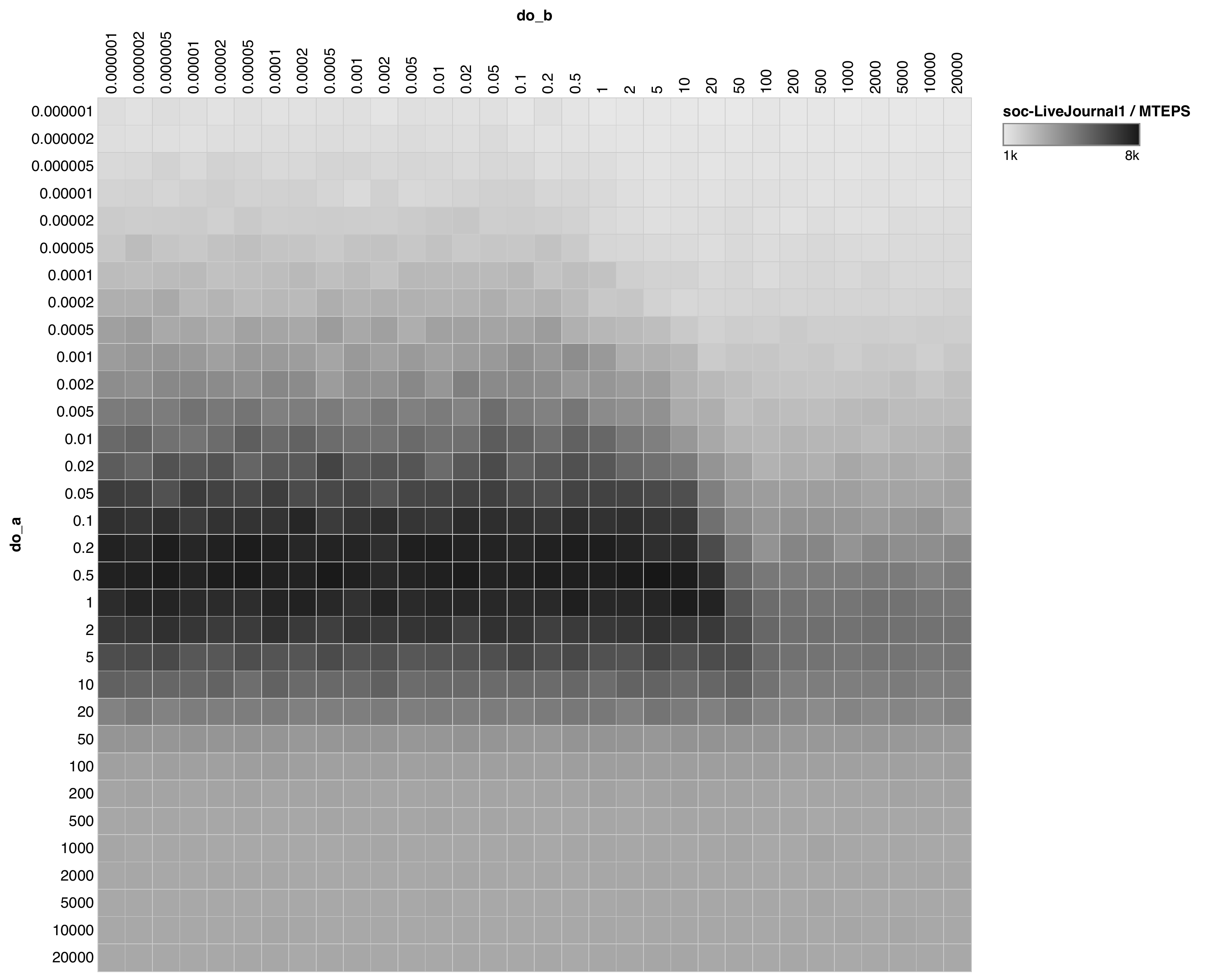}\quad
\includegraphics[width=.3\textwidth]{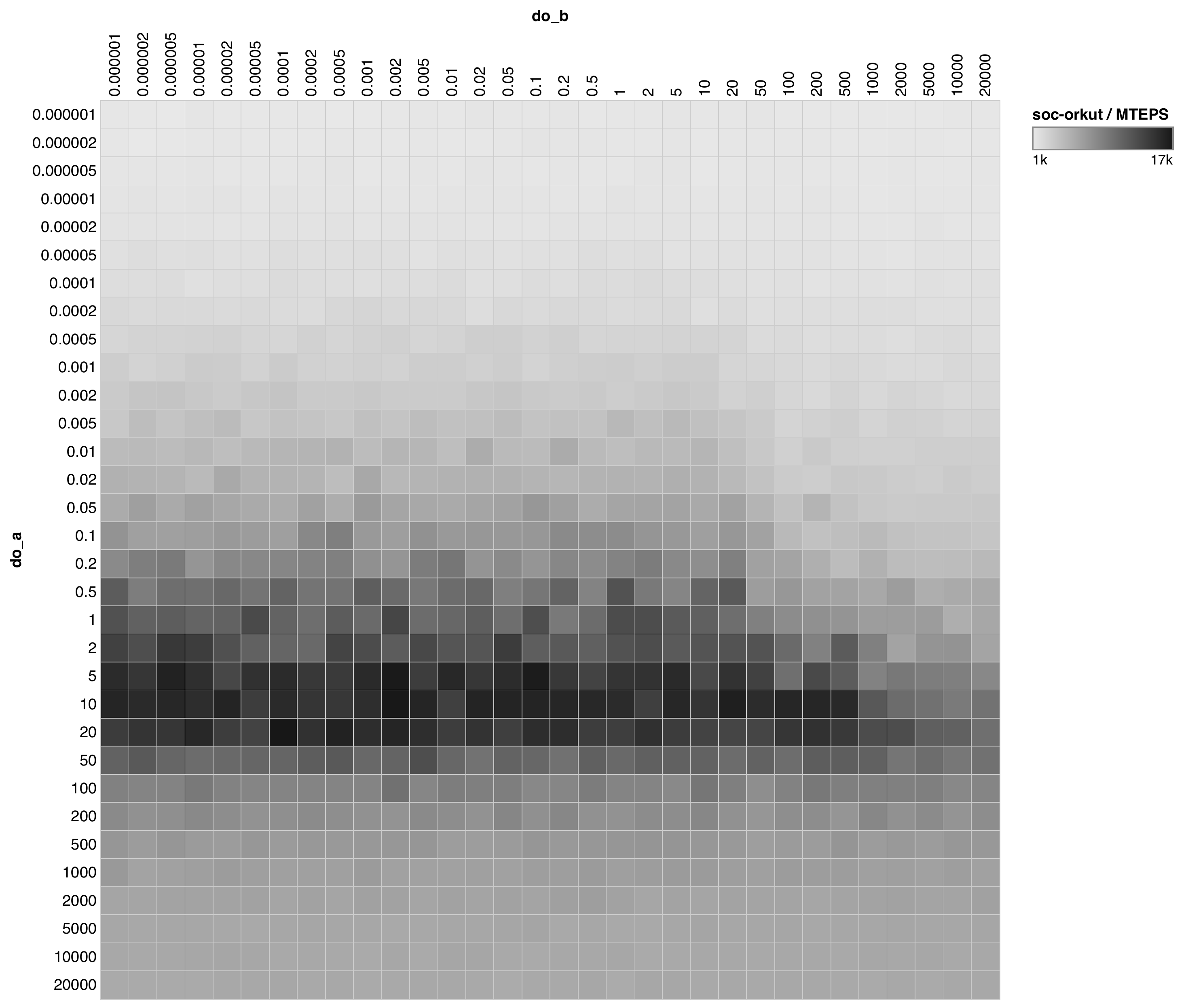}

\caption[Heatmaps of do\_a and do\_b's impact of BFS
performance.]{Heatmaps to show how Gunrock's two direction-optimized
  parameters, \textbf{do\_a}, and \textbf{do\_b} affect BFS
  performance. Each square represents the average of 25 BFS runs, each
  starting at a random node. Darker color means higher TEPS~\@. Three
  datasets of top row are: hollywood-2009, indochina-2004, and
  rmat-n22 (from left to right); three datasets of bottom row are:
  rmat-n23, soc-livejournal, and soc-orkut (from left to right).}
\label{fig:perf_do_heatmap}
\end{figure}

Figure~\ref{fig:perf_do_heatmap} shows performance as a function of
the direction-optimized parameters $do\_a$ and $do\_b$
(Section~\ref{subsec:pull}). In general, R-Mat graphs, social
networks, and web graphs show different patterns due to their
different degree distributions. But every graph has a rectangular
region where performance figures show a discontinuity from the region
outside that rectangle. This is because our two direction-optimized
parameters indirectly affect the number of iterations of push-based
traversal and pull-based traversal. The number of iterations is
discrete and thus forms the rectangular region. As shown in
figure~\ref{fig:perf_do_heatmap}, with one round of BFS execution that
starts with push-based traversal, increasing \textit{do\_a} from a
small to a large value speeds up the switch from push-based to
pull-based traversal, and thus always yield better performance at
first. However, when \textit{do\_a} is too large, it causes pull-based
traversal to start too early and performance drops because usually in
early iterations, small-sized frontiers show better performance on
push-based traversal than pull-based traversal. Parameter
\textit{do\_b} controls when to switch back from pull-based to
push-based traversal. On most graphs, regardless of whether they are
scale-free or mesh-like graphs, keeping a smaller \textit{do\_b} so
that the switch from pull-based to push-based traversal never happens
would help us achieve better performance. However, this is not always
true: in indochina-2004, the location of its rectangular region is at
the lower right, which means either keeping both a larger
\textit{do\_a} and \textit{do\_b} will yield better performance on
this dataset, or the parameter range we have chosen is not wide enough
to show the complete pattern for this dataset. At any rate, it is
clear that the same parameters do not deliver the best performance for
all datasets.

\TwoFig{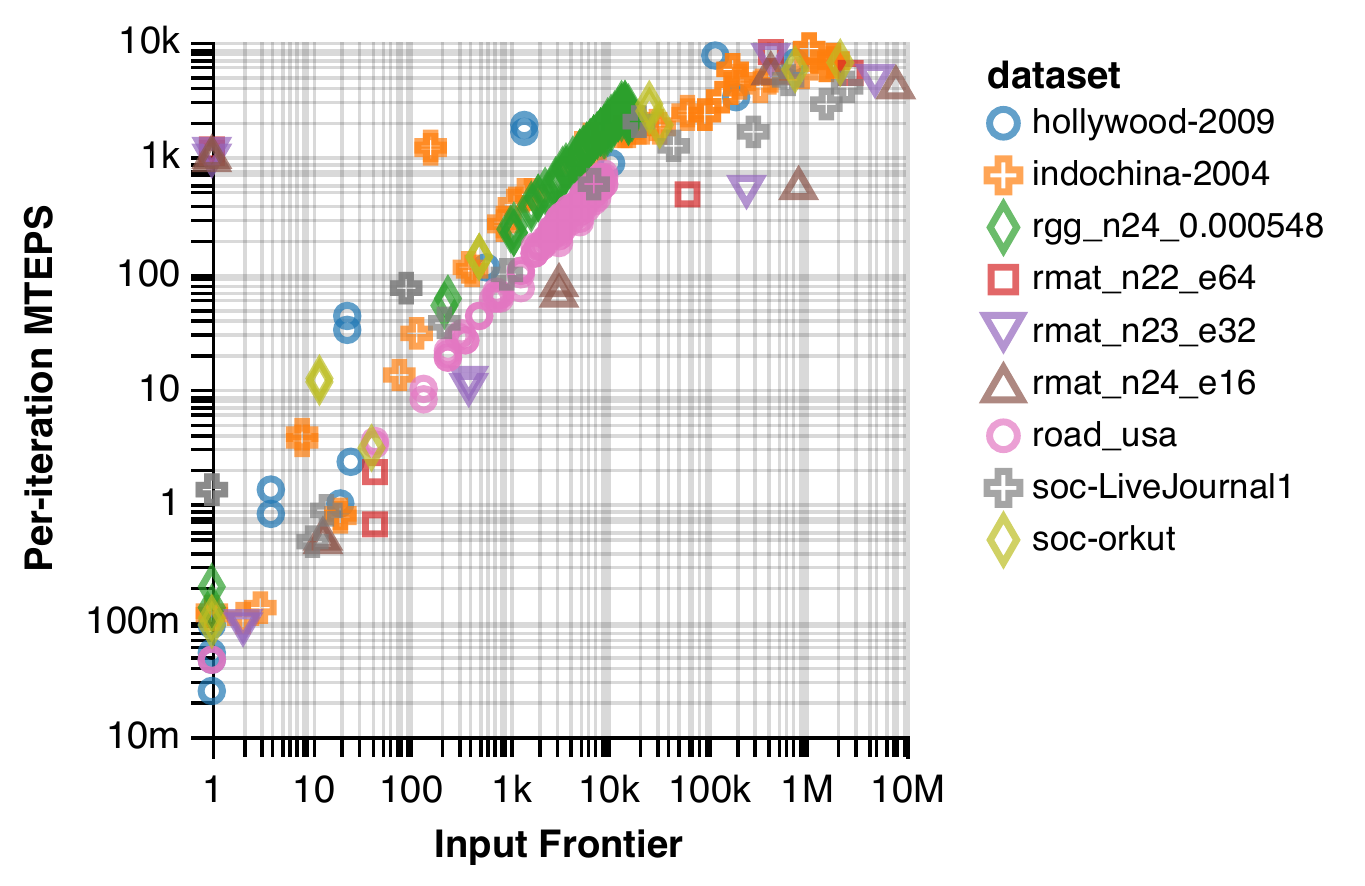}
{Per-iteration advance performance (in MTEPS) vs. input frontier size.}
{fig:perf_input_frontier}
{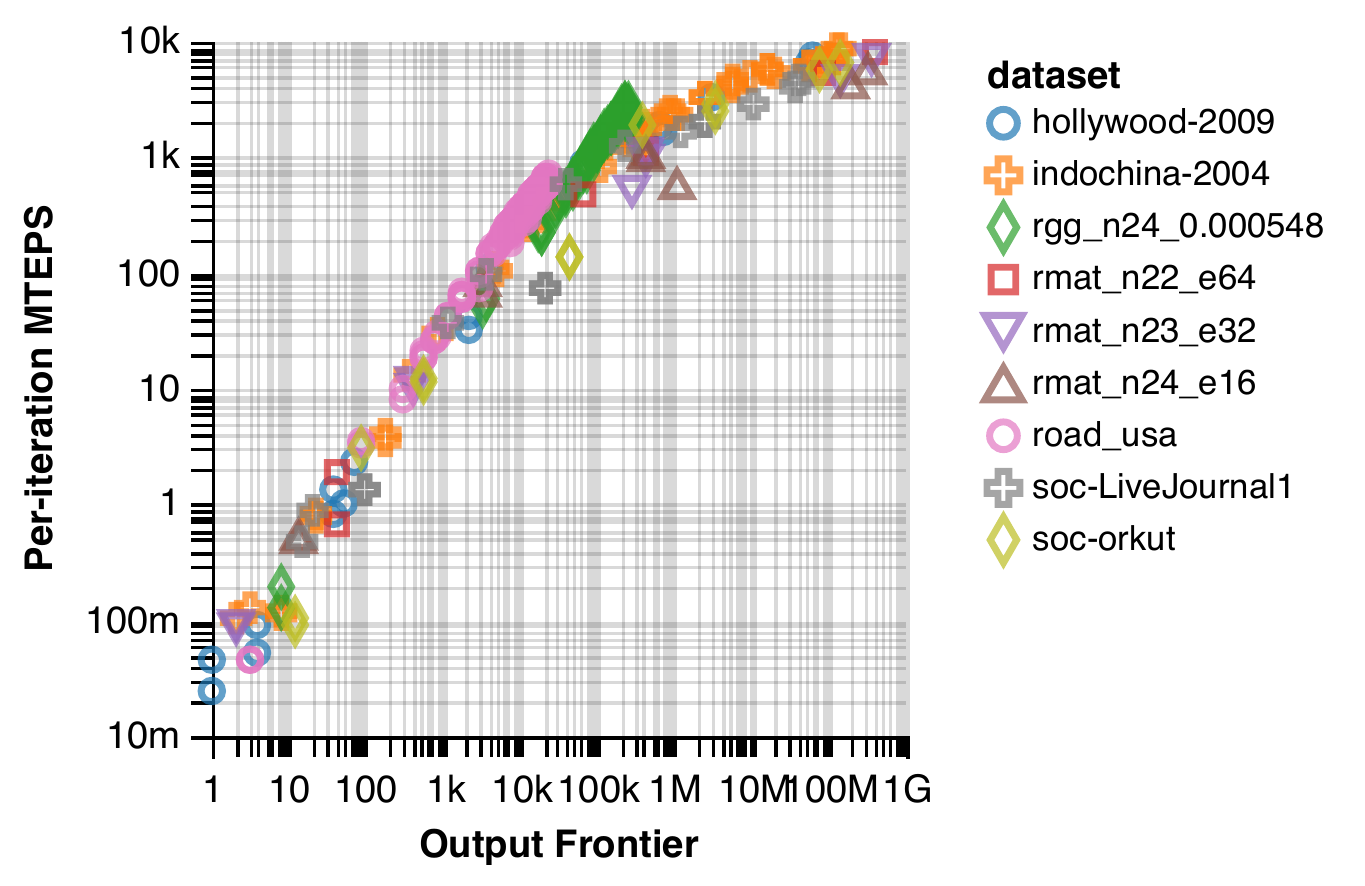}
{Per-iteration advance performance (in MTEPS) vs. output frontier size.}
{fig:perf_output_frontier}

Figures~\ref{fig:perf_input_frontier}
and~\ref{fig:perf_output_frontier} show Gunrock's per-iteration
advance performance as a function of input and output frontier size.
The input frontier figure is noisier because for some scale-free
datasets, a small input frontier can generate a very large output
frontier, causing an outlier in the scatter plot. In these plots, for
the two roadnet graphs (roadnet and rgg), we are using our TWC
strategy, while the other datasets use LB\_CULL\@. This creates two
different types of curve. For datasets that use the LB\_CULL strategy,
the input frontier performance curve reaches a horizontal asymptote
when the input frontier size is larger than 1 million, and the output
frontier performance curve keeps growing even for a billion-element
output frontier. However, the performance increase for datasets that
use LB\_CULL slows down when the output frontier size is larger than
one million. For the two datasets that use the TWC strategy, both the
input frontier performance curve and the output frontier performance
curve are linear, and present better performance than datasets using
LB\_CULL when the output frontier size is larger than 1 million. The
results of this
experiment demonstrate that in general, to achieve good performance,
Gunrock needs a relatively large frontier to fully utilize the
computation resource of a GPU\@.

\subsection{GPU Who-To-Follow Performance Analysis}
We characterize our who-to-follow implementation with different
datasets (Tables~\ref{tab:wtf:dataset}). Table~\ref{tab:wtf:runtimes}
shows the runtimes of this primitive on these datasets. Runtimes are
for GPU computation only and do not include CPU-GPU transfer time. The
wiki-Vote dataset is a real social graph dataset that contains voting
data for Wikipedia administrators; all other datasets are follow
graphs from Twitter and Google
Plus~\cite{Leslovec:2014:SLN,Kwak:2010:WIT}. Twitter09 contains the
complete Twitter follow graph as of 2009; we extract 75\% of its user
size and 50\% of its social relation edge size to form a partial graph
that can fit into the GPU's memory.

\begin{table}
\centering
\caption{Experimental datasets for GPU Who-To-Follow
  algorithm.\label{tab:wtf:dataset}}
\begin{tabular}{*{6}{l}}
\toprule Dataset & Vertices&Edges \\
\midrule wiki-Vote & 7.1k & 103.7k \\
twitter-SNAP & 81.3k & 2.4M \\
gplus-SNAP & 107.6k & 30.5M \\
twitter09 & 30.7M & 680M \\
\bottomrule
\end{tabular}
\end{table}

\subsubsection{Scalability}

\begin{table}
\centering
\caption{GPU WTF runtimes for different graph
  sizes.\label{tab:wtf:runtimes}}
\begin{tabular}{*{6}{r}}
\toprule Time (ms) & wiki-Vote & twitter & gplus & twitter09 \\
\midrule PPR & 0.45 & 0.84 & 4.74 & 832.69\\
CoT & 0.54 & 1.28 & 2.11 & 51.61\\
Money & 2.70 & 5.16 & 18.56 & 158.37\\
Total & 4.37 & 8.36 & 26.57 & 1044.99\\
\bottomrule
\end{tabular}
\end{table}

\begin{figure}
  \centering
  \includegraphics[width=0.9\textwidth]{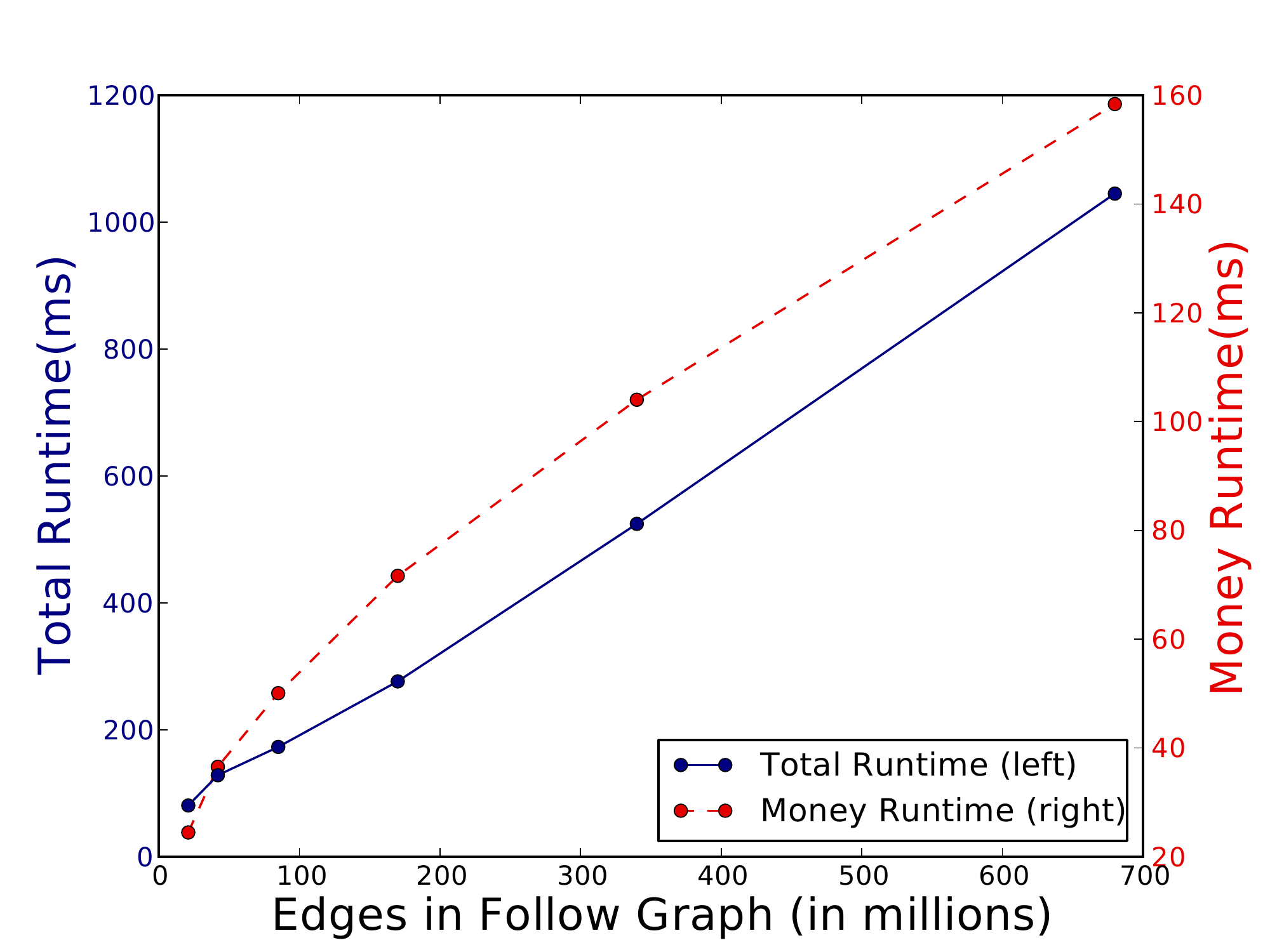}
  \caption{Scalability graph of our GPU recommendation system.\label{fig:runtime-vs-edge-count}}
\end{figure}

In order to test the scalability of our WTF-on-GPU recommendation
system, we ran WTF on six differently-sized subsets of the twitter09
dataset. The results are shown in
Figure~\ref{fig:runtime-vs-edge-count}. We see that the implementation
scales sub-linearly with increasing graph size. As we double the graph
size, the total runtime increases by an average of 1.684x, and the
runtime for Money increases by an average of 1.454x. The reason lies
in our work-efficient parallel implementation. By doing per-vertex
computation exactly once and visiting each edge exactly once, our
parallel algorithm performs linear $O(m+n)$ work. The reason that we
have better scalability for the Money algorithm is that although we
are doubling the graph size each time, we always prune the graph with
a fixed number of nodes that we computed via personalized PageRank
(PPR) and we call ``Circle of Trust'' (CoT). In our implementation we
set the size of CoT to 1000 to match the original Who-To-Follow algorithm.

\subsubsection{Comparison to Cassovary}
We chose to use the Cassovary graph library for our CPU performance
comparison. The results of this comparison are shown in
Table~\ref{tab:cassovary}. Cassovary is a graph library developed at
Twitter. It was the library Twitter used in their first WTF
implementation~\cite{Gupta:2013:WTW}.

\begin{table*}
\centering
\caption{GPU WTF runtimes comparison to Cassovary (C).\label{tab:cassovary}}
\begin{tabular}{*{10}{r}}
\toprule & \multicolumn{2}{c}{wiki-Vote} & \multicolumn{2}{c}{twitter} & \multicolumn{2}{c}{gplus} & \multicolumn{2}{c}{twitter09} \\
\midrule Step (runtime) & C & GPU & C & GPU & C & GPU & C & GPU\\
\midrule PPR (ms) & 418 & 0.45 & 480 & 0.84 & 463 & 4.74 & 884 & 832.69\\
CoT (ms) & 262 & 0.54 & 2173 & 1.28 & 25616 & 2.11 & 2192 & 51.61\\
Money (ms) & 357 & 2.70 & 543 & 5.16 & 2023 & 18.56 & 11216 & 158.37\\
Total (ms) & 1037 & 4.37 & 3196 & 8.36 & 28102 & 26.57 & 14292 & 1044.99\\
Speedup & \multicolumn{2}{r}{235.7} & \multicolumn{2}{r}{380.5} & \multicolumn{2}{r}{1056.5} & \multicolumn{2}{r}{13.7}\\
\bottomrule
\end{tabular}
\end{table*}

We achieve speedups of up to 1000x over Cassovary for the Google Plus
graph, and a speedup of 14x for the 2009 Twitter graph, which is the
most representative dataset for the WTF application. One difference
between the GPU algorithm and the Cassovary algorithm is that we used
the SALSA function that comes with the Cassovary library, instead of
using Twitter's Money algorithm for the final step of the algorithm.
Both are ranking algorithms based on link analysis of bipartite
graphs, and in the original Who-To-Follow paper~\cite{Gupta:2013:WTW},
Gupta et al.\ use a form of SALSA for this step, so this is reasonable
for a comparison.

\subsection{GPU Triangle Counting Performance Analysis}
We compare the performance of our GPU TC to three different exact
triangle counting methods: Green et al.'s state-of-the-art GPU
implementation~\shortcite{Green:2014:FTC} that runs on an NVIDIA K40c
GPU\@, Green et al.'s multicore CPU
implementation~\shortcite{Green:2014:LBC}, and Shun and Tangwongsan's
multicore CPU implementation~\shortcite{Shun:2015:MTC}. Both of the
state-of-the-art CPU implementations are tested on a 40-core shared
memory system with two-way hyper-threading; their results are from
their publications. Our CPU baseline is an implementation based on the
\emph{forward} algorithm by Schank and
Wagner~\shortcite{Schank:2005:FCL}.

\begin{figure}
    \centering
    \begin{minipage}{\textwidth}
    \centering
    \includegraphics[width=0.985\textwidth]{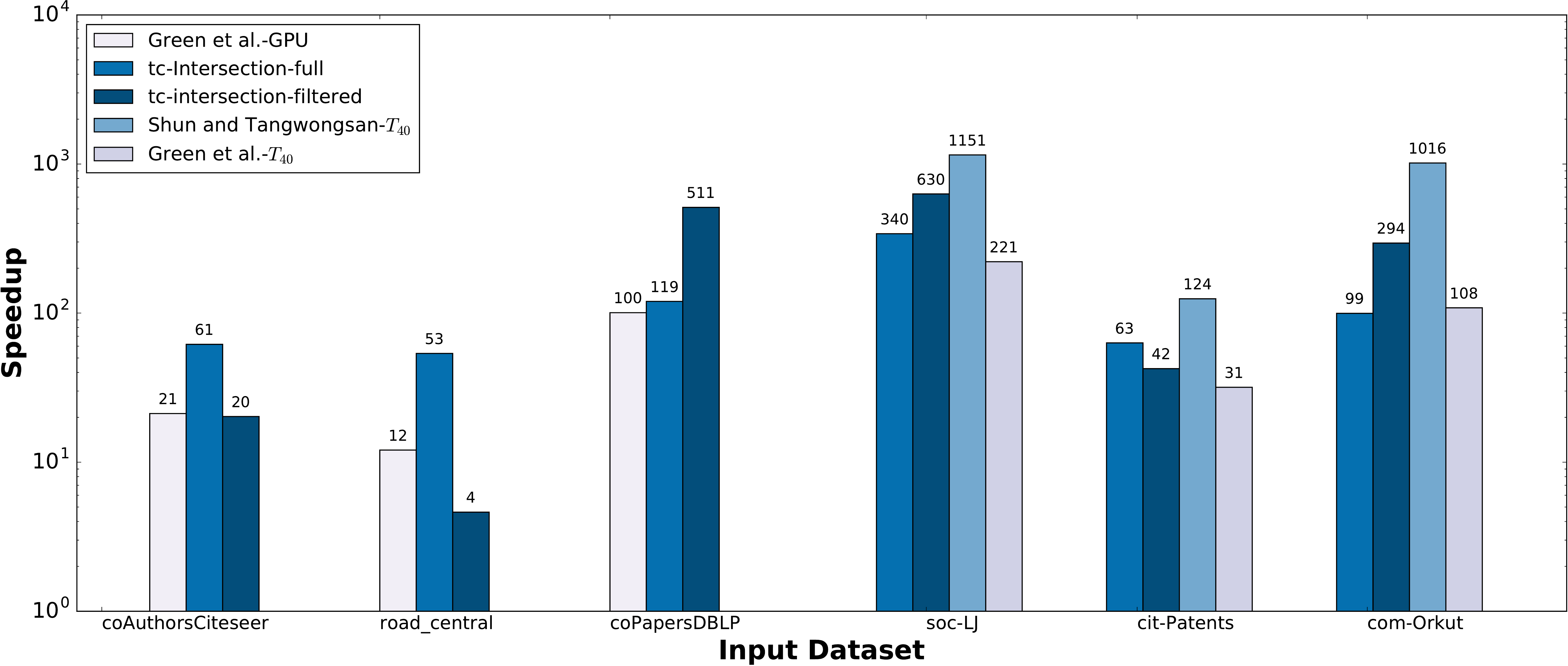}
    \end{minipage}
    \centering
    \begin{adjustbox}{max width=\textwidth}
     \begin{tabular}{*{7}{c}}
      \toprule Dataset &coAuthorsCiteseer&road\_central & coPapersDBLP & soc-LJ & cit-Patents & com-Orkut\\
    \midrule
      time (s) &0.28& 3.254 & 100.7 &564.3 & 9.856 & 1951
    \\ \bottomrule
  \end{tabular}
  \end{adjustbox}
  \caption[Execution-time speedup for our Gunrock TC
  implementations.]{Execution-time speedup (top figure) for our
    Gunrock TC implementations (``tc-intersection-full'' and
    ``tc-intersection-filtered''), Green et al.'s GPU
    implementation~\shortcite{Green:2014:FTC} (``Green et al.-GPU''),
    Shun and Tangwongsan's 40-core CPU
    implementation~\shortcite{Shun:2015:MTC} (``Shun and
    Tangwongsan-$T_{40}$'') and Green et al.'s 40-core CPU
    implementation~\shortcite{Green:2014:LBC} (``Green et
    al.-$T_{40}$''). All are normalized to a baseline CPU
    implementation~\shortcite{Schank:2005:FCL} on six different
    datasets. Baseline runtime (in seconds) given in the
    table.\label{fig:tc_speedup}}
\end{figure}

In general, Gunrock's TC implementation shows better performance than
the state-of-the-art GPU implementations because of our key
optimizations on workload reduction and GPU resource utilization. It
achieves comparable performance to the fastest shared-memory CPU TC
implementation. Gunrock's TC implementation with only the simple
per-thread batch set intersection kernel achieves a 2.51$\times$ and
4.03$\times$ (geometric-mean) speedup as compared to Green et al.'s
CPU~\shortcite{Green:2014:LBC} and GPU~\shortcite{Green:2014:FTC}
implementations respectively. We believe our speedup is the result of
two aspects of our implementation: 1) using filtering in edge list
generation, and reforming the induced subgraph with only the edges not
filtered, effectively reducing five-sixths of the workload; and 2)
dynamic grouping that helps maintain a high GPU resource utilization.
In practice we observe that for scale-free graphs, the last step of
optimization to reform the induced subgraph show a constant speedup
over our intersection method without this step. However, for road
networks and some small scale-free graphs, the overhead of using
segmented reduction will cause a performance drop. We expect further
performance gains from future tuning of launch settings for the
large-neighbor-list intersection kernel, which we do not believe is
quite optimal yet. These optimizations are all transparent to our
data-centric programming model and apply to our segmented intersection
graph operator, which we hope to use in other graph primitives in the
future.

\section{Conclusion}
\label{sec:conc}
Our work on the data-centric programming model for graph processing on
the GPU has enabled us to build a highly programmable,
high-performance graph analytics system, it also opens up various
interesting yet challenging research opportunities.

\subsection{Limitations}
\begin{description}
\item[Single-GPU] Our current data-centric programming model is
  designed and optimized for a single-GPU architecture. We have
  extended our implementation multiple GPUs on one
  node~\cite{Pan:2016:MGA}, but have not yet addressed larger
  machines~\cite{Pearce:2014:FPT}.
\item[Fits-in-memory] Our current single-GPU implementation does not
  support computation on graphs that exceed the size of a GPU's global
  memory.
\item[Static Graph] Our current programming model targets static
  graphs; we have not yet considered streaming graphs or graphs with
  dynamically changing topology.
\item[Limited Compiler Optimizations] Our current implementation does
  not support extensive compiler optimizations such as kernel fusion
  and AST (abstract syntax tree) generation~\cite{Pai:2016:ACT}.
\end{description}

\subsection{Future Work}
Moving forward, there are several aspects we could improve Gunrock,
including architecture, execution model, meta-linguistic abstraction,
performance characterization, core graph operators, new graph
primitives, graph datatype, and usability. We will expand each item
with details in this section.

\subsubsection{Architecture and Execution Model}
To enable large-scale graph analysis where the data do not fit in
single GPU device memory, we see three ways to scale:
\begin{description}
\item[Scale-Out] Gunrock is designed to serve as a standalone
  framework that can be integrated into a multi-GPU single-node graph
  processing system~\cite{Pan:2016:MGA}. When we built this system, we
  identified two interesting research problems: 1)~the impact of
  different partitioning methods on the performance, and 2)~tradeoffs
  between computation and communication for inter-GPU data exchange.
  We hope to extend this framework to multiple nodes and explore these
  two problems more completely, and also investigate dynamic
  repartitioning for load balancing, vertex cut, and heterogeneous
  processing for large-diameter graphs.
\item[Scale-Up] CPU-GPU memory bandwidth is still limited compared to
  intra-GPU and inter-GPU bandwidth. Thus we need to identify the
  situations where out-of-core approaches are necessary and/or useful.
  The current Gunrock framework has not been designed to support
  out-of-core directly, but several other
  projects~\cite{Kyrola:2012:GLG,Shi:2015:OAG} offer interesting
  conclusions that would influence Gunrock's development in terms of
  graph data representation, partitioning, and communication reduction
  to make it suitable for out-of-core computation.
\item[Streaming] Recent work maps the streaming model to GPUs by
  partitioning graphs into edge streams and processing graph analytics
  in parallel~\cite{Roy:2013:XEG,Seo:2015:GGS}. Supporting streaming
  requires more significant changes than out-of-core and raises
  interesting research questions such as how to partition the graph
  and when to synchronize. At its core, we expect Gunrock's
  data-centric programming model maps naturally to streaming process
  where a frontier is an edge stream.
\end{description}

We see two interesting future directions for Gunrock's execution
model:

\begin{description}
\item[Asynchronous] The work of Wang et al.~\shortcite{Wang:2013:ALS}
  views an asynchronous execution model in BSP as the relaxation of
  two synchrony properties: 1)~isolation, meaning within an iteration,
  newly generated messages from other vertices can not be seen; and
  2)~consistency, meaning each thread waits before it proceeds to the
  next iteration until all other threads have finished the process for
  the same iteration. To enable asynchronous execution on current
  Gunrock, we could 1)~use a priority queue in single-node GPU,
  allowing the merging of multiple remote frontiers in multi-node GPU,
  or adding multi-pass micro iterations within one iteration to relax
  the consistency property; or 2)~allow per-vertex/per-edge operations
  to propagate results to others to relax the isolation property. In
  terms of the impact to our data-centric programming model,
  asynchronous execution could significantly accelerate convergence in
  primitives such as SSSP, CC, PR, and LP, by intelligently ordering
  vertex updates and incorporating the most recent updates.
\item[Higher-level task parallelism] Higher-level task parallelism can
  bring more parallelism for several graph primitives. Preliminary
  work~\cite{McLaughlin:2015:FES,Liu:2016:ICB} in this direction
  executes multiple BFS passes simultaneously on GPUs. Applications
  that benefit from adding this additional level of parallelism at the
  streaming multiprocessor level include all-pairs shortest paths and
  BC\@. Usually such higher-level task parallelism requires keeping
  multiple active frontiers and specifying different groups of blocks
  to handle graph workloads on different active frontiers. Gunrock's
  data-centric abstraction can be used to implement this with some
  additional modifications to frontier management and workload
  mapping.
\end{description}

\subsubsection{Meta-Linguistic Abstraction}
\begin{description}
\item[Gunrock as a Backend] Gunrock currently offers C/C++-friendly
  interfaces that make Gunrock callable from Python, Julia, and other
  high-level programming languages. The future work along this path is
  to identify and implement support for Gunrock as a back end to a
  higher-level graph analytics framework (such as TinkerPop or
  NetworkX) or create our own graph query/analysis DSL on top of
  Gunrock.
\item[Add More Backends to Gunrock] Gunrock's current implementation
  is not the only way to implement our data-centric programming model.
  Another interesting research direction is to make other back ends
  support our data-centric programming model. Candidates include
  GraphBLAS (a matrix-based graph processing
  framework)~\cite{Kepner:2016:MFO}, VertexAPI2 (a GAS-style graph
  processing framework)~\cite{Elsen:2013:AVC}, and the EmptyHeaded
  Engine~\cite{Aberger:2015:EBA} (a Boolean algebra
  set-operation-based graph query library).
\end{description}

\subsubsection{Core Graph Operators}
Every graph analytics framework implements its abstraction as a set of
supported operators on a graph. Which operators to choose may be a
function of the hardware architecture and programming language. What
is the right set of graph operators that Gunrock should support? To
Gunrock's original three operators---advance, filter, and compute---we
have added neighborhood reduction and segmented intersection. We see
several other operators as candidates for future Gunrock
implementation:
\begin{description}
\item[Neighborhood Reduction] Neighborhood reduction will visit the
  neighbor list of each item in the input frontier and perform a
  user-specified reduction over each neighbor list. It shares many
  similarities to our advance operator. An efficient neighborhood
  reduction operator would accelerate several graph primitives that
  need to reduce over neighborhoods, such as maximal independent set,
  coloring, and PageRank.
\item[Priority Queue] Using multi-split~\cite{Ashkiani:2016:GM}, we
  can create priority queues with more than one priority level. This
  can be applied to the delta-stepping SSSP algorithm and can also
  serve as an alternative frontier manipulation method in an
  asynchronous execution model.
\item[Sampling] A sampling operator can be viewed as an extension to
  the standard filter. Currently its applications include approximated
  BC, TC, and more approximated graph primitives.
\item[Forming Supervertex] In our current minimum-spanning-tree
  primitive, we have implemented a supervertex-forming phase using a
  series of filter, advance, sort, and prefix-sum. This could be
  integrated into a new operator. Note that currently, we form
  supervertices by rebuilding a new graph that contains supervertices.
  This can be used in hierarchical graph algorithms such as clustering
  and community detection.
\end{description}

An alternative direction would be to offer a set of lower-level
data-parallel primitives including both general primitives such as
prefix-sum and reduction, as well as graph-specific primitives such as
load-balanced search (for evenly distributing edge expanding
workload), and use them to assemble our higher-level graph operators.

Which of these two approaches---implementing many graph operators and
specific optimizations for each operator, and building up graph
operators on top of lower-level data-parallel primitives---do we use
in a GPU graph analytics programming model will affect both the
programmability and performance of the system. In Gunrock, we use a
mixed model where we share common components such as load-balanced
search, prefix-sum, and compact for advance, filter and neighborhood
reduction, while we also include specialized optimizations for
advance, filter, and segmented intersection. An interesting future
research direction is to think of the graph operator abstraction at a
more meta level, and try to build up a hierarchical framework for
designing graph operators that includes not only graph traversal and
computation building blocks such as all five graph operators we have
in Gunrock today, but also memory- and workload-optimization building
blocks that can be plugged into and replaced in our higher-level graph
operator implementation.

\subsubsection{New Graph Primitives}
Our current set of graph primitives are mostly textbook-style
algorithms. One future work is to expand this list to more complex
graph primitives spanning from network analysis, machine learning,
data mining, and graph queries.

\begin{description}
\item[Graph Coloring and Maximal Independent Set] Efficient graph
  matching and coloring algorithms~\cite{Cohen:2012:EGM} can be
  computed by Gunrock more efficiently than previous work. Maximal
  independent set (MIS) algorithms follow the same path. Both
  primitives can be implemented using Gunrock's current operator set
  with new operators such as neighborhood reduction and multi-level
  priority queue potentially improving their performance. These
  primitives can serve as the building blocks for asynchrony and
  task-level parallelism.

\item[Strongly Connected Component] A directed graph is strongly
  connected if any node can reach any other node. This algorithm is
  traditionally a problem related to depth-first search, which is
  considered unsuitable for GPUs. However, Slota et al.'s
  work~\shortcite{Slota:2014:BCP} improved on Barnat et al.'s
  DFS-based work~\shortcite{Barnat:2011:CSC} and avoids DFS by
  combining BFS and graph coloring.

\item[More Traversal-based Algorithms] To make use of our efficient
  graph traversal operators, we can modify our BFS to implement
  several other similar primitives: 1)~\emph{st}-connectivity, which
  simultaneously processes two BFS paths from $s$ and $t$; 2)~A*
  search as a BFS-based path finding and search algorithm;
  3)~Edmonds-Karp as a BFS-based maximum flow algorithm;
  4)~\{reverse\} Cuthill-McKee algorithm, a sparse matrix bandwidth
  reduction algorithm based on BFS; 5)~belief propagation, a
  sum-product message passing algorithm based on BFS; and finally
  6)~radii estimation, a $k$-sample BFS-based algorithm for estimating
  the maximum radius of the graph.

\item[Subgraph Isomorphism] Most optimized subgraph matching algorithms
  on CPUs~\cite{Han:2013:TTU} are based on backtracking strategies which
  follow a recursive routine cannot efficiently be adapted to GPUs.
  Recent subgraph matching on large graphs on the GPU ~\cite{Renz:2015:FSM}
  and distributed memory cloud~\cite{Sun:2012:ESM,Brocheler:2010:CCO},
  both of which use graph traversal extensively,
  and could fit nicely and turn out to work more efficiently in Gunrock's
  framework. However, these methods are memory-bound which suggests a focus
  on more effective filtering and joining methods. Extensions to our current
  segmented intersection operator more flexible input format support for query
  graphs will potentially make subgraph matching in Gunrock perform better.

\item[\emph{k}-SAT] \emph{k}-SAT is a type of Boolean satisfiability
  problem that can be represented as an iterative updating process on
  a bipartite graph. Combining the sampling operator and our efficient
  traversal operator, it potentially fits nicely into Gunrock.

\item[SGD and MCMC] Both Stochastic Gradient Descent in matrix
  completion and Markov Chain Monte Carlo have large amounts of
  parallelism. The former can be presented as either an iterative
  bipartite graph traversal-updating algorithm~\cite{Kaleem:2015:SGD}
  or a conflict-free subgraph parallel updating algorithm. In order to
  decrease the number of conflict updates and increase the data
  re-usage rate, the incomplete matrix is divided into sub-matrices
  where graph coloring can help identify the conflict-free elements in
  those small sub-matrices. Graph coloring can take advantage of
  Gunrock's efficient traversal operator to achieve better
  performance. MCMC's iterative dense update makes it a good candidate
  for matrix-based graph algorithms. However, Gunrock operators can
  also represent this problem.
\end{description}

\subsubsection{New Graph Datatypes}
All graph processing systems on the GPU use CSR and edge lists
internally. We see significant opportunity in designing suitable graph
datatypes and data structures to enable graph primitives on mutable
graphs and matrix-typed graph primitives.

\begin{description}
\item[Mutable Graphs] The meaning of mutable is twofold: mutable by
  primitive, and mutable by input data. \emph{Mutable by algorithm}
  means that the graph primitive changes the graph structure, and
  includes primitives such as MST, community detection, mesh
  refinement, and Karger's mincut. The operations related include
  simple ones such as adding/removing nodes/edges, and more complex
  ones such as forming supervertices. Currently there is no good
  solution for handling general graph mutations on GPUs with
  efficiency. \emph{Mutable by input data} means that we process our
  algorithms on input datasets that change over time (so we would like
  to both incrementally update the graph data structure and
  incrementally update a solution given the incremental change in the
  data structure). We need to provide either approximated results or
  the capability of doing incremental computation.
\item[Adjacency Matrix Form] Matrix-based graph algorithms are also
  widely used for graph processing such as BFS and PageRank. Our
  sparse-matrix sparse-vector operator and its application to
  BFS~\cite{Yang:2015:FSM} has shed light on more applications in
  matrix form such as MIS, PageRank, SSSP, and several spectral
  methods, which are used in algorithms like collaborative filtering.
\item[Optimized Graph Topology Format]The work of Wu et
  al.~\cite{Wu:2013:CAA} shows how data reorganization can help with
  memory uncoalescing. However, for graph analytics, we need to design
  better strategies since the memory access indices (node IDs in the
  frontier) change dynamically every iteration. Both
  CuSha~\cite{Khorasani:2014:CVG} and
  Graphicionado~\cite{Ham:2016:GAH} proposed optimizations on memory
  access via edge-list grouping according to source node ID and
  destination node ID\@. Such optimizations, and a well-designed cache
  framework, would reduce random memory access during graph analytics.
  It is also an open question if bandwidth reduction algorithms such
  as reverse Cuthill-McKee would bring better memory locality for
  graph analytics.
\item[Rich-data-on-\{vertices,edges\}] Complex networks often contain
  rich data on vertices and/or edges. Currently Gunrock puts all the
  information on edges/vertices into a data structure on the GPU,
  which is not ideal. Adding the capability of loading partial
  edge/vertex information onto the GPU could enable graph query tasks
  and allow us to support several network analysis primitives that use
  this rich information during their computation.
\end{description}

\subsection{Summary}
Gunrock was born when we spent two months writing a single hardwired
GPU graph primitive. We knew that for GPUs to make an impact in graph
analytics, we had to raise the level of abstraction in building graph
primitives. With this work, we show that with appropriate high-level
programming model and low-level optimizations, parallel graph
analytics on the GPU can be both simple and efficient. More
specifically, this work has achieved its two high-level goals:
\begin{itemize}
\item Our data-centric, frontier-focused programming model has proven
  to map naturally to the GPU, giving us both good performance and
  good flexibility. We have also found that implementing this
  abstraction has allowed us to integrate numerous optimization
  strategies, including multiple load-balancing strategies for
  traversal, direction-optimal traversal, and a two-level priority
  queue. The result is a framework that is general (able to implement
  numerous simple and complex graph primitives), straightforward to
  program (new primitives only take a few hundred lines of code and
  require minimal GPU programming knowledge), and fast (on par with
  hardwired primitives and faster than any other programmable GPU
  graph library).
\item Our open-sourced GPU graph processing library Gunrock provides a
  graph analytics framework for three types of users: 1)~data
  scientists who want to take the advantage of the GPU's superior
  computing power in big data applications; 2)~algorithm designers who
  want to use the existing efficient graph operators in Gunrock to
  create new graph algorithms and applications; and 3)~researchers who
  want to reproduce the results of our research, or make improvements
  to our core components. We hope that in the future, Gunrock will
  serve as a standard benchmark for graph processing on the GPU\@.
\end{itemize}

\section*{Acknowledgments}
\label{sec:acks}
Thanks to our DARPA program managers Wade Shen and Christopher White,
and DARPA business manager Gabriela Araujo, for their support during
this project. Thanks to the Altair and Vega-lite teams in the
Interactive Data Lab at the University of Washington for graphing
help. Joe Mako provided the speedup chart design. We appreciate the
technical assistance, advice, and machine access from many colleagues
at NVIDIA\@: Chandra Cheij, Joe Eaton, Michael Garland, Mark Harris,
Ujval Kapasi, David Luebke, Duane Merrill, Nikolai Sakharnykh, and
Cliff Woolley. Thanks also to our colleagues at Onu Technology: Erich
Elsen, Guha Jayachandran, and Vishal Vaidyanathan.

We gratefully acknowledge the support of the DARPA XDATA program (US
Army award W911QX-12-C-0059); DARPA STTR awards D14PC00023 and
D15PC00010; NSF awards CCF-1017399, OCI-1032859, and CCF-1629657; and
UC Lab Fees Research Program Award 12-LR-238449.

\bibliographystyle{ACM-Reference-Format}
\bibliography{gunrock}


\begin{thebibliography}{00}


\ifx \showCODEN    \undefined \def \showCODEN     #1{\unskip}     \fi
\ifx \showDOI      \undefined \def \showDOI       #1{{\tt DOI:}\penalty0{#1}\ }
  \fi
\ifx \showISBNx    \undefined \def \showISBNx     #1{\unskip}     \fi
\ifx \showISBNxiii \undefined \def \showISBNxiii  #1{\unskip}     \fi
\ifx \showISSN     \undefined \def \showISSN      #1{\unskip}     \fi
\ifx \showLCCN     \undefined \def \showLCCN      #1{\unskip}     \fi
\ifx \shownote     \undefined \def \shownote      #1{#1}          \fi
\ifx \showarticletitle \undefined \def \showarticletitle #1{#1}   \fi
\ifx \showURL      \undefined \def \showURL       #1{#1}          \fi
\providecommand\bibfield[2]{#2}
\providecommand\bibinfo[2]{#2}
\providecommand\natexlab[1]{#1}
\providecommand\showeprint[2][]{arXiv:#2}

\bibitem[\protect\citeauthoryear{Aberger, N{\"{o}}tzli, Olukotun, and
  R{\'{e}}}{Aberger et~al\mbox{.}}{2015}]%
        {Aberger:2015:EBA}
\bibfield{author}{\bibinfo{person}{Christopher~R. Aberger},
  \bibinfo{person}{Andres N{\"{o}}tzli}, \bibinfo{person}{Kunle Olukotun},
  {and} \bibinfo{person}{Christopher R{\'{e}}}.}
  \bibinfo{year}{2015}\natexlab{}.
\newblock \showarticletitle{EmptyHeaded: Boolean Algebra Based Graph
  Processing}.
\newblock \bibinfo{journal}{{\em CoRR\/}}  \bibinfo{volume}{abs/1503.02368}
  (\bibinfo{year}{2015}).
\newblock
\showURL{%
\url{http://arxiv.org/abs/1503.02368}}


\bibitem[\protect\citeauthoryear{Ashkiani, Davidson, Meyer, and Owens}{Ashkiani
  et~al\mbox{.}}{2016}]%
        {Ashkiani:2016:GM}
\bibfield{author}{\bibinfo{person}{Saman Ashkiani}, \bibinfo{person}{Andrew~A.
  Davidson}, \bibinfo{person}{Ulrich Meyer}, {and} \bibinfo{person}{John~D.
  Owens}.} \bibinfo{year}{2016}\natexlab{}.
\newblock \showarticletitle{{GPU} Multisplit}. In \bibinfo{booktitle}{{\em
  Proceedings of the 21st ACM SIGPLAN Symposium on Principles and Practice of
  Parallel Programming}} {\em (\bibinfo{series}{PPoPP 2016})}.
  \bibinfo{pages}{12:1--12:13}.
\newblock
\showDOI{%
\url{http://dx.doi.org/10.1145/2851141.2851169}}


\bibitem[\protect\citeauthoryear{Barnat, Bauch, Brim, and Ceska}{Barnat
  et~al\mbox{.}}{2011}]%
        {Barnat:2011:CSC}
\bibfield{author}{\bibinfo{person}{Jiri Barnat}, \bibinfo{person}{Petr Bauch},
  \bibinfo{person}{Lubos Brim}, {and} \bibinfo{person}{Milan Ceska}.}
  \bibinfo{year}{2011}\natexlab{}.
\newblock \showarticletitle{Computing Strongly Connected Components in Parallel
  on CUDA}. In \bibinfo{booktitle}{{\em Proceedings of the 2011 IEEE
  International Parallel \& Distributed Processing Symposium}} {\em
  (\bibinfo{series}{IPDPS '11})}. \bibinfo{publisher}{IEEE Computer Society},
  \bibinfo{address}{Washington, DC, USA}, \bibinfo{pages}{544--555}.
\newblock
\showISBNx{978-0-7695-4385-7}
\showDOI{%
\url{http://dx.doi.org/10.1109/IPDPS.2011.59}}


\bibitem[\protect\citeauthoryear{Baxter}{Baxter}{2013}]%
        {Baxter:2013:MGM}
\bibfield{author}{\bibinfo{person}{Sean Baxter}.}
  \bibinfo{year}{2013}\natexlab{}.
\newblock \bibinfo{title}{Modern GPU Multisets}.  (\bibinfo{year}{2013}).
\newblock
\newblock
\shownote{\url{https://nvlabs.github.io/moderngpu/sets.html}.}


\bibitem[\protect\citeauthoryear{Baxter}{Baxter}{2016}]%
        {MGPU:2016}
\bibfield{author}{\bibinfo{person}{Sean Baxter}.}
  \bibinfo{year}{2013--2016}\natexlab{}.
\newblock \bibinfo{title}{Moderngpu: Patterns and Behaviors for {GPU}
  Computing}.  (\bibinfo{year}{2013--2016}).
\newblock
\newblock
\shownote{\url{http://moderngpu.github.io/moderngpu}.}


\bibitem[\protect\citeauthoryear{Beamer, Asanovi\'{c}, and Patterson}{Beamer
  et~al\mbox{.}}{2012}]%
        {Beamer:2012:DBS}
\bibfield{author}{\bibinfo{person}{Scott Beamer}, \bibinfo{person}{Krste
  Asanovi\'{c}}, {and} \bibinfo{person}{David Patterson}.}
  \bibinfo{year}{2012}\natexlab{}.
\newblock \showarticletitle{Direction-Optimizing Breadth-First Search}. In
  \bibinfo{booktitle}{{\em Proceedings of the International Conference on High
  Performance Computing, Networking, Storage and Analysis}} {\em
  (\bibinfo{series}{SC '12})}. Article \bibinfo{articleno}{12},
  \bibinfo{numpages}{10}~pages.
\newblock
\showDOI{%
\url{http://dx.doi.org/10.1109/SC.2012.50}}


\bibitem[\protect\citeauthoryear{Billeter, Olsson, and Assarsson}{Billeter
  et~al\mbox{.}}{2009}]%
        {Billeter:2009:ESC}
\bibfield{author}{\bibinfo{person}{Markus Billeter}, \bibinfo{person}{Ola
  Olsson}, {and} \bibinfo{person}{Ulf Assarsson}.}
  \bibinfo{year}{2009}\natexlab{}.
\newblock \showarticletitle{Efficient Stream Compaction on Wide SIMD Many-core
  Architectures}. In \bibinfo{booktitle}{{\em Proceedings of the Conference on
  High Performance Graphics 2009}} {\em (\bibinfo{series}{HPG '09})}.
  \bibinfo{publisher}{ACM}, \bibinfo{address}{New York, NY, USA},
  \bibinfo{pages}{159--166}.
\newblock
\showISBNx{978-1-60558-603-8}
\showDOI{%
\url{http://dx.doi.org/10.1145/1572769.1572795}}


\bibitem[\protect\citeauthoryear{Brandes}{Brandes}{2001}]%
        {Brandes:2001:AFA}
\bibfield{author}{\bibinfo{person}{Ulrik Brandes}.}
  \bibinfo{year}{2001}\natexlab{}.
\newblock \showarticletitle{A faster algorithm for betweenness centrality}.
\newblock \bibinfo{journal}{{\em Journal of Mathematical Sociology\/}}
  \bibinfo{volume}{25}, \bibinfo{number}{2} (\bibinfo{year}{2001}),
  \bibinfo{pages}{163--177}.
\newblock
\showDOI{%
\url{http://dx.doi.org/10.1080/0022250X.2001.9990249}}


\bibitem[\protect\citeauthoryear{Br\"{o}cheler, Pugliese, and
  Subrahmanian}{Br\"{o}cheler et~al\mbox{.}}{2010}]%
        {Brocheler:2010:CCO}
\bibfield{author}{\bibinfo{person}{M. Br\"{o}cheler}, \bibinfo{person}{A.
  Pugliese}, {and} \bibinfo{person}{V.~S. Subrahmanian}.}
  \bibinfo{year}{2010}\natexlab{}.
\newblock \showarticletitle{{COSI}: Cloud Oriented Subgraph Identification in
  Massive Social Networks}. In \bibinfo{booktitle}{{\em Proceedings of the 2010
  International Conference on Advances in Social Networks Analysis and Mining}}
  {\em (\bibinfo{series}{ASONAM '10})}. \bibinfo{publisher}{IEEE Computer
  Society}, \bibinfo{address}{Washington, DC, USA}, \bibinfo{pages}{248--255}.
\newblock
\showISBNx{978-0-7695-4138-9}
\showDOI{%
\url{http://dx.doi.org/10.1109/ASONAM.2010.80}}


\bibitem[\protect\citeauthoryear{Burtscher, Nasre, and Pingali}{Burtscher
  et~al\mbox{.}}{2012a}]%
        {Burtscher:2012:QSI}
\bibfield{author}{\bibinfo{person}{Martin Burtscher}, \bibinfo{person}{Rupesh
  Nasre}, {and} \bibinfo{person}{Keshav Pingali}.}
  \bibinfo{year}{2012}\natexlab{a}.
\newblock \showarticletitle{A Quantitative Study of Irregular Programs on
  {GPU}s}. In \bibinfo{booktitle}{{\em Proceedings of the 2012 IEEE
  International Symposium on Workload Characterization (IISWC)}} {\em
  (\bibinfo{series}{IISWC '12})}. \bibinfo{publisher}{IEEE Computer Society},
  \bibinfo{address}{Washington, DC, USA}, \bibinfo{pages}{141--151}.
\newblock
\showISBNx{978-1-4673-4531-6}
\showDOI{%
\url{http://dx.doi.org/10.1109/IISWC.2012.6402918}}


\bibitem[\protect\citeauthoryear{Burtscher, Nasre, and Pingali}{Burtscher
  et~al\mbox{.}}{2012b}]%
        {Burtscher:2012:AQS}
\bibfield{author}{\bibinfo{person}{Martin Burtscher}, \bibinfo{person}{Rupesh
  Nasre}, {and} \bibinfo{person}{Keshav Pingali}.}
  \bibinfo{year}{2012}\natexlab{b}.
\newblock \showarticletitle{A Quantitative Study of Irregular Programs on
  {GPU}s}. In \bibinfo{booktitle}{{\em IEEE International Symposium on Workload
  Characterization}} {\em (\bibinfo{series}{IISWC-2012})}.
  \bibinfo{pages}{141--151}.
\newblock
\showDOI{%
\url{http://dx.doi.org/10.1109/IISWC.2012.6402918}}


\bibitem[\protect\citeauthoryear{Busato and Bombieri}{Busato and
  Bombieri}{2015}]%
        {Busato:2015:BAE}
\bibfield{author}{\bibinfo{person}{F. Busato} {and} \bibinfo{person}{N.
  Bombieri}.} \bibinfo{year}{2015}\natexlab{}.
\newblock \showarticletitle{{BFS}-4{K}: An Efficient Implementation of {BFS}
  for Kepler {GPU} Architectures}.
\newblock \bibinfo{journal}{{\em IEEE Transactions on Parallel and Distributed
  Systems\/}} \bibinfo{volume}{26}, \bibinfo{number}{7} (\bibinfo{date}{July}
  \bibinfo{year}{2015}), \bibinfo{pages}{1826--1838}.
\newblock
\showISSN{1045-9219}
\showDOI{%
\url{http://dx.doi.org/10.1109/TPDS.2014.2330597}}


\bibitem[\protect\citeauthoryear{Cederman and Tsigas}{Cederman and
  Tsigas}{2008}]%
        {Cederman:2008:ODL}
\bibfield{author}{\bibinfo{person}{Daniel Cederman} {and}
  \bibinfo{person}{Philippas Tsigas}.} \bibinfo{year}{2008}\natexlab{}.
\newblock \showarticletitle{On Dynamic Load-Balancing on Graphics Processors}.
  In \bibinfo{booktitle}{{\em Graphics Hardware \textln{2008}}}.
  \bibinfo{pages}{57--64}.
\newblock
\showDOI{%
\url{http://dx.doi.org/10.2312/EGGH/EGGH08/057-064}}


\bibitem[\protect\citeauthoryear{Chen, Luo, Liu, Zhang, He, Wang, Li, Chen, Xu,
  Sun, and Temam}{Chen et~al\mbox{.}}{2016}]%
        {Ham:2016:GAH}
\bibfield{author}{\bibinfo{person}{Yunji Chen}, \bibinfo{person}{Tao Luo},
  \bibinfo{person}{Shaoli Liu}, \bibinfo{person}{Shijin Zhang},
  \bibinfo{person}{Liqiang He}, \bibinfo{person}{Jia Wang},
  \bibinfo{person}{Ling Li}, \bibinfo{person}{Tianshi Chen},
  \bibinfo{person}{Zhiwei Xu}, \bibinfo{person}{Ninghui Sun}, {and}
  \bibinfo{person}{Olivier Temam}.} \bibinfo{year}{2016}\natexlab{}.
\newblock \showarticletitle{Graphicionado: A High-Performance and
  Energy-Efficient Accelerator for Graph Analytics}. In
  \bibinfo{booktitle}{{\em Proceedings of the 49th Annual IEEE/ACM
  International Symposium on Microarchitecture}} {\em
  (\bibinfo{series}{MICRO-49})}. \bibinfo{publisher}{IEEE Computer Society},
  \bibinfo{address}{Washington, DC, USA}.
\newblock


\bibitem[\protect\citeauthoryear{Cohen and CastonGuay}{Cohen and
  CastonGuay}{2012}]%
        {Cohen:2012:EGM}
\bibfield{author}{\bibinfo{person}{Jonathan Cohen} {and}
  \bibinfo{person}{Patrice CastonGuay}.} \bibinfo{year}{2012}\natexlab{}.
\newblock \showarticletitle{Efficient Graph Matching and Coloring on the
  {GPU}}.
\newblock \bibinfo{journal}{{\em GPU Technology Conference\/}}
  (\bibinfo{date}{March} \bibinfo{year}{2012}).
\newblock
\showURL{%
\url{http://on-demand.gputechconf.com/gtc/2012/presentations/S0332-Efficient-Graph-Matching-and-Coloring-on-GPUs.pdf}}


\bibitem[\protect\citeauthoryear{Davidson, Baxter, Garland, and Owens}{Davidson
  et~al\mbox{.}}{2014}]%
        {Davidson:2014:WPG}
\bibfield{author}{\bibinfo{person}{Andrew Davidson}, \bibinfo{person}{Sean
  Baxter}, \bibinfo{person}{Michael Garland}, {and} \bibinfo{person}{John~D.
  Owens}.} \bibinfo{year}{2014}\natexlab{}.
\newblock \showarticletitle{Work-Efficient Parallel {GPU} Methods for Single
  Source Shortest Paths}. In \bibinfo{booktitle}{{\em Proceedings of the 28th
  IEEE International Parallel and Distributed Processing Symposium}} {\em
  (\bibinfo{series}{IPDPS 2014})}. \bibinfo{pages}{349--359}.
\newblock
\showDOI{%
\url{http://dx.doi.org/10.1109/IPDPS.2014.45}}


\bibitem[\protect\citeauthoryear{Delling, Goldberg, Nowatzyk, and
  Werneck}{Delling et~al\mbox{.}}{2010}]%
        {Delling:2010:PHS}
\bibfield{author}{\bibinfo{person}{Daniel Delling}, \bibinfo{person}{Andrew~V.
  Goldberg}, \bibinfo{person}{Andreas Nowatzyk}, {and}
  \bibinfo{person}{Renato~F. Werneck}.} \bibinfo{year}{2010}\natexlab{}.
\newblock \showarticletitle{{PHAST}: Hardware-accelerated shortest path trees}.
\newblock \bibinfo{journal}{{\it J. Parallel and Distrib. Comput.}}
  \bibinfo{volume}{73} (\bibinfo{date}{Sept.} \bibinfo{year}{2010}),
  \bibinfo{pages}{940--952}.
\newblock
\showDOI{%
\url{http://dx.doi.org/10.1016/j.jpdc.2012.02.007}}


\bibitem[\protect\citeauthoryear{Elsen and Vaidyanathan}{Elsen and
  Vaidyanathan}{2013}]%
        {Elsen:2013:AVC}
\bibfield{author}{\bibinfo{person}{Erich Elsen} {and} \bibinfo{person}{Vishal
  Vaidyanathan}.} \bibinfo{year}{2013}\natexlab{}.
\newblock \bibinfo{title}{A vertex-centric {CUDA/C++} {API} for large graph
  analytics on {GPU}s using the Gather-Apply-Scatter abstraction}.
\newblock   (\bibinfo{year}{2013}).
\newblock
\newblock
\shownote{\url{http://www.github.com/RoyalCaliber/vertexAPI2}.}


\bibitem[\protect\citeauthoryear{Fu, Personick, and Thompson}{Fu
  et~al\mbox{.}}{2014}]%
        {Fu:2014:MAH}
\bibfield{author}{\bibinfo{person}{Zhisong Fu}, \bibinfo{person}{Michael
  Personick}, {and} \bibinfo{person}{Bryan Thompson}.}
  \bibinfo{year}{2014}\natexlab{}.
\newblock \showarticletitle{{MapGraph}: A High Level {API} for Fast Development
  of High Performance Graph Analytics on {GPU}s}. In \bibinfo{booktitle}{{\em
  Proceedings of the Workshop on GRAph Data Management Experiences and
  Systems}} {\em (\bibinfo{series}{GRADES '14})}. Article
  \bibinfo{articleno}{2}, \bibinfo{numpages}{6}~pages.
\newblock
\showDOI{%
\url{http://dx.doi.org/10.1145/2621934.2621936}}


\bibitem[\protect\citeauthoryear{Geil, Wang, and Owens}{Geil
  et~al\mbox{.}}{2014}]%
        {Geil:2014:WGC}
\bibfield{author}{\bibinfo{person}{Afton Geil}, \bibinfo{person}{Yangzihao
  Wang}, {and} \bibinfo{person}{John~D. Owens}.}
  \bibinfo{year}{2014}\natexlab{}.
\newblock \showarticletitle{{WTF}, {GPU}! {C}omputing {T}witter's Who-To-Follow
  on the {GPU}}. In \bibinfo{booktitle}{{\em Proceedings of the Second ACM
  Conference on Online Social Networks}} {\em (\bibinfo{series}{COSN '14})}.
  \bibinfo{pages}{63--68}.
\newblock
\showDOI{%
\url{http://dx.doi.org/10.1145/2660460.2660481}}


\bibitem[\protect\citeauthoryear{Gharaibeh, Reza, Santos-Neto, Costa, Sallinen,
  and Ripeanu}{Gharaibeh et~al\mbox{.}}{2014}]%
        {Gharaibeh:2014:ELS}
\bibfield{author}{\bibinfo{person}{Abdullah Gharaibeh}, \bibinfo{person}{Tahsin
  Reza}, \bibinfo{person}{Elizeu Santos-Neto}, \bibinfo{person}{Lauro~Beltrao
  Costa}, \bibinfo{person}{Scott Sallinen}, {and} \bibinfo{person}{Matei
  Ripeanu}.} \bibinfo{year}{2014}\natexlab{}.
\newblock \showarticletitle{Efficient Large-Scale Graph Processing on Hybrid
  {CPU} and {GPU} Systems}.
\newblock \bibinfo{journal}{{\em CoRR\/}} \bibinfo{volume}{abs/1312.3018},
  \bibinfo{number}{1312.3018v2} (\bibinfo{date}{Dec.} \bibinfo{year}{2014}).
\newblock
\showeprint{1312.3018v2}


\bibitem[\protect\citeauthoryear{Goel, Gupta, Sirois, Wang, Sharma, and
  Gurumurthy}{Goel et~al\mbox{.}}{2015}]%
        {Goel:2015:WST}
\bibfield{author}{\bibinfo{person}{Ashish Goel}, \bibinfo{person}{Pankaj
  Gupta}, \bibinfo{person}{John Sirois}, \bibinfo{person}{Dong Wang},
  \bibinfo{person}{Aneesh Sharma}, {and} \bibinfo{person}{Siva Gurumurthy}.}
  \bibinfo{year}{2015}\natexlab{}.
\newblock \showarticletitle{The {W}ho-{T}o-{F}ollow System at {T}witter:
  Strategy, Algorithms, and Revenue Impact}.
\newblock \bibinfo{journal}{{\em Interfaces\/}} \bibinfo{volume}{45},
  \bibinfo{number}{1} (\bibinfo{date}{Feb.} \bibinfo{year}{2015}),
  \bibinfo{pages}{98--107}.
\newblock
\showISSN{0092-2102}
\showDOI{%
\url{http://dx.doi.org/10.1287/inte.2014.0784}}


\bibitem[\protect\citeauthoryear{Gonzalez, Low, Gu, Bickson, and
  Guestrin}{Gonzalez et~al\mbox{.}}{2012}]%
        {Gonzalez:2012:PDG}
\bibfield{author}{\bibinfo{person}{Joseph~E. Gonzalez},
  \bibinfo{person}{Yucheng Low}, \bibinfo{person}{Haijie Gu},
  \bibinfo{person}{Danny Bickson}, {and} \bibinfo{person}{Carlos Guestrin}.}
  \bibinfo{year}{2012}\natexlab{}.
\newblock \showarticletitle{{PowerGraph}: Distributed Graph-Parallel
  Computation on Natural Graphs}. In \bibinfo{booktitle}{{\em Proceedings of
  the 10th USENIX Conference on Operating Systems Design and Implementation}}
  {\em (\bibinfo{series}{OSDI '12})}. \bibinfo{publisher}{USENIX Association},
  \bibinfo{pages}{17--30}.
\newblock


\bibitem[\protect\citeauthoryear{Gonzalez, Xin, Dave, Crankshaw, Franklin, and
  Stoica}{Gonzalez et~al\mbox{.}}{2014}]%
        {Gonzalez:2014:GGP}
\bibfield{author}{\bibinfo{person}{Joseph~E. Gonzalez},
  \bibinfo{person}{Reynold~S. Xin}, \bibinfo{person}{Ankur Dave},
  \bibinfo{person}{Daniel Crankshaw}, \bibinfo{person}{Michael~J. Franklin},
  {and} \bibinfo{person}{Ion Stoica}.} \bibinfo{year}{2014}\natexlab{}.
\newblock \showarticletitle{{GraphX}: Graph Processing in a Distributed
  Dataflow Framework}. In \bibinfo{booktitle}{{\em Proceedings of the 11th
  USENIX Conference on Operating Systems Design and Implementation}} {\em
  (\bibinfo{series}{OSDI'14})}. \bibinfo{publisher}{USENIX Association},
  \bibinfo{address}{Berkeley, CA, USA}, \bibinfo{pages}{599--613}.
\newblock
\showISBNx{978-1-931971-16-4}
\showURL{%
\url{http://dl.acm.org/citation.cfm?id=2685048.2685096}}


\bibitem[\protect\citeauthoryear{Green, McColl, and Bader}{Green
  et~al\mbox{.}}{2012}]%
        {Green:2012:GMP}
\bibfield{author}{\bibinfo{person}{Oded Green}, \bibinfo{person}{Robert
  McColl}, {and} \bibinfo{person}{David~A. Bader}.}
  \bibinfo{year}{2012}\natexlab{}.
\newblock \showarticletitle{{GPU} Merge Path: A {GPU} Merging Algorithm}. In
  \bibinfo{booktitle}{{\em Proceedings of the 26th ACM International Conference
  on Supercomputing}} {\em (\bibinfo{series}{ICS '12})}.
  \bibinfo{publisher}{ACM}, \bibinfo{address}{New York, NY, USA},
  \bibinfo{pages}{331--340}.
\newblock
\showISBNx{978-1-4503-1316-2}
\showDOI{%
\url{http://dx.doi.org/10.1145/2304576.2304621}}


\bibitem[\protect\citeauthoryear{Green, Mungu\'{\i}a, and Bader}{Green
  et~al\mbox{.}}{2014a}]%
        {Green:2014:LBC}
\bibfield{author}{\bibinfo{person}{Oded Green},
  \bibinfo{person}{Llu\'{\i}s-Miquel Mungu\'{\i}a}, {and}
  \bibinfo{person}{David~A. Bader}.} \bibinfo{year}{2014}\natexlab{a}.
\newblock \showarticletitle{Load Balanced Clustering Coefficients}. In
  \bibinfo{booktitle}{{\em Proceedings of the First Workshop on Parallel
  Programming for Analytics Applications}} {\em (\bibinfo{series}{PPAA '14})}.
  \bibinfo{pages}{3--10}.
\newblock
\showISBNx{978-1-4503-2654-4}
\showDOI{%
\url{http://dx.doi.org/10.1145/2567634.2567635}}


\bibitem[\protect\citeauthoryear{Green, Yalamanchili, and Mungu\'{\i}a}{Green
  et~al\mbox{.}}{2014b}]%
        {Green:2014:FTC}
\bibfield{author}{\bibinfo{person}{Oded Green}, \bibinfo{person}{Pavan
  Yalamanchili}, {and} \bibinfo{person}{Llu\'{\i}s-Miquel Mungu\'{\i}a}.}
  \bibinfo{year}{2014}\natexlab{b}.
\newblock \showarticletitle{Fast Triangle Counting on the {GPU}}. In
  \bibinfo{booktitle}{{\em Proceedings of the Fourth Workshop on Irregular
  Applications: Architectures and Algorithms}} {\em (\bibinfo{series}{IA3
  '14})}. \bibinfo{pages}{1--8}.
\newblock
\showISBNx{978-1-4799-7056-8}
\showDOI{%
\url{http://dx.doi.org/10.1109/IA3.2014.7}}


\bibitem[\protect\citeauthoryear{Gregor and Lumsdaine}{Gregor and
  Lumsdaine}{2005}]%
        {Gregor:2005:PBG}
\bibfield{author}{\bibinfo{person}{Douglas Gregor} {and}
  \bibinfo{person}{Andrew Lumsdaine}.} \bibinfo{year}{2005}\natexlab{}.
\newblock \showarticletitle{The Parallel {BGL}: A Generic Library for
  Distributed Graph Computations}. In \bibinfo{booktitle}{{\em Parallel
  Object-Oriented Scientific Computing (POOSC)}}.
\newblock


\bibitem[\protect\citeauthoryear{Greiner}{Greiner}{1994}]%
        {Greiner:1994:CPA}
\bibfield{author}{\bibinfo{person}{John Greiner}.}
  \bibinfo{year}{1994}\natexlab{}.
\newblock \showarticletitle{A Comparison of Parallel Algorithms for Connected
  Components}. In \bibinfo{booktitle}{{\em Proceedings of the Sixth Annual ACM
  Symposium on Parallel Algorithms and Architectures}} {\em
  (\bibinfo{series}{SPAA '94})}. \bibinfo{pages}{16--25}.
\newblock
\showDOI{%
\url{http://dx.doi.org/10.1145/181014.181021}}


\bibitem[\protect\citeauthoryear{Gupta, Goel, Lin, Sharma, Wang, and
  Zadeh}{Gupta et~al\mbox{.}}{2013}]%
        {Gupta:2013:WTW}
\bibfield{author}{\bibinfo{person}{Pankaj Gupta}, \bibinfo{person}{Ashish
  Goel}, \bibinfo{person}{Jimmy Lin}, \bibinfo{person}{Aneesh Sharma},
  \bibinfo{person}{Dong Wang}, {and} \bibinfo{person}{Reza Zadeh}.}
  \bibinfo{year}{2013}\natexlab{}.
\newblock \showarticletitle{{WTF}: The Who to Follow Service at {T}witter}. In
  \bibinfo{booktitle}{{\em Proceedings of the International Conference on the
  World Wide Web}}. \bibinfo{pages}{505--514}.
\newblock


\bibitem[\protect\citeauthoryear{Han, Lee, and Lee}{Han et~al\mbox{.}}{2013}]%
        {Han:2013:TTU}
\bibfield{author}{\bibinfo{person}{Wook-Shin Han}, \bibinfo{person}{Jinsoo
  Lee}, {and} \bibinfo{person}{Jeong-Hoon Lee}.}
  \bibinfo{year}{2013}\natexlab{}.
\newblock \showarticletitle{Turboiso: Towards Ultrafast and Robust Subgraph
  Isomorphism Search in Large Graph Databases}. In \bibinfo{booktitle}{{\em
  Proceedings of the 2013 ACM SIGMOD International Conference on Management of
  Data}} {\em (\bibinfo{series}{SIGMOD '13})}. \bibinfo{publisher}{ACM},
  \bibinfo{address}{New York, NY, USA}, \bibinfo{pages}{337--348}.
\newblock
\showISBNx{978-1-4503-2037-5}
\showDOI{%
\url{http://dx.doi.org/10.1145/2463676.2465300}}


\bibitem[\protect\citeauthoryear{Harish and Narayanan}{Harish and
  Narayanan}{2007}]%
        {Harish:2007:ALG}
\bibfield{author}{\bibinfo{person}{Pawan Harish} {and} \bibinfo{person}{P.~J.
  Narayanan}.} \bibinfo{year}{2007}\natexlab{}.
\newblock \showarticletitle{Accelerating large graph algorithms on the {GPU}
  using {CUDA}}. In \bibinfo{booktitle}{{\em Proceedings of the 14th
  International Conference on High Performance Computing}} {\em
  (\bibinfo{series}{HiPC'07})}. \bibinfo{publisher}{Springer-Verlag},
  \bibinfo{address}{Berlin, Heidelberg}, \bibinfo{pages}{197--208}.
\newblock
\showDOI{%
\url{http://dx.doi.org/10.1007/978-3-540-77220-0_21}}


\bibitem[\protect\citeauthoryear{Harris, Owens, Sengupta, Zhang, and
  Davidson}{Harris et~al\mbox{.}}{2016}]%
        {Harris:2016:CUDPP}
\bibfield{author}{\bibinfo{person}{Mark Harris}, \bibinfo{person}{John~D.
  Owens}, \bibinfo{person}{Shubho Sengupta}, \bibinfo{person}{Yao Zhang}, {and}
  \bibinfo{person}{Andrew Davidson}.} \bibinfo{year}{2009--2016}\natexlab{}.
\newblock \bibinfo{title}{{CUDPP}: {CUDA} Data Parallel Primitives Library}.
  (\bibinfo{year}{2009--2016}).
\newblock
\newblock
\shownote{\url{http://cudpp.github.io/}.}


\bibitem[\protect\citeauthoryear{Hong, Chafi, Sedlar, and Olukotun}{Hong
  et~al\mbox{.}}{2012}]%
        {Hong:2012:GDE}
\bibfield{author}{\bibinfo{person}{Sungpack Hong}, \bibinfo{person}{Hassan
  Chafi}, \bibinfo{person}{Edic Sedlar}, {and} \bibinfo{person}{Kunle
  Olukotun}.} \bibinfo{year}{2012}\natexlab{}.
\newblock \showarticletitle{{Green-Marl}: A {DSL} for Easy and Efficient Graph
  Analysis}. In \bibinfo{booktitle}{{\em Proceedings of the Seventeenth
  International Conference on Architectural Support for Programming Languages
  and Operating Systems}} {\em (\bibinfo{series}{ASPLOS XVII})}.
  \bibinfo{pages}{349--362}.
\newblock
\showDOI{%
\url{http://dx.doi.org/10.1145/2189750.2151013}}


\bibitem[\protect\citeauthoryear{Hong, Kim, Oguntebi, and Olukotun}{Hong
  et~al\mbox{.}}{2011}]%
        {Hong:2011:ACG}
\bibfield{author}{\bibinfo{person}{Sungpack Hong}, \bibinfo{person}{Sang~Kyun
  Kim}, \bibinfo{person}{Tayo Oguntebi}, {and} \bibinfo{person}{Kunle
  Olukotun}.} \bibinfo{year}{2011}\natexlab{}.
\newblock \showarticletitle{Accelerating {CUDA} Graph Algorithms at Maximum
  Warp}. In \bibinfo{booktitle}{{\em Proceedings of the 16th ACM Symposium on
  Principles and Practice of Parallel Programming}} {\em
  (\bibinfo{series}{PPoPP '11})}. \bibinfo{pages}{267--276}.
\newblock
\showDOI{%
\url{http://dx.doi.org/10.1145/1941553.1941590}}


\bibitem[\protect\citeauthoryear{Jia, Lu, Hoberock, Garland, and Hart}{Jia
  et~al\mbox{.}}{2011}]%
        {Jia:2011:ENP}
\bibfield{author}{\bibinfo{person}{Yuntao Jia}, \bibinfo{person}{Victor Lu},
  \bibinfo{person}{Jared Hoberock}, \bibinfo{person}{Michael Garland}, {and}
  \bibinfo{person}{John~C. Hart}.} \bibinfo{year}{2011}\natexlab{}.
\newblock \showarticletitle{Edge v.\ Node Parallelism for Graph Centrality
  Metrics}.
\newblock In \bibinfo{booktitle}{{\em GPU Computing Gems Jade Edition}},
  \bibfield{editor}{\bibinfo{person}{Wen{-mei}~W. Hwu}} (Ed.).
  \bibinfo{publisher}{Morgan Kaufmann}, Chapter~2, \bibinfo{pages}{15--28}.
\newblock
\showDOI{%
\url{http://dx.doi.org/10.1016/B978-0-12-385963-1.00002-2}}


\bibitem[\protect\citeauthoryear{Kaleem, Pai, and Pingali}{Kaleem
  et~al\mbox{.}}{2015}]%
        {Kaleem:2015:SGD}
\bibfield{author}{\bibinfo{person}{Rashid Kaleem}, \bibinfo{person}{Sreepathi
  Pai}, {and} \bibinfo{person}{Keshav Pingali}.}
  \bibinfo{year}{2015}\natexlab{}.
\newblock \showarticletitle{Stochastic Gradient Descent on {GPU}s}. In
  \bibinfo{booktitle}{{\em Proceedings of the 8th Workshop on General Purpose
  Processing Using GPUs}} {\em (\bibinfo{series}{GPGPU 2015})}.
  \bibinfo{publisher}{ACM}, \bibinfo{address}{New York, NY, USA},
  \bibinfo{pages}{81--89}.
\newblock
\showISBNx{978-1-4503-3407-5}
\showDOI{%
\url{http://dx.doi.org/10.1145/2716282.2716289}}


\bibitem[\protect\citeauthoryear{Keckler, Dally, Khailany, Garland, and
  Glasco}{Keckler et~al\mbox{.}}{2011}]%
        {Keckler:2011:GAT}
\bibfield{author}{\bibinfo{person}{Stephen~W. Keckler},
  \bibinfo{person}{William~J. Dally}, \bibinfo{person}{Brucek Khailany},
  \bibinfo{person}{Michael Garland}, {and} \bibinfo{person}{David Glasco}.}
  \bibinfo{year}{2011}\natexlab{}.
\newblock \showarticletitle{{GPU}s and the Future of Parallel Computing}.
\newblock \bibinfo{journal}{{\em IEEE Micro\/}} \bibinfo{volume}{31},
  \bibinfo{number}{5} (\bibinfo{date}{Sept.} \bibinfo{year}{2011}),
  \bibinfo{pages}{7--17}.
\newblock
\showDOI{%
\url{http://dx.doi.org/10.1109/MM.2011.89}}


\bibitem[\protect\citeauthoryear{Kepner, Aaltonen, Bader, Bulu\c{c},
  Franchetti, Gilbert, Hutchison, Kumar, Lumsdaine, Meyerhenke, McMillan,
  Moreira, Owens, Yang, Zalewski, and Mattson}{Kepner et~al\mbox{.}}{2016}]%
        {Kepner:2016:MFO}
\bibfield{author}{\bibinfo{person}{Jeremy Kepner}, \bibinfo{person}{Peter
  Aaltonen}, \bibinfo{person}{David Bader}, \bibinfo{person}{Ayd\i{}n
  Bulu\c{c}}, \bibinfo{person}{Franz Franchetti}, \bibinfo{person}{John
  Gilbert}, \bibinfo{person}{Dylan Hutchison}, \bibinfo{person}{Manoj Kumar},
  \bibinfo{person}{Andrew Lumsdaine}, \bibinfo{person}{Henning Meyerhenke},
  \bibinfo{person}{Scott McMillan}, \bibinfo{person}{Jose Moreira},
  \bibinfo{person}{John~D. Owens}, \bibinfo{person}{Carl Yang},
  \bibinfo{person}{Marcin Zalewski}, {and} \bibinfo{person}{Timothy Mattson}.}
  \bibinfo{year}{2016}\natexlab{}.
\newblock \showarticletitle{Mathematical Foundations of the {GraphBLAS}}. In
  \bibinfo{booktitle}{{\em Proceedings of the IEEE High Performance Extreme
  Computing Conference}}.
\newblock
\showDOI{%
\url{http://dx.doi.org/10.1109/HPEC.2016.7761646}}


\bibitem[\protect\citeauthoryear{Khorasani, Vora, Gupta, and Bhuyan}{Khorasani
  et~al\mbox{.}}{2014}]%
        {Khorasani:2014:CVG}
\bibfield{author}{\bibinfo{person}{Farzad Khorasani}, \bibinfo{person}{Keval
  Vora}, \bibinfo{person}{Rajiv Gupta}, {and} \bibinfo{person}{Laxmi~N.
  Bhuyan}.} \bibinfo{year}{2014}\natexlab{}.
\newblock \showarticletitle{{CuSha}: Vertex-centric Graph Processing on
  {GPUs}}. In \bibinfo{booktitle}{{\em Proceedings of the 23rd International
  Symposium on High-performance Parallel and Distributed Computing}} {\em
  (\bibinfo{series}{HPDC '14})}. \bibinfo{pages}{239--252}.
\newblock
\showISBNx{978-1-4503-2749-7}
\showDOI{%
\url{http://dx.doi.org/10.1145/2600212.2600227}}


\bibitem[\protect\citeauthoryear{Kwak, Lee, Park, and Moon}{Kwak
  et~al\mbox{.}}{2010}]%
        {Kwak:2010:WIT}
\bibfield{author}{\bibinfo{person}{Haewoon Kwak}, \bibinfo{person}{Changhyun
  Lee}, \bibinfo{person}{Hosung Park}, {and} \bibinfo{person}{Sue Moon}.}
  \bibinfo{year}{2010}\natexlab{}.
\newblock \showarticletitle{{W}hat is {T}witter, a social network or a news
  media?}. In \bibinfo{booktitle}{{\em Proceedings of the International
  Conference on the World Wide Web}}. \bibinfo{pages}{591--600}.
\newblock
\showISBNx{978-1-60558-799-8}
\showDOI{%
\url{http://dx.doi.org/10.1145/1772690.1772751}}


\bibitem[\protect\citeauthoryear{Kyrola, Blelloch, and Guestrin}{Kyrola
  et~al\mbox{.}}{2012}]%
        {Kyrola:2012:GLG}
\bibfield{author}{\bibinfo{person}{Aapo Kyrola}, \bibinfo{person}{Guy
  Blelloch}, {and} \bibinfo{person}{Carlos Guestrin}.}
  \bibinfo{year}{2012}\natexlab{}.
\newblock \showarticletitle{{GraphChi}: Large-scale Graph Computation on Just a
  {PC}}. In \bibinfo{booktitle}{{\em Proceedings of the 10th USENIX Conference
  on Operating Systems Design and Implementation}} {\em
  (\bibinfo{series}{OSDI'12})}. \bibinfo{publisher}{USENIX Association},
  \bibinfo{address}{Berkeley, CA, USA}, \bibinfo{pages}{31--46}.
\newblock
\showISBNx{978-1-931971-96-6}
\showURL{%
\url{http://dl.acm.org/citation.cfm?id=2387880.2387884}}


\bibitem[\protect\citeauthoryear{Leskovec}{Leskovec}{2016}]%
        {Leslovec:2014:SLN}
\bibfield{author}{\bibinfo{person}{Jure Leskovec}.}
  \bibinfo{year}{2009--2016}\natexlab{}.
\newblock \bibinfo{title}{{SNAP}: Stanford Large Network Dataset Collection}.
  (\bibinfo{year}{2009--2016}).
\newblock
\newblock
\shownote{\url{http://snap.stanford.edu/data/}.}


\bibitem[\protect\citeauthoryear{Liu and Huang}{Liu and Huang}{2015}]%
        {Liu:2015:EBG}
\bibfield{author}{\bibinfo{person}{Hang Liu} {and} \bibinfo{person}{H.~Howie
  Huang}.} \bibinfo{year}{2015}\natexlab{}.
\newblock \showarticletitle{Enterprise: Breadth-first Graph Traversal on
  {GPU}s}. In \bibinfo{booktitle}{{\em Proceedings of the International
  Conference for High Performance Computing, Networking, Storage and Analysis}}
  {\em (\bibinfo{series}{SC '15})}. \bibinfo{publisher}{ACM},
  \bibinfo{address}{New York, NY, USA}, Article \bibinfo{articleno}{68},
  \bibinfo{numpages}{12}~pages.
\newblock
\showISBNx{978-1-4503-3723-6}
\showDOI{%
\url{http://dx.doi.org/10.1145/2807591.2807594}}


\bibitem[\protect\citeauthoryear{Liu, Huang, and Hu}{Liu et~al\mbox{.}}{2016}]%
        {Liu:2016:ICB}
\bibfield{author}{\bibinfo{person}{Hang Liu}, \bibinfo{person}{H.~Howie Huang},
  {and} \bibinfo{person}{Yang Hu}.} \bibinfo{year}{2016}\natexlab{}.
\newblock \showarticletitle{i{BFS}: Concurrent Breadth-First Search on {GPU}s}.
  In \bibinfo{booktitle}{{\em Proceedings of the 2016 International Conference
  on Management of Data}} {\em (\bibinfo{series}{SIGMOD '16})}.
  \bibinfo{publisher}{ACM}, \bibinfo{address}{New York, NY, USA},
  \bibinfo{pages}{403--416}.
\newblock
\showISBNx{978-1-4503-3531-7}
\showDOI{%
\url{http://dx.doi.org/10.1145/2882903.2882959}}


\bibitem[\protect\citeauthoryear{Low, Gonzalez, Kyrola, Bickson, Guestrin, and
  Hellerstein}{Low et~al\mbox{.}}{2010}]%
        {Low:2010:GAN}
\bibfield{author}{\bibinfo{person}{Yucheng Low}, \bibinfo{person}{Joseph
  Gonzalez}, \bibinfo{person}{Aapo Kyrola}, \bibinfo{person}{Danny Bickson},
  \bibinfo{person}{Carlos Guestrin}, {and} \bibinfo{person}{Joseph~M.
  Hellerstein}.} \bibinfo{year}{2010}\natexlab{}.
\newblock \showarticletitle{{GraphLab}: A New Parallel Framework for Machine
  Learning}. In \bibinfo{booktitle}{{\em Proceedings of the Twenty-Sixth Annual
  Conference on Uncertainty in Artificial Intelligence}} {\em
  (\bibinfo{series}{UAI-10})}. \bibinfo{pages}{340--349}.
\newblock


\bibitem[\protect\citeauthoryear{Malewicz, Austern, Bik, Dehnert, Horn, Leiser,
  and Czajkowski}{Malewicz et~al\mbox{.}}{2010}]%
        {Malewicz:2010:PSL}
\bibfield{author}{\bibinfo{person}{Grzegorz Malewicz},
  \bibinfo{person}{Matthew~H. Austern}, \bibinfo{person}{Aart J.~C. Bik},
  \bibinfo{person}{James~C. Dehnert}, \bibinfo{person}{Ilan Horn},
  \bibinfo{person}{Naty Leiser}, {and} \bibinfo{person}{Grzegorz Czajkowski}.}
  \bibinfo{year}{2010}\natexlab{}.
\newblock \showarticletitle{Pregel: A System for Large-scale Graph Processing}.
  In \bibinfo{booktitle}{{\em Proceedings of the 2010 ACM SIGMOD International
  Conference on Management of Data}} {\em (\bibinfo{series}{SIGMOD '10})}.
  \bibinfo{pages}{135--146}.
\newblock
\showDOI{%
\url{http://dx.doi.org/10.1145/1807167.1807184}}


\bibitem[\protect\citeauthoryear{McColl, Ediger, Poovey, Campbell, and
  Bader}{McColl et~al\mbox{.}}{2014}]%
        {McColl:2014:APE}
\bibfield{author}{\bibinfo{person}{Robert~Campbell McColl},
  \bibinfo{person}{David Ediger}, \bibinfo{person}{Jason Poovey},
  \bibinfo{person}{Dan Campbell}, {and} \bibinfo{person}{David~A. Bader}.}
  \bibinfo{year}{2014}\natexlab{}.
\newblock \showarticletitle{A Performance Evaluation of Open Source Graph
  Databases}. In \bibinfo{booktitle}{{\em Proceedings of the First Workshop on
  Parallel Programming for Analytics Applications}} {\em (\bibinfo{series}{PPAA
  '14})}. \bibinfo{pages}{11--18}.
\newblock
\showDOI{%
\url{http://dx.doi.org/10.1145/2567634.2567638}}


\bibitem[\protect\citeauthoryear{McLaughlin and Bader}{McLaughlin and
  Bader}{2014}]%
        {McLaughlin:2014:SAH}
\bibfield{author}{\bibinfo{person}{Adam McLaughlin} {and}
  \bibinfo{person}{David~A. Bader}.} \bibinfo{year}{2014}\natexlab{}.
\newblock \showarticletitle{Scalable and High Performance Betweenness
  Centrality on the {GPU}}. In \bibinfo{booktitle}{{\em Proceedings of the
  International Conference for High Performance Computing, Networking, Storage
  and Analysis}} {\em (\bibinfo{series}{SC14})}. \bibinfo{pages}{572--583}.
\newblock
\showISBNx{978-1-4799-5500-8}
\showDOI{%
\url{http://dx.doi.org/10.1109/SC.2014.52}}


\bibitem[\protect\citeauthoryear{McLaughlin and Bader}{McLaughlin and
  Bader}{2015}]%
        {McLaughlin:2015:FES}
\bibfield{author}{\bibinfo{person}{A. McLaughlin} {and} \bibinfo{person}{D.~A.
  Bader}.} \bibinfo{year}{2015}\natexlab{}.
\newblock \showarticletitle{Fast Execution of Simultaneous Breadth-First
  Searches on Sparse Graphs}. In \bibinfo{booktitle}{{\em 2015 IEEE 21st
  International Conference on Parallel and Distributed Systems (ICPADS)}}.
  \bibinfo{pages}{9--18}.
\newblock
\showDOI{%
\url{http://dx.doi.org/10.1109/ICPADS.2015.10}}


\bibitem[\protect\citeauthoryear{McLaughlin, Riedy, and Bader}{McLaughlin
  et~al\mbox{.}}{2015}]%
        {McLaughlin:2015:AFE}
\bibfield{author}{\bibinfo{person}{A. McLaughlin}, \bibinfo{person}{J. Riedy},
  {and} \bibinfo{person}{D.~A. Bader}.} \bibinfo{year}{2015}\natexlab{}.
\newblock \showarticletitle{A fast, energy-efficient abstraction for
  simultaneous breadth-first searches}. In \bibinfo{booktitle}{{\em 2015 IEEE
  High Performance Extreme Computing Conference (HPEC)}}.
  \bibinfo{pages}{1--6}.
\newblock
\showDOI{%
\url{http://dx.doi.org/10.1109/HPEC.2015.7322466}}


\bibitem[\protect\citeauthoryear{Merrill, Garland, and Grimshaw}{Merrill
  et~al\mbox{.}}{2012}]%
        {Merrill:2012:SGG}
\bibfield{author}{\bibinfo{person}{Duane Merrill}, \bibinfo{person}{Michael
  Garland}, {and} \bibinfo{person}{Andrew Grimshaw}.}
  \bibinfo{year}{2012}\natexlab{}.
\newblock \showarticletitle{Scalable {GPU} Graph Traversal}. In
  \bibinfo{booktitle}{{\em Proceedings of the 17th ACM SIGPLAN Symposium on
  Principles and Practice of Parallel Programming}} {\em
  (\bibinfo{series}{PPoPP '12})}. \bibinfo{pages}{117--128}.
\newblock
\showDOI{%
\url{http://dx.doi.org/10.1145/2145816.2145832}}


\bibitem[\protect\citeauthoryear{Meyer and Sanders}{Meyer and Sanders}{2003}]%
        {Meyer:2003:DAP}
\bibfield{author}{\bibinfo{person}{U. Meyer} {and} \bibinfo{person}{P.
  Sanders}.} \bibinfo{year}{2003}\natexlab{}.
\newblock \showarticletitle{{$\Delta$}-stepping: a parallelizable shortest path
  algorithm}.
\newblock \bibinfo{journal}{{\em Journal of Algorithms\/}}
  \bibinfo{volume}{49}, \bibinfo{number}{1} (\bibinfo{date}{Oct.}
  \bibinfo{year}{2003}), \bibinfo{pages}{114--152}.
\newblock
\showDOI{%
\url{http://dx.doi.org/10.1016/S0196-6774(03)00076-2}}
\newblock
\shownote{1998 European Symposium on Algorithms.}


\bibitem[\protect\citeauthoryear{Nguyen, Lenharth, and Pingali}{Nguyen
  et~al\mbox{.}}{2013}]%
        {Nguyen:2013:ALI}
\bibfield{author}{\bibinfo{person}{Donald Nguyen}, \bibinfo{person}{Andrew
  Lenharth}, {and} \bibinfo{person}{Keshav Pingali}.}
  \bibinfo{year}{2013}\natexlab{}.
\newblock \showarticletitle{A Lightweight Infrastructure for Graph Analytics}.
  In \bibinfo{booktitle}{{\em Proceedings of ACM Symposium on Operating Systems
  Principles}} {\em (\bibinfo{series}{SOSP '13})}. \bibinfo{pages}{456--471}.
\newblock
\showDOI{%
\url{http://dx.doi.org/10.1145/2517349.2522739}}


\bibitem[\protect\citeauthoryear{Pai and Pingali}{Pai and Pingali}{2016}]%
        {Pai:2016:ACT}
\bibfield{author}{\bibinfo{person}{Sreepathi Pai} {and} \bibinfo{person}{Keshav
  Pingali}.} \bibinfo{year}{2016}\natexlab{}.
\newblock \showarticletitle{A Compiler for Throughput Optimization of Graph
  Algorithms on {GPU}s}.
\newblock \bibinfo{journal}{{\em SIGPLAN Not.\/}} \bibinfo{volume}{51},
  \bibinfo{number}{10} (\bibinfo{date}{Oct.} \bibinfo{year}{2016}),
  \bibinfo{pages}{1--19}.
\newblock
\showISSN{0362-1340}
\showDOI{%
\url{http://dx.doi.org/10.1145/3022671.2984015}}


\bibitem[\protect\citeauthoryear{Pan, Wang, Wu, Yang, and Owens}{Pan
  et~al\mbox{.}}{2016}]%
        {Pan:2016:MGA}
\bibfield{author}{\bibinfo{person}{Yuechao Pan}, \bibinfo{person}{Yangzihao
  Wang}, \bibinfo{person}{Yuduo Wu}, \bibinfo{person}{Carl Yang}, {and}
  \bibinfo{person}{John~D. Owens}.} \bibinfo{year}{2016}\natexlab{}.
\newblock \showarticletitle{{M}ulti-{GPU} Graph Analytics}.
\newblock \bibinfo{journal}{{\em CoRR\/}} \bibinfo{volume}{abs/1504.04804},
  \bibinfo{number}{1504.04804v3} (\bibinfo{date}{April} \bibinfo{year}{2016}).
\newblock
\showeprint[arxiv]{cs.DC/1504.04804v3}


\bibitem[\protect\citeauthoryear{Pande and Bader}{Pande and Bader}{2011}]%
        {Pande:2011:CBC}
\bibfield{author}{\bibinfo{person}{Pushkar~R. Pande} {and}
  \bibinfo{person}{David~A. Bader}.} \bibinfo{year}{2011}\natexlab{}.
\newblock \showarticletitle{Computing Betweenness Centrality for Small World
  Networks on a {GPU}}. In \bibinfo{booktitle}{{\em 2011 IEEE Conference on
  High Performance Embedded Computing}}.
\newblock


\bibitem[\protect\citeauthoryear{Pearce, Gokhale, and Amato}{Pearce
  et~al\mbox{.}}{2014}]%
        {Pearce:2014:FPT}
\bibfield{author}{\bibinfo{person}{Roger Pearce}, \bibinfo{person}{Maya
  Gokhale}, {and} \bibinfo{person}{Nancy~M. Amato}.}
  \bibinfo{year}{2014}\natexlab{}.
\newblock \showarticletitle{Faster Parallel Traversal of Scale Free Graphs at
  Extreme Scale with Vertex Delegates}. In \bibinfo{booktitle}{{\em Proceedings
  of the International Conference for High Performance Computing, Networking,
  Storage and Analysis}} {\em (\bibinfo{series}{SC '14})}.
  \bibinfo{publisher}{IEEE Press}, \bibinfo{address}{Piscataway, NJ, USA},
  \bibinfo{pages}{549--559}.
\newblock
\showISBNx{978-1-4799-5500-8}
\showDOI{%
\url{http://dx.doi.org/10.1109/SC.2014.50}}


\bibitem[\protect\citeauthoryear{Pingali, Nguyen, Kulkarni, Burtscher, Hassaan,
  Kaleem, Lee, Lenharth, Manevich, M{\'e}ndez-Lojo, Prountzos, and Sui}{Pingali
  et~al\mbox{.}}{2011}]%
        {Pingali:2011:TTO}
\bibfield{author}{\bibinfo{person}{Keshav Pingali}, \bibinfo{person}{Donald
  Nguyen}, \bibinfo{person}{Milind Kulkarni}, \bibinfo{person}{Martin
  Burtscher}, \bibinfo{person}{M.~Amber Hassaan}, \bibinfo{person}{Rashid
  Kaleem}, \bibinfo{person}{Tsung-Hsien Lee}, \bibinfo{person}{Andrew
  Lenharth}, \bibinfo{person}{Roman Manevich}, \bibinfo{person}{Mario
  M{\'e}ndez-Lojo}, \bibinfo{person}{Dimitrios Prountzos}, {and}
  \bibinfo{person}{Xin Sui}.} \bibinfo{year}{2011}\natexlab{}.
\newblock \showarticletitle{The Tao of Parallelism in Algorithms}. In
  \bibinfo{booktitle}{{\em Proceedings of the 32nd ACM SIGPLAN Conference on
  Programming Language Design and Implementation}} {\em (\bibinfo{series}{PLDI
  '11})}. \bibinfo{pages}{12--25}.
\newblock
\showDOI{%
\url{http://dx.doi.org/10.1145/1993498.1993501}}


\bibitem[\protect\citeauthoryear{Polak}{Polak}{2015}]%
        {Polak:2015:CTL}
\bibfield{author}{\bibinfo{person}{Adam Polak}.}
  \bibinfo{year}{2015}\natexlab{}.
\newblock \showarticletitle{Counting Triangles in Large Graphs on {GPU}}.
\newblock \bibinfo{journal}{{\em CoRR\/}}  \bibinfo{volume}{abs/1503.00576}
  (\bibinfo{year}{2015}).
\newblock
\showURL{%
\url{http://arxiv.org/abs/1503.00576}}


\bibitem[\protect\citeauthoryear{Roy, Mihailovic, and Zwaenepoel}{Roy
  et~al\mbox{.}}{2013}]%
        {Roy:2013:XEG}
\bibfield{author}{\bibinfo{person}{Amitabha Roy}, \bibinfo{person}{Ivo
  Mihailovic}, {and} \bibinfo{person}{Willy Zwaenepoel}.}
  \bibinfo{year}{2013}\natexlab{}.
\newblock \showarticletitle{X-{S}tream: Edge-centric Graph Processing using
  Streaming Partitions}. In \bibinfo{booktitle}{{\em Proceedings of the
  Twenty-Fourth ACM Symposium on Operating Systems Principles}}. ACM,
  \bibinfo{pages}{472--488}.
\newblock


\bibitem[\protect\citeauthoryear{Salihoglu and Widom}{Salihoglu and
  Widom}{2014}]%
        {Salihoglu:2014:HHP}
\bibfield{author}{\bibinfo{person}{Semih Salihoglu} {and}
  \bibinfo{person}{Jennifer Widom}.} \bibinfo{year}{2014}\natexlab{}.
\newblock \showarticletitle{{HelP}: High-level Primitives For Large-Scale Graph
  Processing}. In \bibinfo{booktitle}{{\em Proceedings of the Workshop on GRAph
  Data Management Experiences and Systems}} {\em (\bibinfo{series}{GRADES
  '14})}. Article \bibinfo{articleno}{3}, \bibinfo{numpages}{6}~pages.
\newblock
\showDOI{%
\url{http://dx.doi.org/10.1145/2621934.2621938}}


\bibitem[\protect\citeauthoryear{Sallinen, Gharaibeh, and Ripeanu}{Sallinen
  et~al\mbox{.}}{2015}]%
        {Sallinen:2015:ADB}
\bibfield{author}{\bibinfo{person}{Scott Sallinen}, \bibinfo{person}{Abdullah
  Gharaibeh}, {and} \bibinfo{person}{Matei Ripeanu}.}
  \bibinfo{year}{2015}\natexlab{}.
\newblock \showarticletitle{Accelerating Direction-Optimized Breadth First
  Search on Hybrid Architectures}.
\newblock \bibinfo{journal}{{\em CoRR\/}} \bibinfo{volume}{abs/1503.04359},
  \bibinfo{number}{1503.04359v1} (\bibinfo{date}{March} \bibinfo{year}{2015}).
\newblock
\showeprint[arxiv]{cs.DC/1503.04359v1}


\bibitem[\protect\citeauthoryear{Sariy\"{u}ce, Kaya, Saule, and
  \c{C}ataly\"{u}rek}{Sariy\"{u}ce et~al\mbox{.}}{2013}]%
        {Sariyuce:2013:BCO}
\bibfield{author}{\bibinfo{person}{Ahmet~Erdem Sariy\"{u}ce},
  \bibinfo{person}{Kamer Kaya}, \bibinfo{person}{Erik Saule}, {and}
  \bibinfo{person}{\"{U}mit~V. \c{C}ataly\"{u}rek}.}
  \bibinfo{year}{2013}\natexlab{}.
\newblock \showarticletitle{Betweenness Centrality on {GPU}s and Heterogeneous
  Architectures}. In \bibinfo{booktitle}{{\em Proceedings of the 6th Workshop
  on General Purpose Processor Using Graphics Processing Units}} {\em
  (\bibinfo{series}{GPGPU-6})}. \bibinfo{pages}{76--85}.
\newblock
\showDOI{%
\url{http://dx.doi.org/10.1145/2458523.2458531}}


\bibitem[\protect\citeauthoryear{Schank and Wagner}{Schank and Wagner}{2005}]%
        {Schank:2005:FCL}
\bibfield{author}{\bibinfo{person}{Thomas Schank} {and}
  \bibinfo{person}{Dorothea Wagner}.} \bibinfo{year}{2005}\natexlab{}.
\newblock \showarticletitle{Finding, Counting and Listing All Triangles in
  Large Graphs, an Experimental Study}. In \bibinfo{booktitle}{{\em Proceedings
  of the 4th International Conference on Experimental and Efficient
  Algorithms}} {\em (\bibinfo{series}{WEA'05})}. \bibinfo{pages}{606--609}.
\newblock
\showISBNx{3-540-25920-1, 978-3-540-25920-6}
\showDOI{%
\url{http://dx.doi.org/10.1007/11427186_54}}


\bibitem[\protect\citeauthoryear{Seo, Kim, and Kim}{Seo et~al\mbox{.}}{2015}]%
        {Seo:2015:GGS}
\bibfield{author}{\bibinfo{person}{Hyunseok Seo}, \bibinfo{person}{Jinwook
  Kim}, {and} \bibinfo{person}{Min-Soo Kim}.} \bibinfo{year}{2015}\natexlab{}.
\newblock \showarticletitle{{GS}tream: A Graph Streaming Processing Method for
  Large-scale Graphs on {GPU}s}. In \bibinfo{booktitle}{{\em Proceedings of the
  20th ACM SIGPLAN Symposium on Principles and Practice of Parallel
  Programming}} {\em (\bibinfo{series}{PPoPP 2015})}. \bibinfo{publisher}{ACM},
  \bibinfo{address}{New York, NY, USA}, \bibinfo{pages}{253--254}.
\newblock
\showISBNx{978-1-4503-3205-7}
\showDOI{%
\url{http://dx.doi.org/10.1145/2688500.2688526}}


\bibitem[\protect\citeauthoryear{Shi, Liang, Di, He, Jin, Lu, Wang, Luo, and
  Zhong}{Shi et~al\mbox{.}}{2015}]%
        {Shi:2015:OAG}
\bibfield{author}{\bibinfo{person}{Xuanhua Shi}, \bibinfo{person}{Junling
  Liang}, \bibinfo{person}{Sheng Di}, \bibinfo{person}{Bingsheng He},
  \bibinfo{person}{Hai Jin}, \bibinfo{person}{Lu Lu}, \bibinfo{person}{Zhixiang
  Wang}, \bibinfo{person}{Xuan Luo}, {and} \bibinfo{person}{Jianlong Zhong}.}
  \bibinfo{year}{2015}\natexlab{}.
\newblock \showarticletitle{Optimization of Asynchronous Graph Processing on
  {GPU} with Hybrid Coloring Model}. In \bibinfo{booktitle}{{\em Proceedings of
  the 20th ACM SIGPLAN Symposium on Principles and Practice of Parallel
  Programming}} {\em (\bibinfo{series}{PPoPP 2015})}.
  \bibinfo{pages}{271--272}.
\newblock
\showISBNx{978-1-4503-3205-7}
\showDOI{%
\url{http://dx.doi.org/10.1145/2688500.2688542}}


\bibitem[\protect\citeauthoryear{Shun and Blelloch}{Shun and Blelloch}{2013}]%
        {Shun:2013:LAL}
\bibfield{author}{\bibinfo{person}{Julian Shun} {and} \bibinfo{person}{Guy~E.
  Blelloch}.} \bibinfo{year}{2013}\natexlab{}.
\newblock \showarticletitle{Ligra: a lightweight graph processing framework for
  shared memory}. In \bibinfo{booktitle}{{\em Proceedings of the 18th ACM
  SIGPLAN Symposium on Principles and Practice of Parallel Programming}} {\em
  (\bibinfo{series}{PPoPP '13})}. \bibinfo{pages}{135--146}.
\newblock
\showDOI{%
\url{http://dx.doi.org/10.1145/2442516.2442530}}


\bibitem[\protect\citeauthoryear{Shun and Tangwongsan}{Shun and
  Tangwongsan}{2015}]%
        {Shun:2015:MTC}
\bibfield{author}{\bibinfo{person}{J. Shun} {and} \bibinfo{person}{K.
  Tangwongsan}.} \bibinfo{year}{2015}\natexlab{}.
\newblock \showarticletitle{Multicore Triangle Computations Without Tuning}. In
  \bibinfo{booktitle}{{\em IEEE 31st International Conference on Data
  Engineering}}. \bibinfo{pages}{149--160}.
\newblock
\showISSN{1063-6382}
\showDOI{%
\url{http://dx.doi.org/10.1109/ICDE.2015.7113280}}


\bibitem[\protect\citeauthoryear{Siek, Lee, and Lumsdaine}{Siek
  et~al\mbox{.}}{2001}]%
        {Siek:2001:TBG}
\bibfield{author}{\bibinfo{person}{Jeremy~G. Siek}, \bibinfo{person}{Lie-Quan
  Lee}, {and} \bibinfo{person}{Andrew Lumsdaine}.}
  \bibinfo{year}{2001}\natexlab{}.
\newblock \bibinfo{booktitle}{{\em The {B}oost Graph Library: User Guide and
  Reference Manual}}.
\newblock \bibinfo{publisher}{Addison-Wesley}.
\newblock


\bibitem[\protect\citeauthoryear{Slota, Rajamanickam, and Madduri}{Slota
  et~al\mbox{.}}{2014}]%
        {Slota:2014:BCP}
\bibfield{author}{\bibinfo{person}{G.~M. Slota}, \bibinfo{person}{S.
  Rajamanickam}, {and} \bibinfo{person}{K. Madduri}.}
  \bibinfo{year}{2014}\natexlab{}.
\newblock \showarticletitle{{BFS} and Coloring-Based Parallel Algorithms for
  Strongly Connected Components and Related Problems}. In
  \bibinfo{booktitle}{{\em 2014 IEEE 28th International Parallel and
  Distributed Processing Symposium}}. \bibinfo{pages}{550--559}.
\newblock
\showISSN{1530-2075}
\showDOI{%
\url{http://dx.doi.org/10.1109/IPDPS.2014.64}}


\bibitem[\protect\citeauthoryear{Soman, Kishore, and Narayanan}{Soman
  et~al\mbox{.}}{2010}]%
        {Soman:2010:AFG}
\bibfield{author}{\bibinfo{person}{Jyothish Soman}, \bibinfo{person}{Kothapalli
  Kishore}, {and} \bibinfo{person}{P~J Narayanan}.}
  \bibinfo{year}{2010}\natexlab{}.
\newblock \showarticletitle{A Fast {GPU} Algorithm for Graph Connectivity}. In
  \bibinfo{booktitle}{{\em 24th IEEE International Symposium on Parallel and
  Distributed Processing, Workshops and PhD Forum}} {\em
  (\bibinfo{series}{IPDPSW 2010})}. \bibinfo{pages}{1--8}.
\newblock
\showDOI{%
\url{http://dx.doi.org/10.1109/IPDPSW.2010.5470817}}


\bibitem[\protect\citeauthoryear{Sun, Wang, Wang, Shao, and Li}{Sun
  et~al\mbox{.}}{2012}]%
        {Sun:2012:ESM}
\bibfield{author}{\bibinfo{person}{Zhao Sun}, \bibinfo{person}{Hongzhi Wang},
  \bibinfo{person}{Haixun Wang}, \bibinfo{person}{Bin Shao}, {and}
  \bibinfo{person}{Jianzhong Li}.} \bibinfo{year}{2012}\natexlab{}.
\newblock \showarticletitle{Efficient Subgraph Matching on Billion Node
  Graphs}.
\newblock \bibinfo{journal}{{\em Proc. VLDB Endow.\/}} \bibinfo{volume}{5},
  \bibinfo{number}{9} (\bibinfo{date}{May} \bibinfo{year}{2012}),
  \bibinfo{pages}{788--799}.
\newblock
\showISSN{2150-8097}
\showDOI{%
\url{http://dx.doi.org/10.14778/2311906.2311907}}


\bibitem[\protect\citeauthoryear{Tran, Kim, and He}{Tran et~al\mbox{.}}{2015}]%
        {Renz:2015:FSM}
\bibfield{author}{\bibinfo{person}{Ha-Nguyen Tran}, \bibinfo{person}{Jung-jae
  Kim}, {and} \bibinfo{person}{Bingsheng He}.} \bibinfo{year}{2015}\natexlab{}.
\newblock \showarticletitle{Fast Subgraph Matching on Large Graphs using
  Graphics Processors}.
\newblock In \bibinfo{booktitle}{{\em Database Systems for Advanced
  Applications}}, \bibfield{editor}{\bibinfo{person}{Matthias Renz},
  \bibinfo{person}{Cyrus Shahabi}, \bibinfo{person}{Xiaofang Zhou}, {and}
  \bibinfo{person}{Muhammad~Aamir Cheema}} (Eds.). \bibinfo{series}{Lecture
  Notes in Computer Science}, Vol.~\bibinfo{volume}{9049}.
  \bibinfo{publisher}{Springer International Publishing},
  \bibinfo{pages}{299--315}.
\newblock
\showISBNx{978-3-319-18119-6}
\showDOI{%
\url{http://dx.doi.org/10.1007/978-3-319-18120-2_18}}


\bibitem[\protect\citeauthoryear{Tzeng, Lloyd, and Owens}{Tzeng
  et~al\mbox{.}}{2012}]%
        {Tzeng:2012:AGT}
\bibfield{author}{\bibinfo{person}{Stanley Tzeng}, \bibinfo{person}{Brandon
  Lloyd}, {and} \bibinfo{person}{John~D. Owens}.}
  \bibinfo{year}{2012}\natexlab{}.
\newblock \showarticletitle{A {GPU} Task-Parallel Model with Dependency
  Resolution}.
\newblock \bibinfo{journal}{{\em IEEE Computer\/}} \bibinfo{volume}{45},
  \bibinfo{number}{8} (\bibinfo{date}{Aug.} \bibinfo{year}{2012}),
  \bibinfo{pages}{34--41}.
\newblock
\showDOI{%
\url{http://dx.doi.org/10.1109/MC.2012.255}}


\bibitem[\protect\citeauthoryear{Wang, Xie, Demers, and Gehrke}{Wang
  et~al\mbox{.}}{2013}]%
        {Wang:2013:ALS}
\bibfield{author}{\bibinfo{person}{Guozhang Wang}, \bibinfo{person}{Wenlei
  Xie}, \bibinfo{person}{Alan~J. Demers}, {and} \bibinfo{person}{Johannes
  Gehrke}.} \bibinfo{year}{2013}\natexlab{}.
\newblock \showarticletitle{Asynchronous Large-Scale Graph Processing Made
  Easy}. In \bibinfo{booktitle}{{\em CIDR}}.
  \bibinfo{publisher}{www.cidrdb.org}.
\newblock


\bibitem[\protect\citeauthoryear{Wang, Wang, Yang, and Owens}{Wang
  et~al\mbox{.}}{2016b}]%
        {Wang:2016:ACS}
\bibfield{author}{\bibinfo{person}{Leyuan Wang}, \bibinfo{person}{Yangzihao
  Wang}, \bibinfo{person}{Carl Yang}, {and} \bibinfo{person}{John~D. Owens}.}
  \bibinfo{year}{2016}\natexlab{b}.
\newblock \showarticletitle{A Comparative Study on Exact Triangle Counting
  Algorithms on the {GPU}}. In \bibinfo{booktitle}{{\em Proceedings of the 1st
  High Performance Graph Processing Workshop}} {\em (\bibinfo{series}{HPGP
  '16})}. \bibinfo{pages}{1--8}.
\newblock
\showDOI{%
\url{http://dx.doi.org/10.1145/2915516.2915521}}


\bibitem[\protect\citeauthoryear{Wang, Davidson, Pan, Wu, Riffel, and
  Owens}{Wang et~al\mbox{.}}{2016a}]%
        {Wang:2016:GAH}
\bibfield{author}{\bibinfo{person}{Yangzihao Wang}, \bibinfo{person}{Andrew
  Davidson}, \bibinfo{person}{Yuechao Pan}, \bibinfo{person}{Yuduo Wu},
  \bibinfo{person}{Andy Riffel}, {and} \bibinfo{person}{John~D. Owens}.}
  \bibinfo{year}{2016}\natexlab{a}.
\newblock \showarticletitle{{G}unrock: A High-Performance Graph Processing
  Library on the {GPU}}. In \bibinfo{booktitle}{{\em Proceedings of the 21st
  ACM SIGPLAN Symposium on Principles and Practice of Parallel Programming}}
  {\em (\bibinfo{series}{PPoPP 2016})}. \bibinfo{pages}{11:1--11:12}.
\newblock
\showDOI{%
\url{http://dx.doi.org/10.1145/2851141.2851145}}


\bibitem[\protect\citeauthoryear{Wu, Zhao, Zhang, Jiang, and Shen}{Wu
  et~al\mbox{.}}{2013}]%
        {Wu:2013:CAA}
\bibfield{author}{\bibinfo{person}{Bo Wu}, \bibinfo{person}{Zhijia Zhao},
  \bibinfo{person}{Eddy~Zheng Zhang}, \bibinfo{person}{Yunlian Jiang}, {and}
  \bibinfo{person}{Xipeng Shen}.} \bibinfo{year}{2013}\natexlab{}.
\newblock \showarticletitle{Complexity Analysis and Algorithm Design for
  Reorganizing Data to Minimize Non-coalesced Memory Accesses on GPU}. In
  \bibinfo{booktitle}{{\em Proceedings of the 18th ACM SIGPLAN Symposium on
  Principles and Practice of Parallel Programming}} {\em
  (\bibinfo{series}{PPoPP '13})}. \bibinfo{publisher}{ACM},
  \bibinfo{address}{New York, NY, USA}, \bibinfo{pages}{57--68}.
\newblock
\showISBNx{978-1-4503-1922-5}
\showDOI{%
\url{http://dx.doi.org/10.1145/2442516.2442523}}


\bibitem[\protect\citeauthoryear{Wu, Wang, Pan, Yang, and Owens}{Wu
  et~al\mbox{.}}{2015}]%
        {Wu:2015:PCF}
\bibfield{author}{\bibinfo{person}{Yuduo Wu}, \bibinfo{person}{Yangzihao Wang},
  \bibinfo{person}{Yuechao Pan}, \bibinfo{person}{Carl Yang}, {and}
  \bibinfo{person}{John~D. Owens}.} \bibinfo{year}{2015}\natexlab{}.
\newblock \showarticletitle{Performance Characterization for High-Level
  Programming Models for {GPU} Graph Analytics}. In \bibinfo{booktitle}{{\em
  IEEE International Symposium on Workload Characterization}} {\em
  (\bibinfo{series}{IISWC-2015})}. \bibinfo{pages}{66--75}.
\newblock
\showDOI{%
\url{http://dx.doi.org/10.1109/IISWC.2015.13}}


\bibitem[\protect\citeauthoryear{Yang, Wang, and Owens}{Yang
  et~al\mbox{.}}{2015}]%
        {Yang:2015:FSM}
\bibfield{author}{\bibinfo{person}{Carl Yang}, \bibinfo{person}{Yangzihao
  Wang}, {and} \bibinfo{person}{John~D. Owens}.}
  \bibinfo{year}{2015}\natexlab{}.
\newblock \showarticletitle{Fast Sparse Matrix and Sparse Vector Multiplication
  Algorithm on the {GPU}}. In \bibinfo{booktitle}{{\em Graph Algorithms
  Building Blocks}} {\em (\bibinfo{series}{GABB 2015})}.
  \bibinfo{pages}{841--847}.
\newblock
\showDOI{%
\url{http://dx.doi.org/10.1109/IPDPSW.2015.77}}


\bibitem[\protect\citeauthoryear{Zhong and He}{Zhong and He}{2014}]%
        {Zhong:2014:MSG}
\bibfield{author}{\bibinfo{person}{Jianlong Zhong} {and}
  \bibinfo{person}{Bingsheng He}.} \bibinfo{year}{2014}\natexlab{}.
\newblock \showarticletitle{Medusa: Simplified Graph Processing on {GPU}s}.
\newblock \bibinfo{journal}{{\em IEEE Transactions on Parallel and Distributed
  Systems\/}} \bibinfo{volume}{25}, \bibinfo{number}{6} (\bibinfo{date}{June}
  \bibinfo{year}{2014}), \bibinfo{pages}{1543--1552}.
\newblock
\showDOI{%
\url{http://dx.doi.org/10.1109/TPDS.2013.111}}


\end{thebibliography}

\end{document}